\documentclass[fleqn,usenatbib,useAMS]{mnras}

\usepackage{graphicx}	
\usepackage{amsmath}	
\usepackage{amssymb}	
\usepackage{multicol}        
\usepackage{bm}		
\usepackage{pdflscape}	
\usepackage{caption}
\usepackage[table]{xcolor}

\usepackage[T1]{fontenc}
\usepackage{ae,aecompl}

\usepackage{newtxtext,newtxmath}

\title[MHD Torus]
{Evolution of MHD Torus and Mass Outflow Around Spinning AGN}

\author[ Aktar, Pan, \& Okuda]{Ramiz Aktar$^{1}$ \thanks{Contact e-mail: \href{ramizaktar@gmail.com}{ramizaktar@gmail.com}}, Kuo-Chuan Pan$^1$$^2$$^3$, and Toru Okuda$^4$%
\\
$^{1}${Department of Physics and Institute of Astronomy, National Tsing Hua University, 30013 Hsinchu, Taiwan}\\
$^{2}${Center for Theory and Computation, National Tsing Hua University, Hsinchu 30013, Taiwan}\\
$^{3}${Physics Division, National Center for Theoretical Sciences, National Taiwan University, Taipei 10617, Taiwan}\\
$^{4}${Hakodate Campus, Hokkaido University of Education, Hachiman-Cho 1-2, Hakodate
040-8567, Japan}
}

\pubyear{2022}

\begin{document}
\label{firstpage}
\pagerange{\pageref{firstpage}--\pageref{lastpage}}
\maketitle

\begin{abstract}

We perform axisymmetric, two-dimensional magnetohydrodynamic (MHD) simulations to investigate accretion flows around spinning AGN. To mimic the space-time geometry of spinning black holes, we consider effective Kerr potential, and the mass of the black holes is $10^8 M_{\odot}$. We initialize the accretion disc with a magnetized torus by adopting the toroidal component of the magnetic vector potential. The initial magnetic field strength is set by using the plasma beta parameter ($\beta_0$). We observe self-consistent turbulence generated by magneto rotational instability (MRI) in the disc. The MRI turbulence transports angular momentum in the disc, resulting in an angular momentum distribution that approaches a Keplerian distribution. We investigate the effect of the magnetic field on the dynamics of the torus and associated mass outflow from the disc around a maximally spinning black hole $(a_k = 0.99)$. For the purpose of our analysis, we investigate the magnetic state of our simulation model. The model $\beta_0 = 10$ indicates the behaviour similar to the ``magnetically arrested disk (MAD)'' state, and all the other low magnetic model remains in the SANE state. We observe that mass outflow rates are significantly enhanced with the increased magnetic field in the disc. We find a positive correlation between the magnetic field and mass outflow rates. We also investigate the effect of black hole spin on the magnetized torus evolution. However, we have not found any significant effect of black hole spin on mass outflows in our model. Finally, we discuss the possible astrophysical applications of our simulation results.

\end{abstract}



\begingroup
\let\clearpage\relax
\endgroup
\newpage

\section{Introduction} \label{sec:intro}

Active galactic nuclei (AGNs) are ubiquitously observed at every galaxy's center. In general, accretion on to AGN may produce highly radiative luminous events across the entire electromagnetic spectrum. However, AGN does not always show high radiative emission, and surprisingly they spend most of the time in a quiescence state. Based on the observed optical/UV spectral properties, AGNs are classified as Type 1 and Type 2. Type 1 AGNs shows both broad band ($\gtrsim 1000$ km s$^{-1}$) and narrow band ($\lesssim 1000$ km s$^{-1}$) emission lines in optical/UV. On the other hand, AGNs only manifest narrow bands are known as Type 2. In recent years, it has been observed that AGN-type changes between Type 1 and Type 2, and these are called changing-look AGNs (CLAGNs) \citep{LaMassa-etal15, MacLeod-etal16, Ricci-etal16}. In this regard, the low luminosity AGNs (LLAGNs) are defined as luminosity $0.1\%$ L$_{\rm Edd}$, where L$_{\rm Edd}$ is the Eddington luminosity. These low luminosity AGNs can be explained based on radiatively inefficient accretion flows (RIAFs) onto AGNs \citep{Narayan-etal95, Yuan-Narayan14}. RIAF models are basically hot, optically thin, advection-dominated and include some other effects such as outflows and convection \citep{Yuan-Narayan14}. The radiation energy is much smaller compared to the thermal energy in RIAF model.


Further, astrophysical jets and mass outflows are commonly observed in black hole X-ray binaries (BH-XRBs) and AGNs \citep{Mirabel-etal92, Fender-Gallo-14} and reference therein. Several theoretical studies are carried out to investigate jets and mass outflows around black holes starting from the Penrose process \citep{Penrose-69}. Penrose process first provides an explanation of the extraction of energy from the infalling matter into the rotating black hole. Later, the seminal paper by \citet{Blandford-Znajek77} (BZ) showed that jet energy could be extracted from the rotational energy from large-scale magnetic fields around spinning black holes. Subsequently, \citep{Blandford-Payne82} (BP) pointed out that the matter can also leave the surface of the accretion disc due to magneto-centrifugal acceleration. Several simulation studies confirm the role of black hole spin in generating powerful jets around black holes \citep{Tchekhovskoy-etal-10, Tchekhovskoy-etal11, Tchekhovskoy-McKinney-12, Narayan-etal-22}. There are several other models have been proposed to address the mass outflow around black holes. In this regard, the advection-dominated inflow-outflow solutions (ADIOS) have been proposed in the literature. In this model, the inward decrease of mass accretion rate is due to the mass loss as outflow at every radius of the disc $(\dot{M}_{\rm acc}\propto r^s; 0\leq s<1)$ \citep{Blandford-Begelman-99, Becker-etal-01, Blandford-Begelman-04, Xue-Wang-05, Begelman-12, Yuan-etal-12}. Simultaneously, convection-dominated accretion flow (CDAF) has been introduced to explain the mass outflow from hot HD and MHD accretion flows \citep{Narayan-etal00, Quataert-Gruzinov-00}. In this model, the accreting gas moves in and out with convective eddies, and this motion provides the fluxes of inflowing and outflowing matter. It is also believed that inward angular momentum is transported by convection and outward transport by viscous stresses. Numerical simulations also investigated the mass outflow, which is in support of the CDAF model \citep{Igumenshchev-Abramowicz-99, Stone-etal-99, Igumenshchev-etal-00, Igumenshchev-etal-03}. On the other hand, accretion shock-driven mass outflow studies also investigated in analytical as well simulation study and their astrophysical applications around black hole \citep{Chattopadhyay-Das07, Kumar-Chattopadhyay13, Das-etal14, Okuda-Das15, Aktar-etal15, Aktar-etal17, Okuda-etal19, Kim-etal-19, Okuda-etal22, Okuda-etal23}. 


It is noteworthy to mention that in recent years, an important development has been achieved in the accretion disc theory for highly magnetized accretion flows recognized as ``magnetically arrested disk (MAD)'' \citep{Narayan-etal-03}. \citet{Tchekhovskoy-etal11} first showed in their pioneering work that hot accretion flow in MAD state can launch a very powerful jet based on GRMHD simulation around a spinning black hole. They also showed that the jet carries more power than the accretion energy for a highly spinning black hole. This is because of the extraction of rotational energy from the black hole via the BZ process. Subsequently, several numerical studies confirm the existence of the MAD state along with the ``standard and normal evolution (SANE)" state in highly magnetized flow \citep{Narayan-etal12, McKinney-etal-12, McKinney-etal-15, Dihingia-etal21, Dihingia-etal22, Chatterjee-Narayan-22, Dihingia-etal23, Dhang-etal23, Hong-Xuan-etal23}. Moreover, high angular resolution polarization observations of M87 by the Event Horizon Telescope observe that the accretion flow in this system is likely to be in the MAD state \citep{Event-Horizon-etal-21}.

One of the fundamental questions in accretion disc physics is how the angular momentum transfer in the disc. Initially, the seminal paper by \citet{Shakura-Sunyaev73} proposed the `$\alpha$-disc' model. However, the origin of `{\it ad-hoc}' viscosity is still questionable in this model. Conversely, it has been widely accepted in recent years that the mechanism of angular momentum transport in the accretion flows is the magnetorotational instability (MRI) \citep{Balbus-Hawley91, Balbus-Hawley98}. This instability is set to increase the initial magnetic field exponentially in the accreting gas until the magnetohydrodynamic (MHD) turbulence develops in the system. In this turbulent state, the Maxwell and Reynolds stress transport angular momentum outwards and causes inward mass accretion. To understand the nonlinear turbulence state of MRI, numerical simulation studies of accretion flows are inevitable. Several numerical simulations also show that the Maxwell stress always dominates over the Reynolds stress by a factor of several in MHD flows. Over the years, several MHD simulations have been carried out to investigate accretion flows around black holes considering magnetized torus. In this context, \citet{Hawley-00} investigate global three-dimensional MHD simulations of non-radiative accretion flows. In this work, \citet{Hawley-00} showed that mass accretion is primarily driven by Maxwell stress, and it is enhanced by MRI. On the other hand, \citet{Machida-etal00} showed that the magnetic field is enhanced by MRI and buoyantly escapes from the disc to form a magnetically active disc corona. There are several other MHD simulations have been carried out considering non-radiative as well as radiative accretion flows around black holes in the literature considering pseudo-Newtonian potential \citep{Paczyński-Witta80} around non-spinning black hole \citep{Kuwabara-etal00, Hawley-Krolik01, Stone-Pringle01, Hawley-etal02, Kuwabara-etal05, Ohsuga-etal09, Ohsuga-Mineshige11, Igarashi-etal20}. Moreover, it has been investigated that global MHD simulation is more realistic than global HD simulation because MHD simulations self-consistently generate shear stress via MRI turbulence. On the other hand, in HD simulation, one needs an `{\it ad hoc}' viscosity to generate shear stress in the flow. In recent years, with the considerable development of numerical advancement, general relativistic simulations of MHD torus around black holes have been investigated \citep{Gammie-etal03, De-Villiers-etal03a, De-Villiers-etal03b, Tchekhovskoy-etal11, Narayan-etal12, Dihingia-etal21, Chatterjee-Narayan-22, Dihingia-etal22, Narayan-etal-22, Dhang-etal23,  Curd-Narayan23, Hong-Xuan-etal23}.


In this paper, we investigate the time-dependent accretion and associated mass outflows considering MHD torus around spinning AGNs. Here, the space-time geometry is modeled around a spinning black hole using effective Kerr potential derived by \citet{Dihingia-etal18b}. This effective potential quite accurately mimics space-time geometry around Kerr black hole having spin $0\leq a_k < 1$. \citet{Dihingia-etal18b} also showed that the analytical transonic solutions using this effective potential in semi-relativistic flow are in excellent close agreement with general relativistic results even in the maximally spinning regime. During accretion, the accreting gas moves inward with the increase of magnetic pressure. Also, the gas expands in the vertical direction below and above the torus via MRI turbulence and carries a significant amount of magnetic field with it as magnetized mass outflow \citep{Machida-etal00, Hawley-Krolik01, Hawley-Balbus02}. In general, torus evolution around the black hole is primarily determined by three parameters: the strength of the initial magnetic field, flow angular momentum, and spin of the black hole. First, we investigate the magnetic state of the accretion flow by calculating normalized magnetic flux for our simulation model \citep{Tchekhovskoy-etal11}. Further, we examine the effect of these three parameters on the accretion and associated mass outflow around AGNs. More specifically, we investigate the correlation between the magnetic field and spin on the mass outflow around black hole. We also estimate bremsstrahlung luminosity under optically thin disc approximation from our model and investigate the effect of basic parameters on luminosity variation.

We organize the paper as follows. In section \ref{numerical-method}, we present the description of the numerical model and governing equations. In section \ref{results}, we discuss the results of our model in detail. Finally, we draw the concluding remarks in section \ref{conclusion}.

\section{Numerical Method}
\label{numerical-method}

We perform axisymmetric two-dimensional MHD simulations using the publicaly available numerical simulation package PLUTO\footnote{\url{http://plutocode.ph.unito.it}} \citep{Migone-etal07}. We adopt the unit system as $G = M_{\rm BH} = c =1$, where $G$, $M_{\rm BH}$ and $c$ are the gravitational constant, the mass of the black hole and the speed of light, respectively. In this unit system, we measure distance, velocity, and time as $r_g = GM_{\rm BH}/c^2$, $c$ and $t_g = GM_{\rm BH}/c^3$, respectively. Here, we consider no explicit resistivity in the flow. We also ignore any radiation transport and loss in our model.

\subsection{Governing Equations}

In this paper, we write MHD governing equations in cylindrical coordinates $(r,\phi, z)$. The governing equations are as follows\\
\begin{align}\label{radial_eq}
& \frac{\partial \rho}{\partial t} + \nabla \cdot (\rho \bm{v}) =0,\\
& \frac{\partial (\rho \bm{v})}{\partial t} + \nabla \cdot (\rho \bm{v} \bm{v} - \bm{B}\bm{B}) + \nabla P_t = - \rho \nabla \Phi, \\
& \frac{\partial E }{\partial t} + \nabla \cdot [(E+P_t)\bm{v} - (\bm{v}.\bm{B})\bm{B}]  = -\rho \bm{v}.\nabla \Phi,\\
& \frac{\partial \bm{B}}{\partial t} + \nabla \cdot (\bm{v}\bm{B}-\bm{B}\bm{v}) =0,
\end{align}
where $\rho$, $\bm{v}$, and $\bm{B}$ are the mass density, fluid velocity, and magnetic field, respectively. $P_t = P_{\rm gas} + B^2/2$ is the total pressure comprised with gas pressure $(P_{\rm gas})$ and magnetic pressure $(B^2/2)$. $\Phi$ represents the gravitational potential in the presence of a black hole. The Bernoulli parameter (Be) can be obtained as
\begin{align} \label{Bernoulli_eqn}
Be = \frac{P_{\rm gas}}{\gamma -1} + \frac{1}{2} \rho v^2 + \frac{1}{2} B^2 + \Phi^{\rm eff} -1,
\end{align}
where $\gamma$ is the adiabatic index. Here, we assume the adiabatic equation of state $P_{\rm gas} = \rho \epsilon (\gamma -1)$. Here, $\epsilon$ is the specific internal energy. Also the adiabatic sound speed is defined as $c_s = \sqrt{\frac{\gamma P_{\rm gas}}{\rho}}$. We subtract the rest mass energy of the gaseous matter from the total energy to obtain the Bernoulli parameter \citep{Narayan-Yi94, Narayan-etal12}. Here, $\Phi^{\rm eff}$ is the effective Kerr potential as described in the sub-section \ref{Kerr_pot}.

\subsection{Graviational potential}
\label{Kerr_pot}

The gravitational potential around spinning black holes is modeled using \citet{Dihingia-etal18b} effective Kerr potential. The effective Kerr potential is given by
 \begin{align}
 \Phi^{\rm eff}\left(r,z,a_k, \lambda\right)= \frac{1}{2} \ln \left(\frac{A (2 R - \Sigma) r^2 - 4 a_k^2 r ^4}{\Sigma \lambda \left(\Sigma \lambda  R^2 + 4 a_k r ^2 R - 2 \lambda  R^3\right)- A \Sigma r ^2 }\right),
 \end{align}
 where $R=\sqrt{r^2+z^2}$ = spherical radial distance, $\Delta=a_k^2+R^2-2 R$, $\Sigma=\frac{a_k^2 z^2}{R^2}+R^2$ and $A=\left(a_k^2+R^2\right)^2-\frac{a_k^2 r^2 \Delta}{R^2}$. Here, $\lambda$ and $a_k$ are specific flow angular momentum and black hole spin, respectively. It is to be noted that the gravitational and centrifugal potential is coupled in this potential. Therefore, we modify the radial momentum conservation equations in the PLUTO code to incorporate the effective Kerr potential. The Keplerian angular momentum can be obtained in the equatorial plane $(z \rightarrow 0)$ as
\begin{align} \label{kep_ang}
 \lambda_K = \sqrt{r^3 \frac{\partial \Phi_{\rm eff}}{\partial r}|_{\lambda \rightarrow 0}}.
\end{align}
The angular frequency is $\Omega = \lambda/r^2$ and the circular orbital period at distance $r$ is $P_{\rm orb} = 2 \pi \Omega^{-1}$.

\subsection{Accretion Torus set up}
\label{torus_set_up}

We consider an equilibrium accretion disc around a spinning black hole. The black hole is surrounded by a hot corona \citep{Matsumoto-etal96, Hawley-00, Hawley-Krolik01}. The equilibrium torus solutions can be obtained by adopting the Newtonian analog of relativistic tori \citep{Abramowicz-etal78}. The density distribution of the torus can be obtained considering constant angular momentum flow $(\lambda = \rm constant)$ as \citep{Matsumoto-etal96, Hawley-00, Kuwabara-etal05}
\begin{align}
\Phi^{\rm eff}\left(r,z, a_k, \lambda\right) + \frac{\gamma}{\gamma -1}\frac{P_{\rm gas}}{\rho}={\cal C} = {\rm constant}.
\end{align}
The constant `$\mathcal{C}$' can be determined considering zero-gas pressure surface $(P_{\rm gas} \rightarrow 0)$ at $r = r_{\rm min}$ at the equatorial plane. Here, $r_{\rm min}$ represents the inner edge of the torus. Using adiabatic equation of state $P_{\rm gas} = K \rho^{\gamma}$, the density distribution inside the torus can be determined as 
\begin{align}
 \rho=\left[\frac{\gamma-1}{K\gamma}\left({\cal C}-\Phi^{\rm eff}\left(r,z, a_k, \lambda\right)\right)\right]^{\frac{1}{\gamma -1}},   
\end{align}
where $K$ can be determined considering the density maximum surface $(\rho_{\rm max})$ at $r = r_{\rm max}$ at the equatorial plane and is given by
\begin{align}
 K=\frac{\gamma - 1}{\gamma}\left[\mathcal{C}-\Phi^{\rm eff}\left(r_{\rm max},0, a_k, \lambda\right)\right]\frac{1}{\rho_{\rm max}^{\gamma-1}}.   
\end{align}
Now we assume that the density distribution outside the torus is isothermal, non-rotating, and high-temperature halo surrounding the black holes \citep{Matsumoto-etal96, Kuwabara-etal05}. For the halo, the density distribution is assumed to be in hydrostatic equilibrium as
\begin{align} \label{hydro_equi}
c_s^2\frac{\nabla \rho}{\rho}=-\nabla\Phi^{\rm eff}\left(r,z, a_k, \lambda \rightarrow 0\right),    
\end{align}
where $c_s$ is the sound speed. Integrating equation (\ref{hydro_equi}) from maximum density of the torus $(r_{\max}, 0)$ to the outside torus $(r,z)$ upto halo is given by 
\begin{align} \label{eqn_density_halo}
\rho={\eta}\rho_{\rm max}\exp\left[\left(\Phi^{\rm eff}\left(r_{\rm max},z, a_k, \lambda \rightarrow 0\right)-\Phi^{\rm eff}\left(r,z, a_k, \lambda \rightarrow 0\right)\right){\cal H}\right],    
\end{align}
where, $\cal{H}$ = $1/c_s^2$ and for representative case, we consider $\cal{H}$ equal to 2 throughout the simulation. Here, we also introduce a constant factor $\eta$ which represents the density fraction between the density of the halo to the maximum torus density.

\begin{figure}
	\begin{center}
		\includegraphics[width=0.5\textwidth]{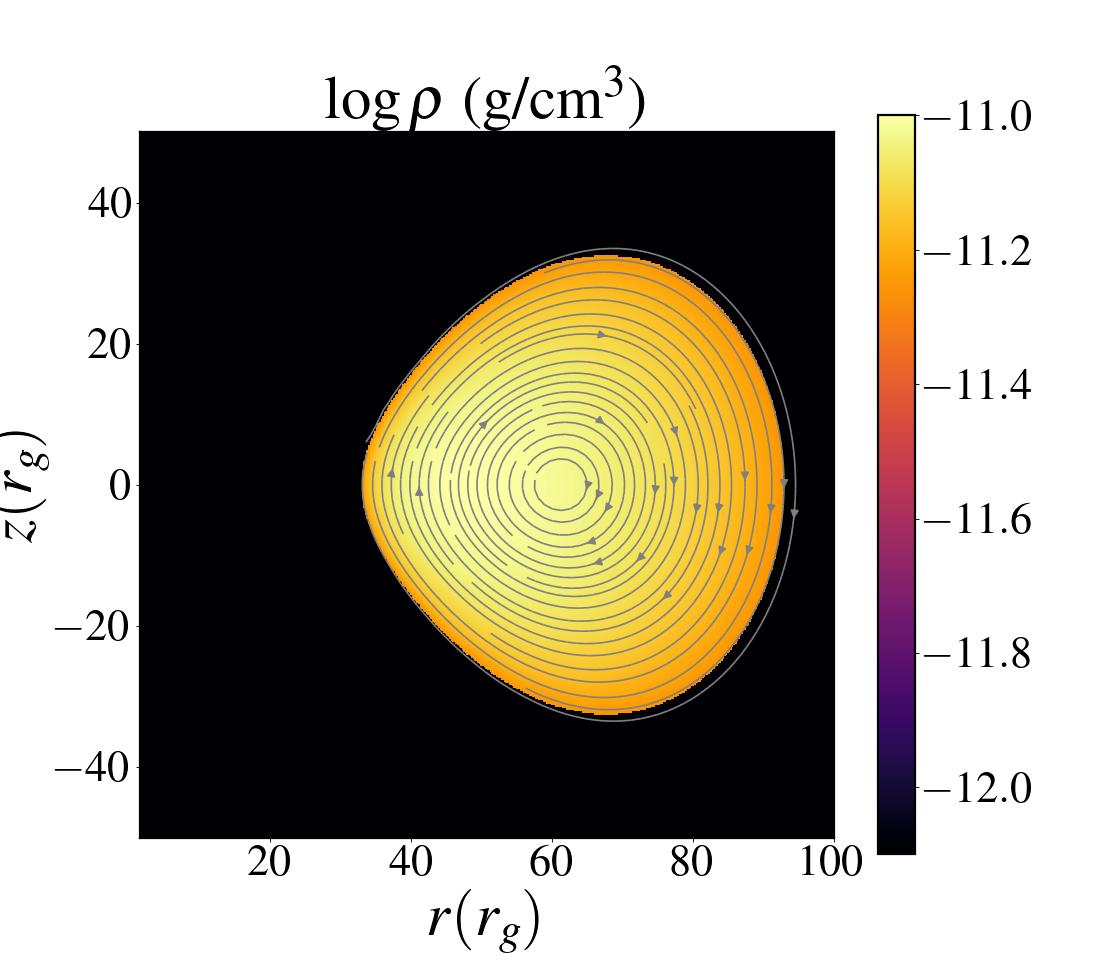} 
	\end{center}
	\caption{Initial equilibrium torus profile of density ($\log \rho$) and magnetic
field lines (grey lines) for $a_k = 0.99$ and $\beta_0 = 10$. We consider the initial maximum density at $r=50 r_g$. See the text for details.}
	\label{Figure_1}
\end{figure}

\begin{figure*}
	\begin{center}

		\includegraphics[width=0.26\textwidth]{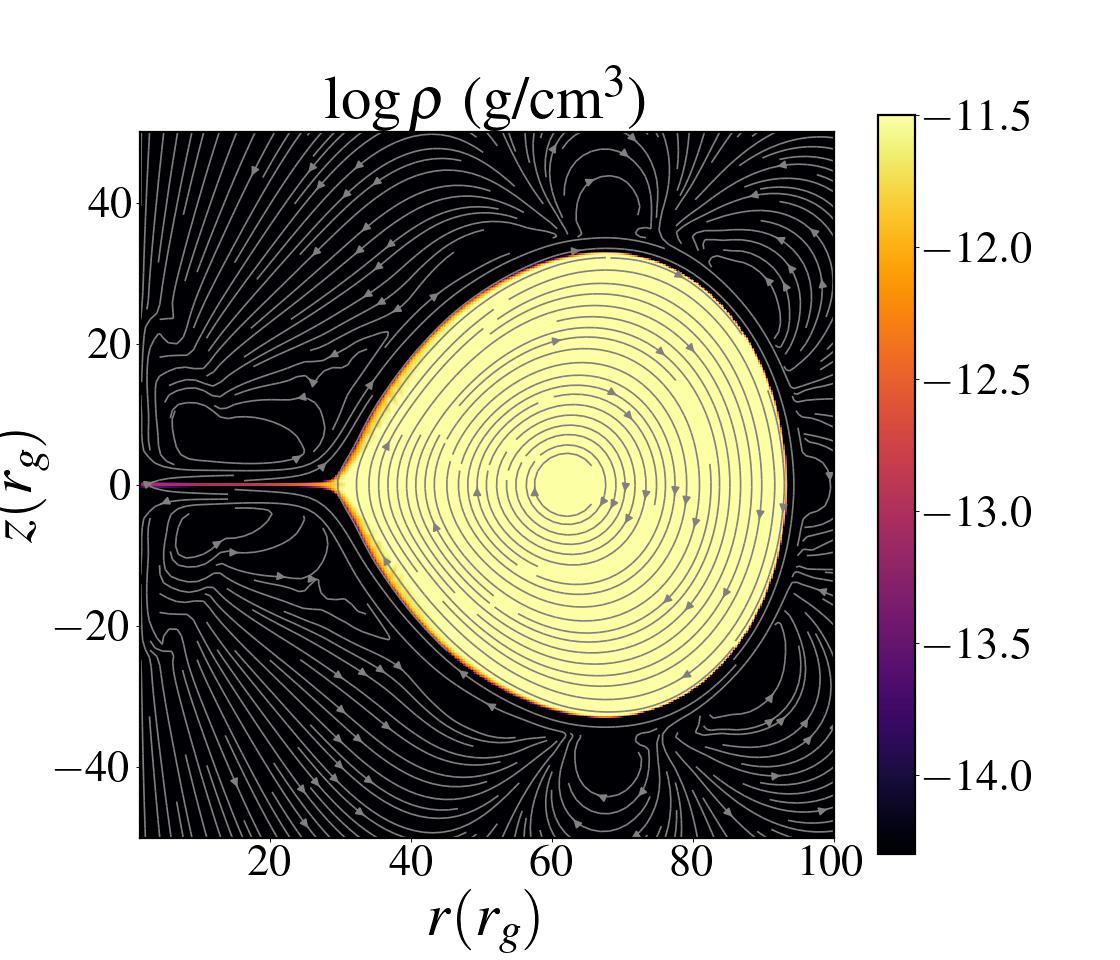} 
        \hskip -4 mm
        \includegraphics[width=0.26\textwidth]{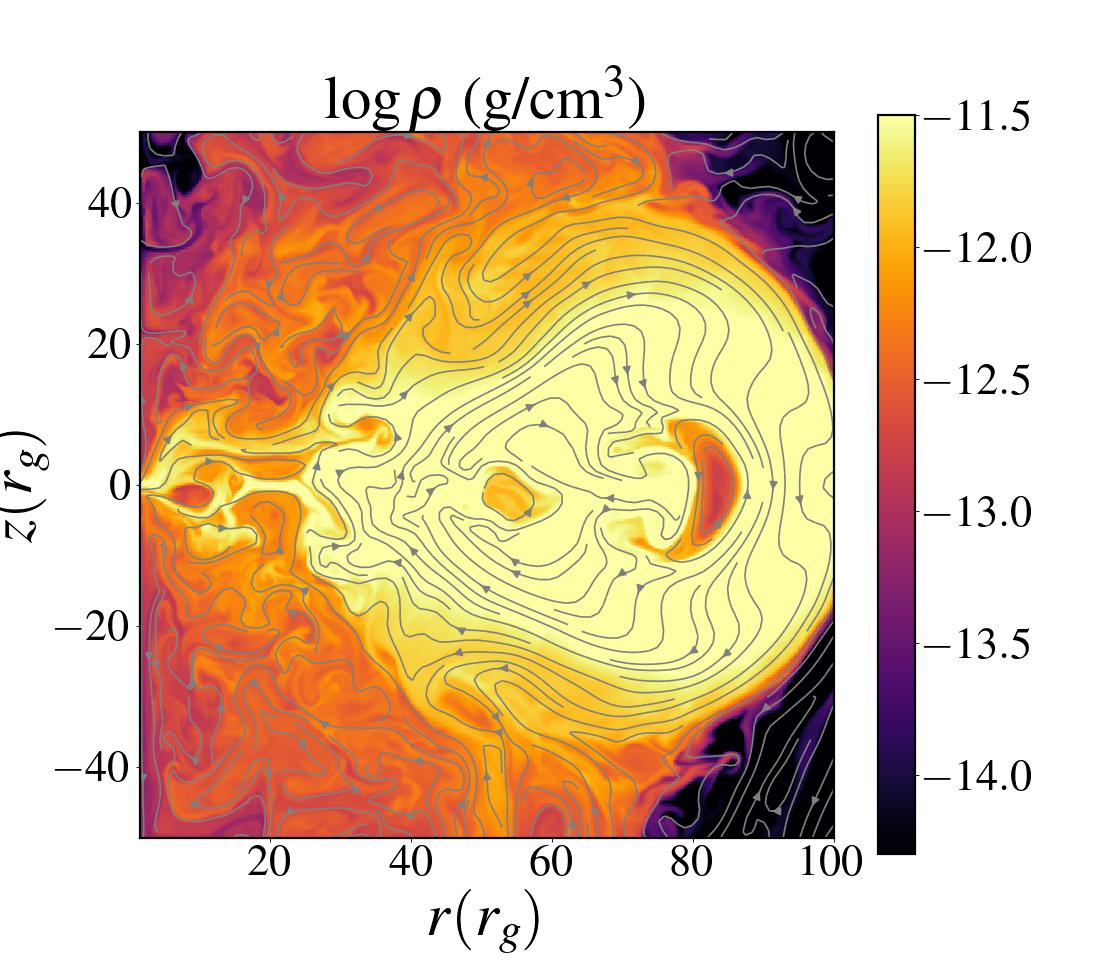} 
        \hskip -4 mm
        \includegraphics[width=0.26\textwidth]{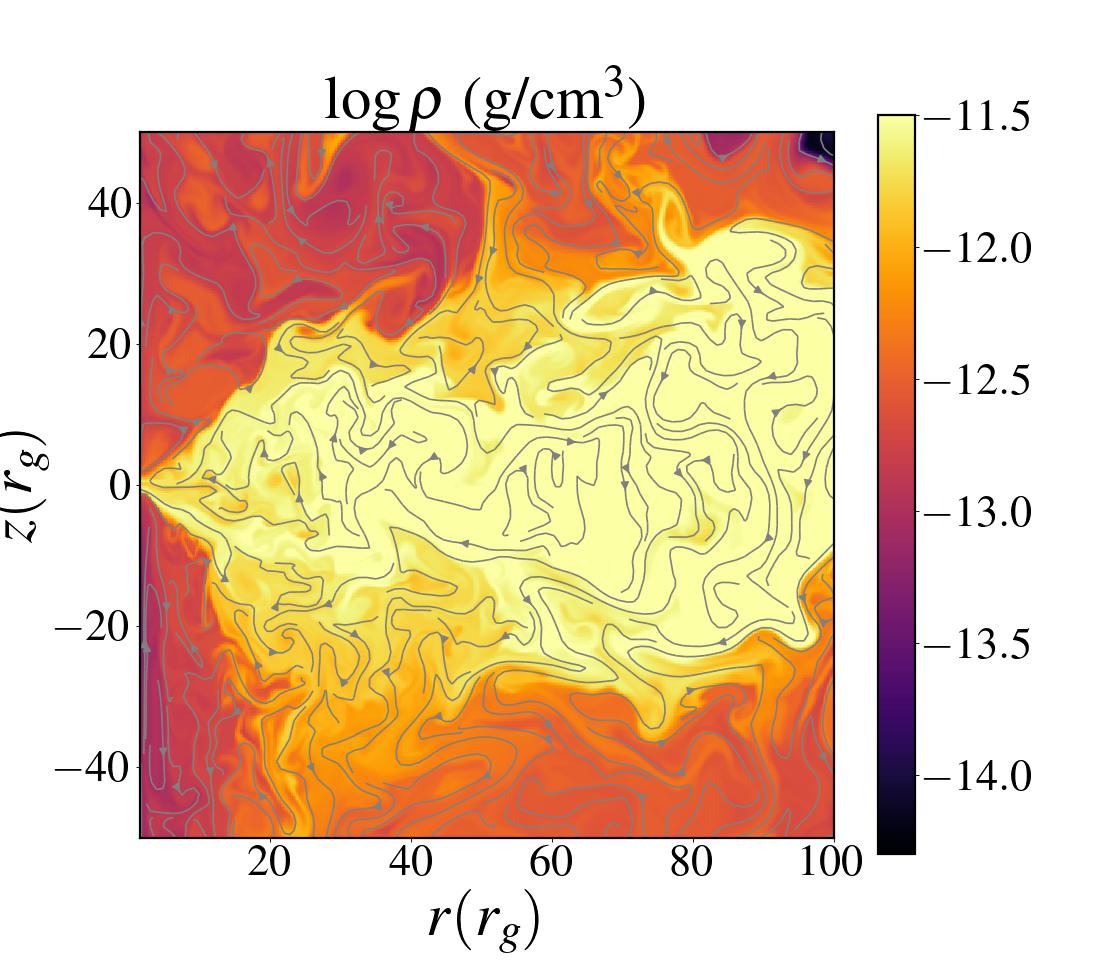} 
        \hskip -4 mm
        \includegraphics[width=0.26\textwidth]{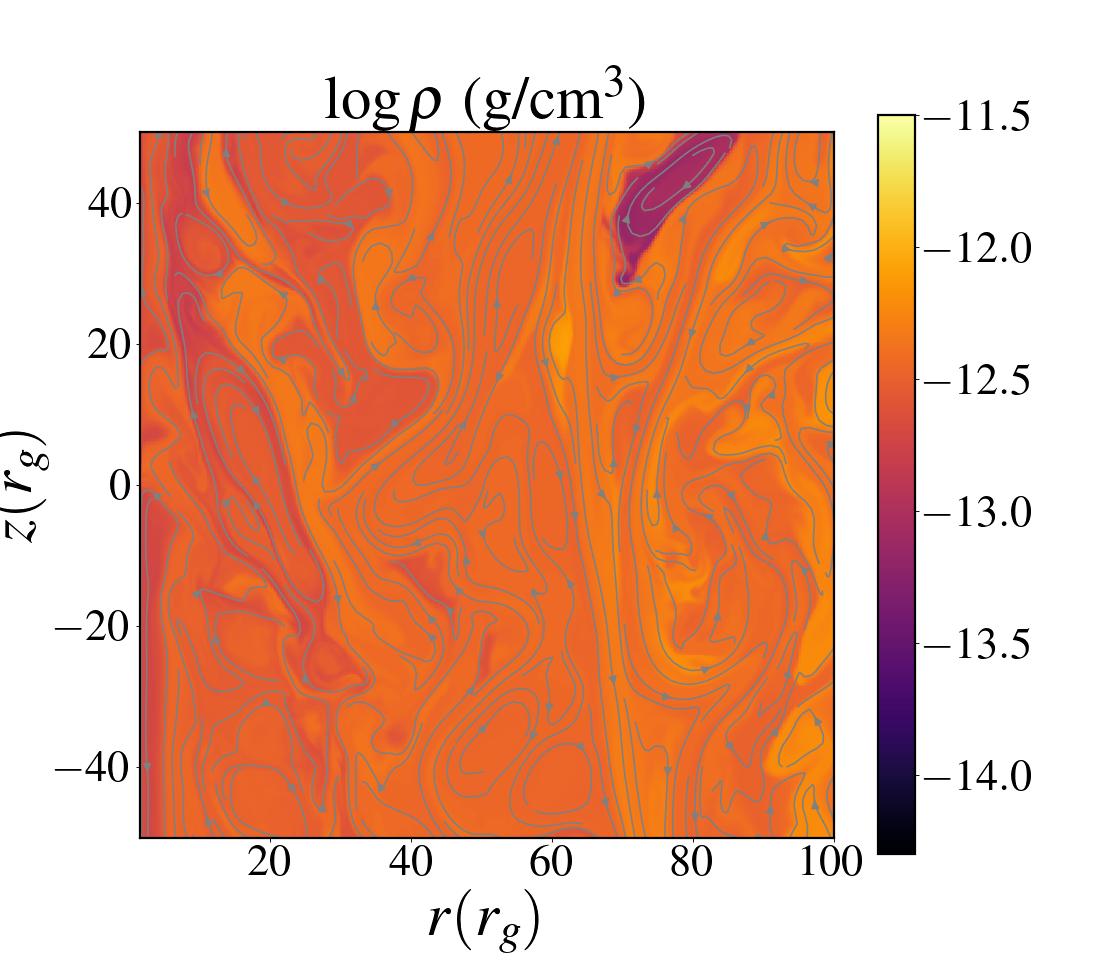} 
    
        \includegraphics[width=0.26\textwidth]{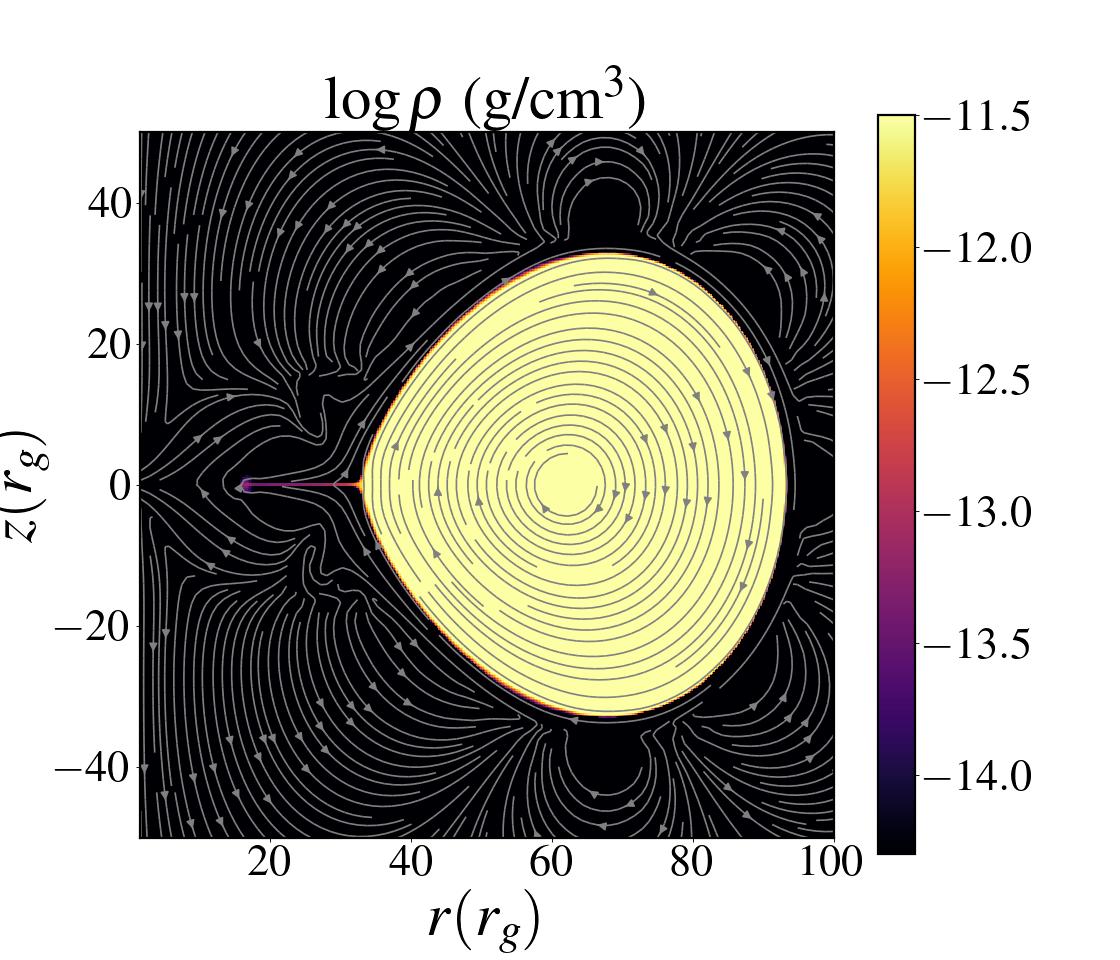} 
        \hskip -4 mm
        \includegraphics[width=0.26\textwidth]{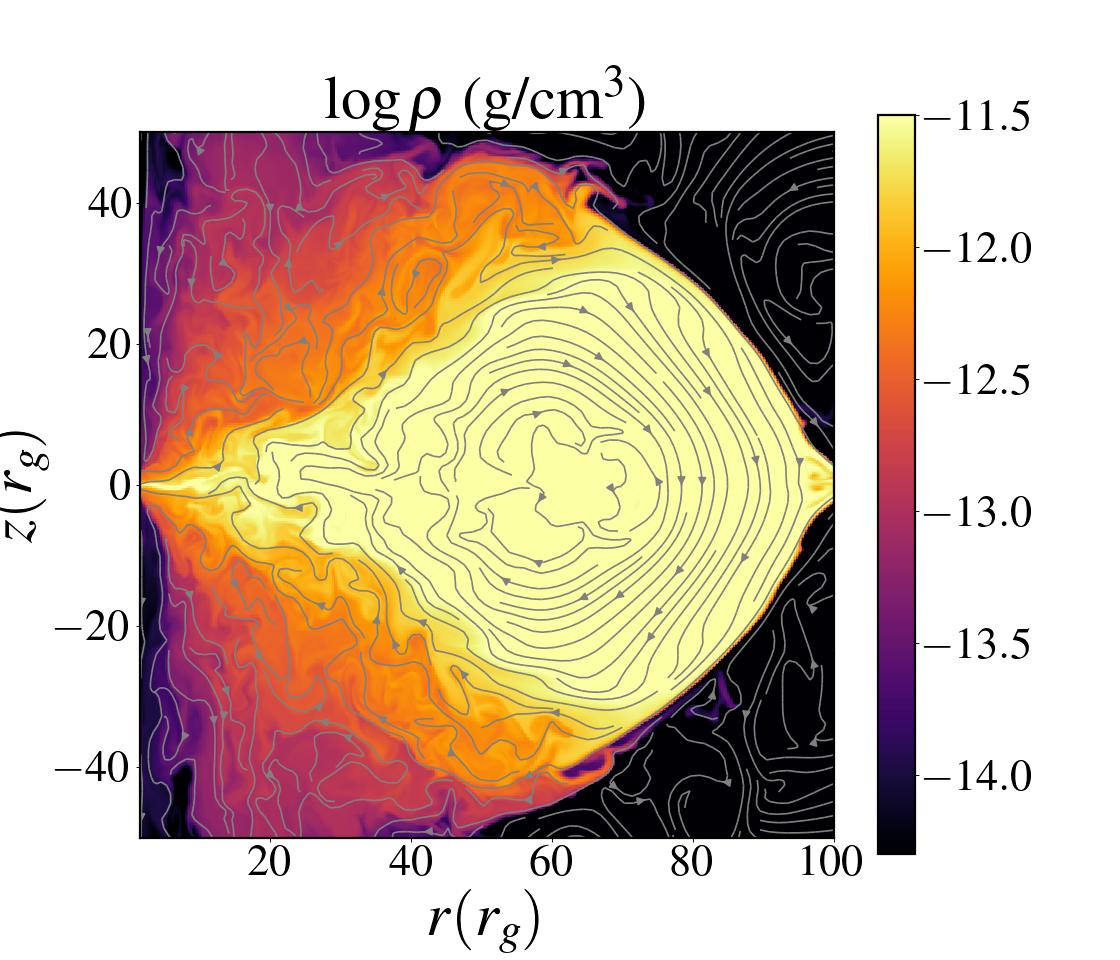} 
        \hskip -4 mm
        \includegraphics[width=0.26\textwidth]{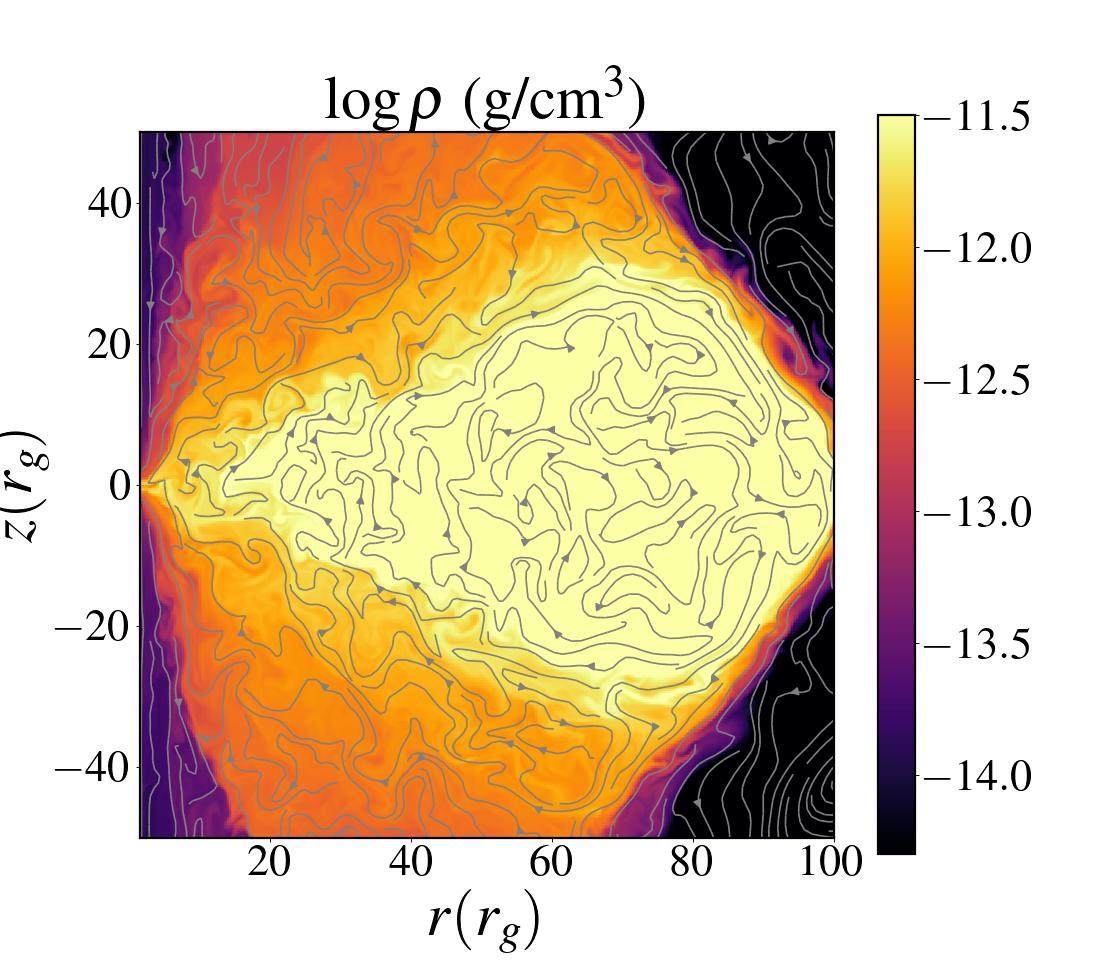} 
        \hskip -4 mm
        \includegraphics[width=0.26\textwidth]{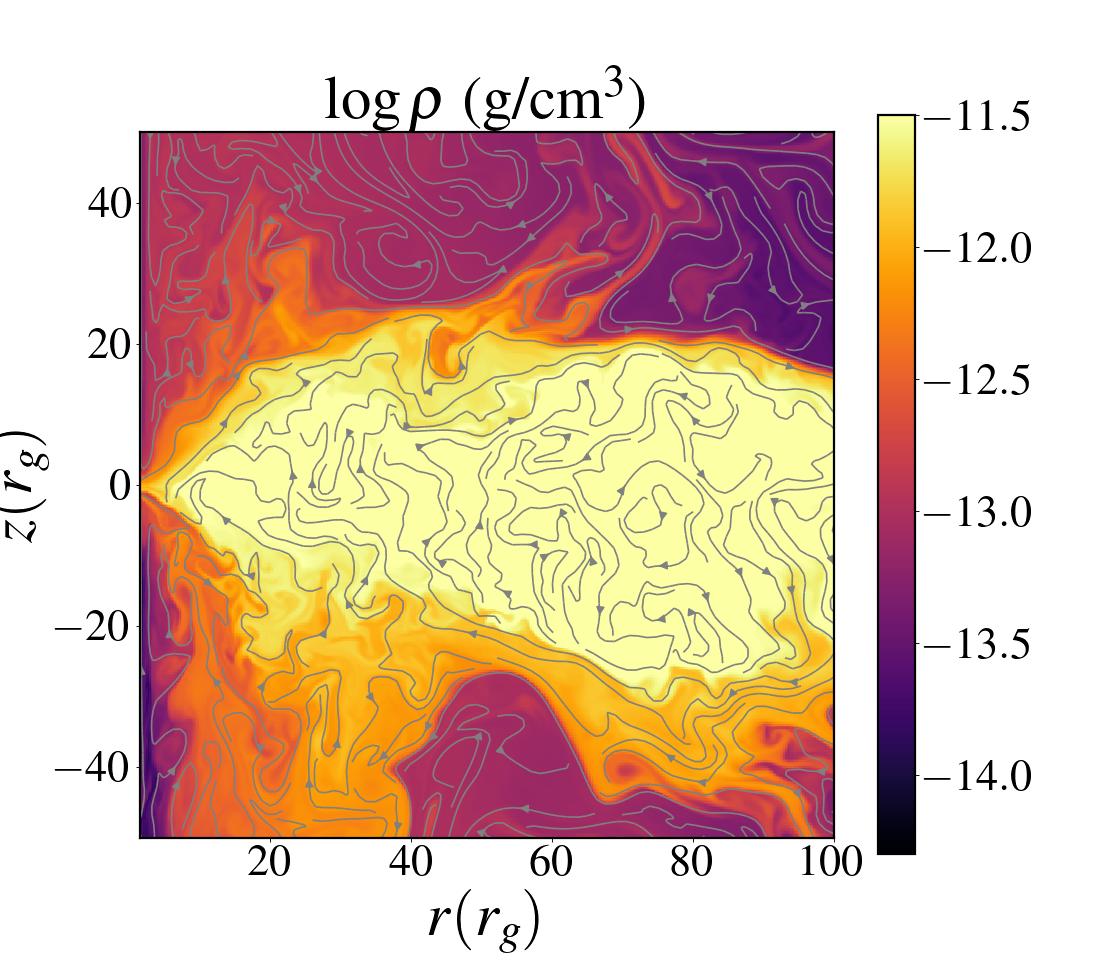} 

        \includegraphics[width=0.26\textwidth]{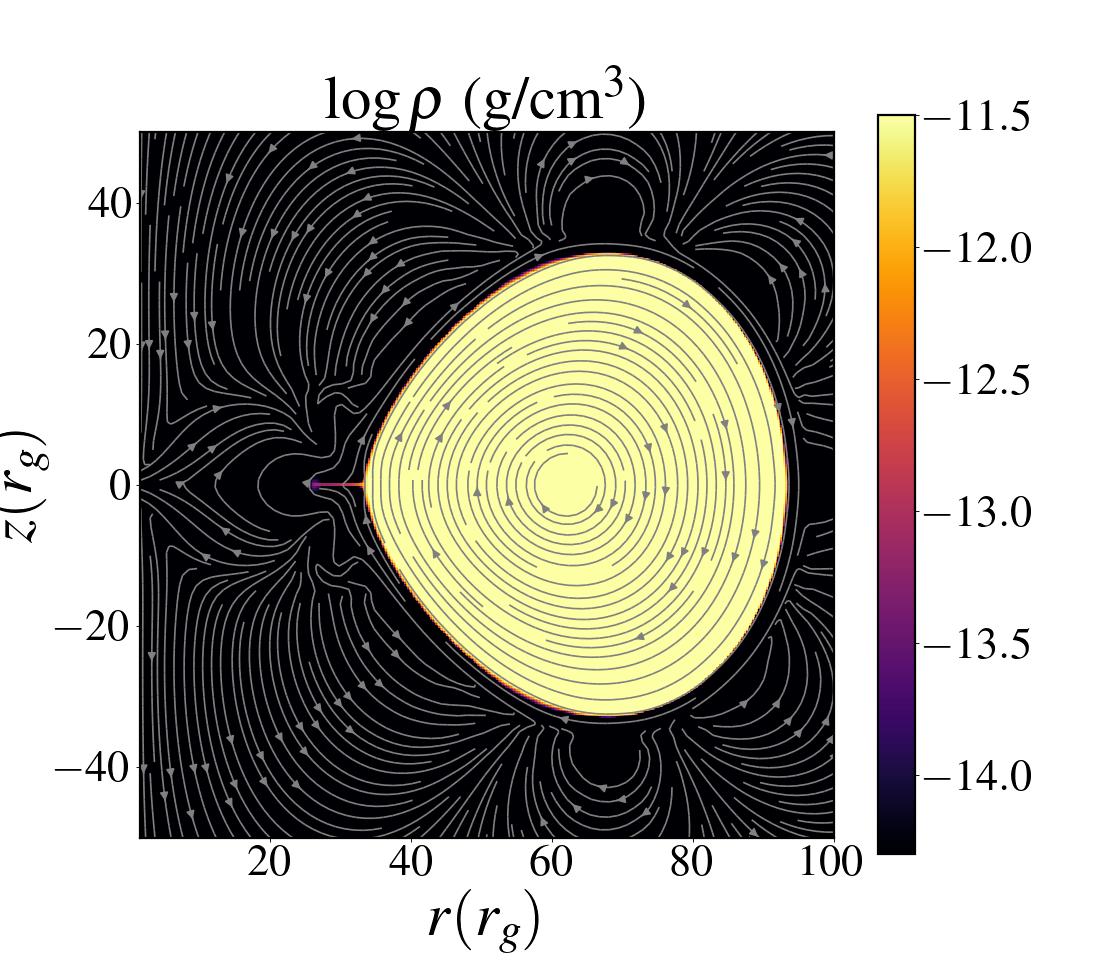} 
        \hskip -4 mm
        \includegraphics[width=0.26\textwidth]{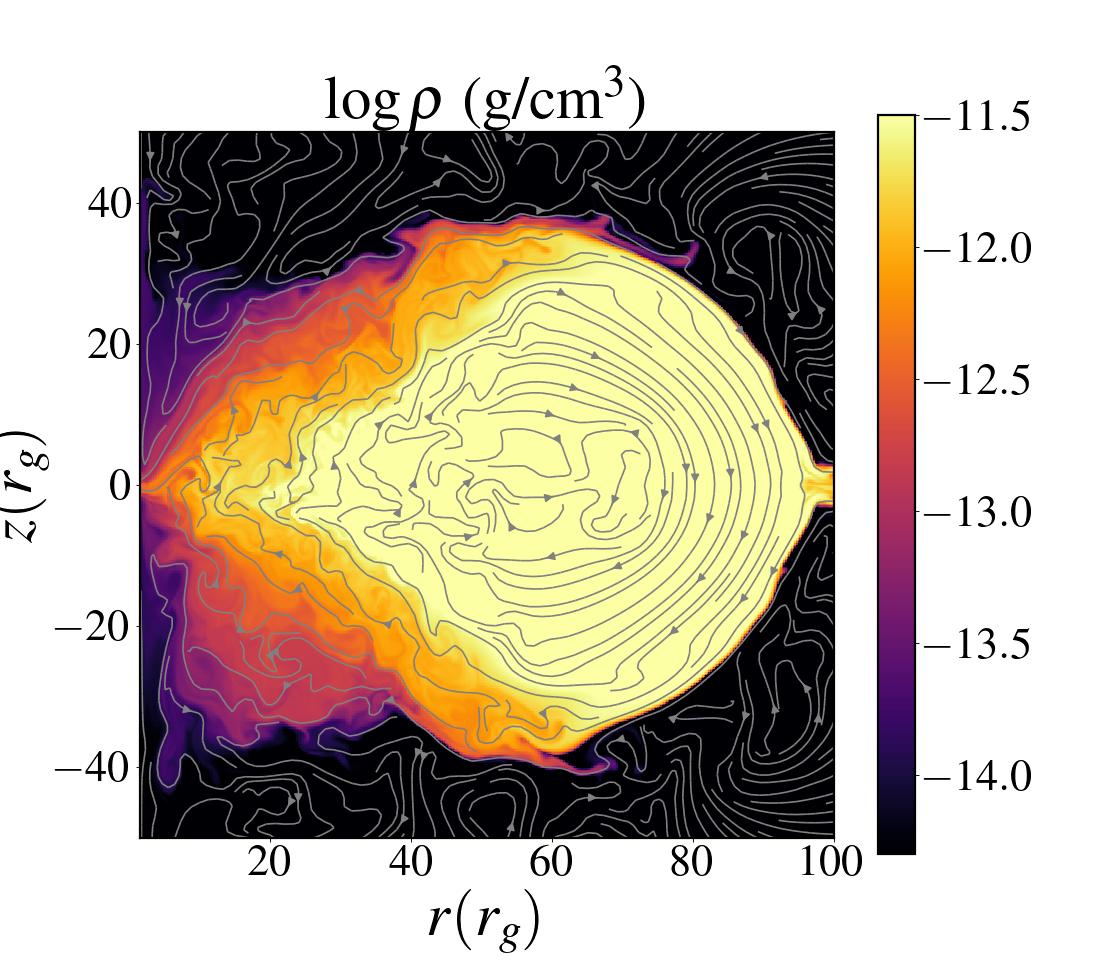} 
        \hskip -4 mm
        \includegraphics[width=0.26\textwidth]{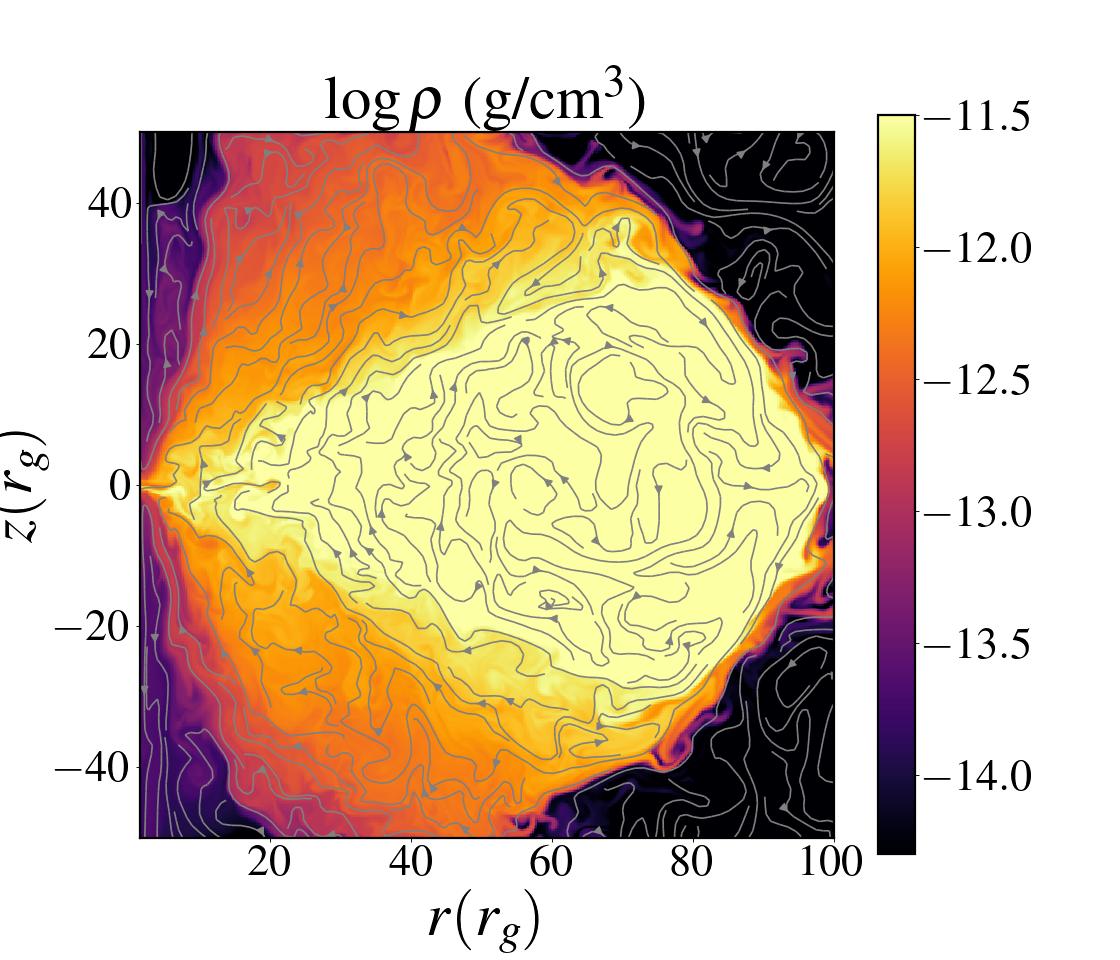} 
        \hskip -4 mm
        \includegraphics[width=0.26\textwidth]{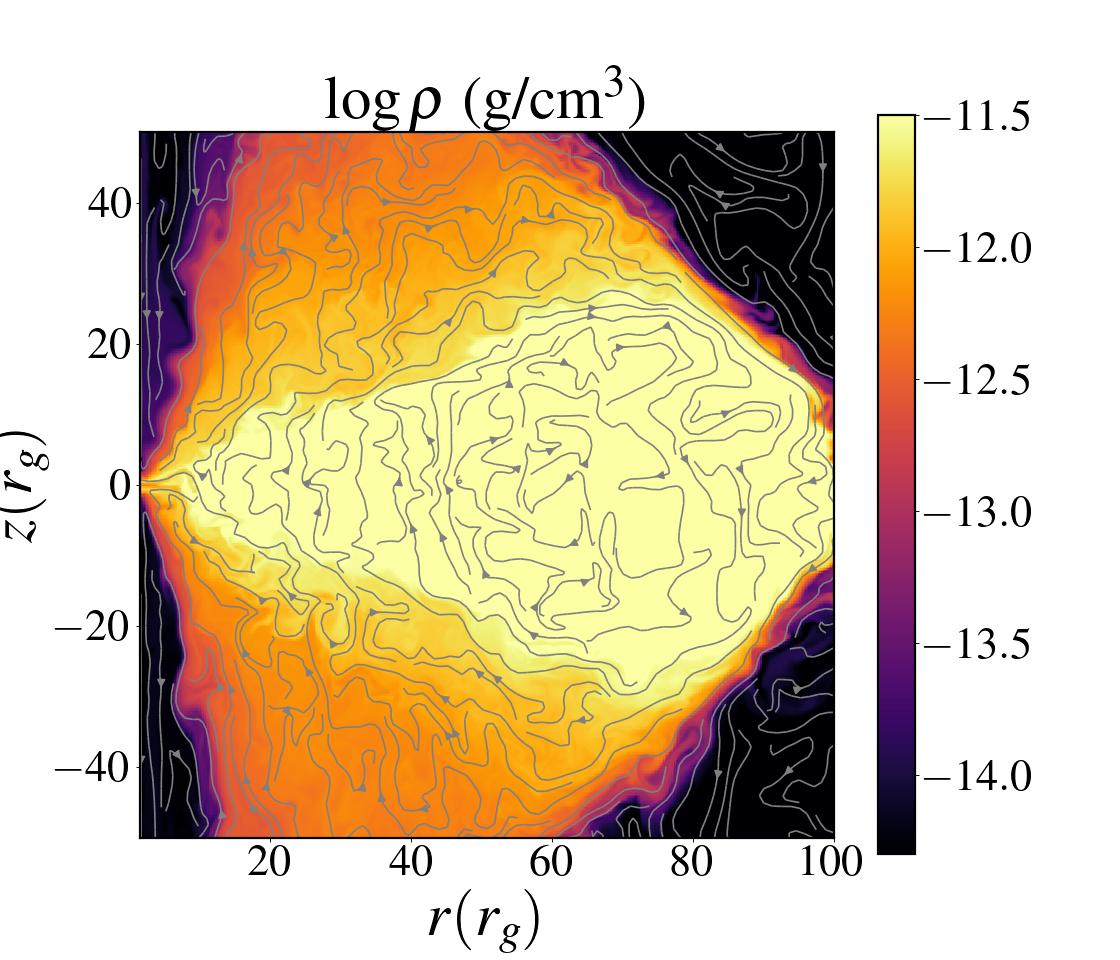} 
        
        \includegraphics[width=0.26\textwidth]{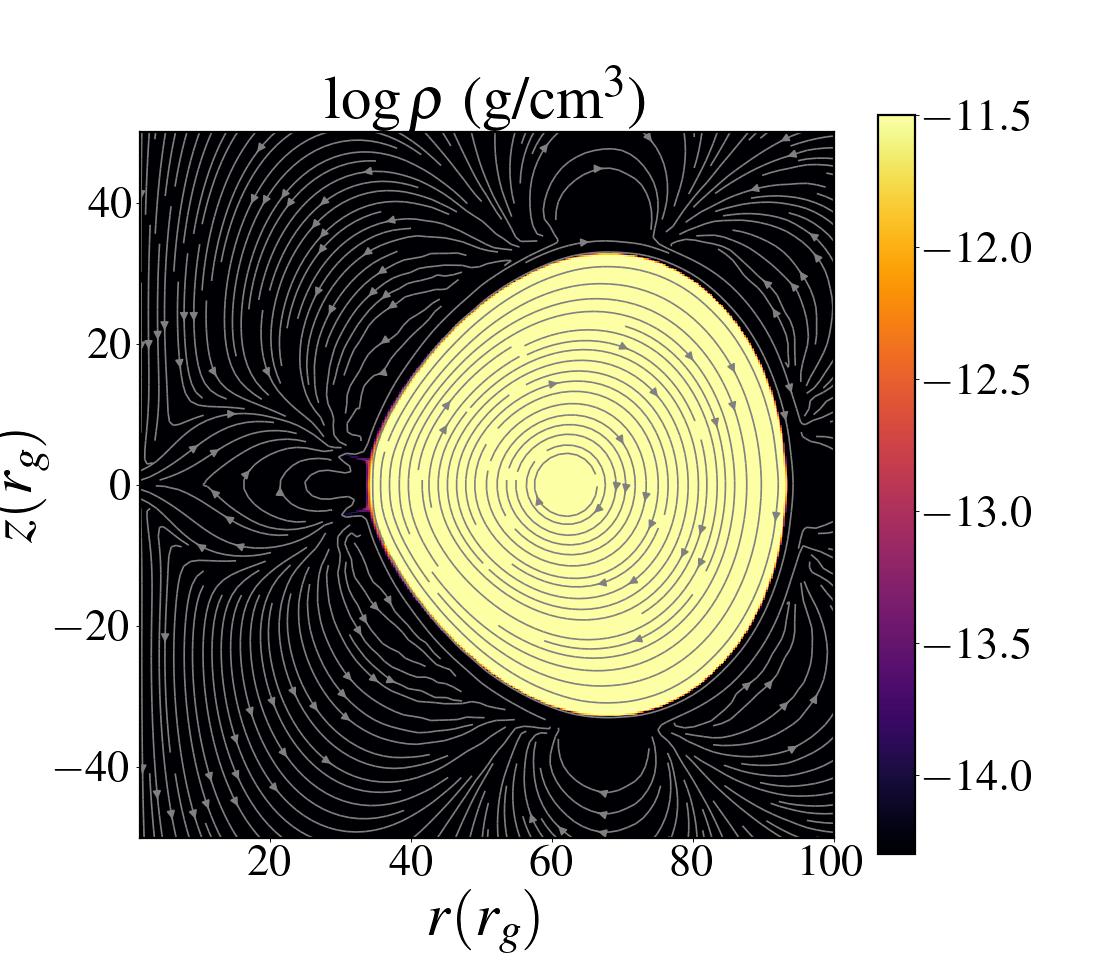} 
        \hskip -4 mm
        \includegraphics[width=0.26\textwidth]{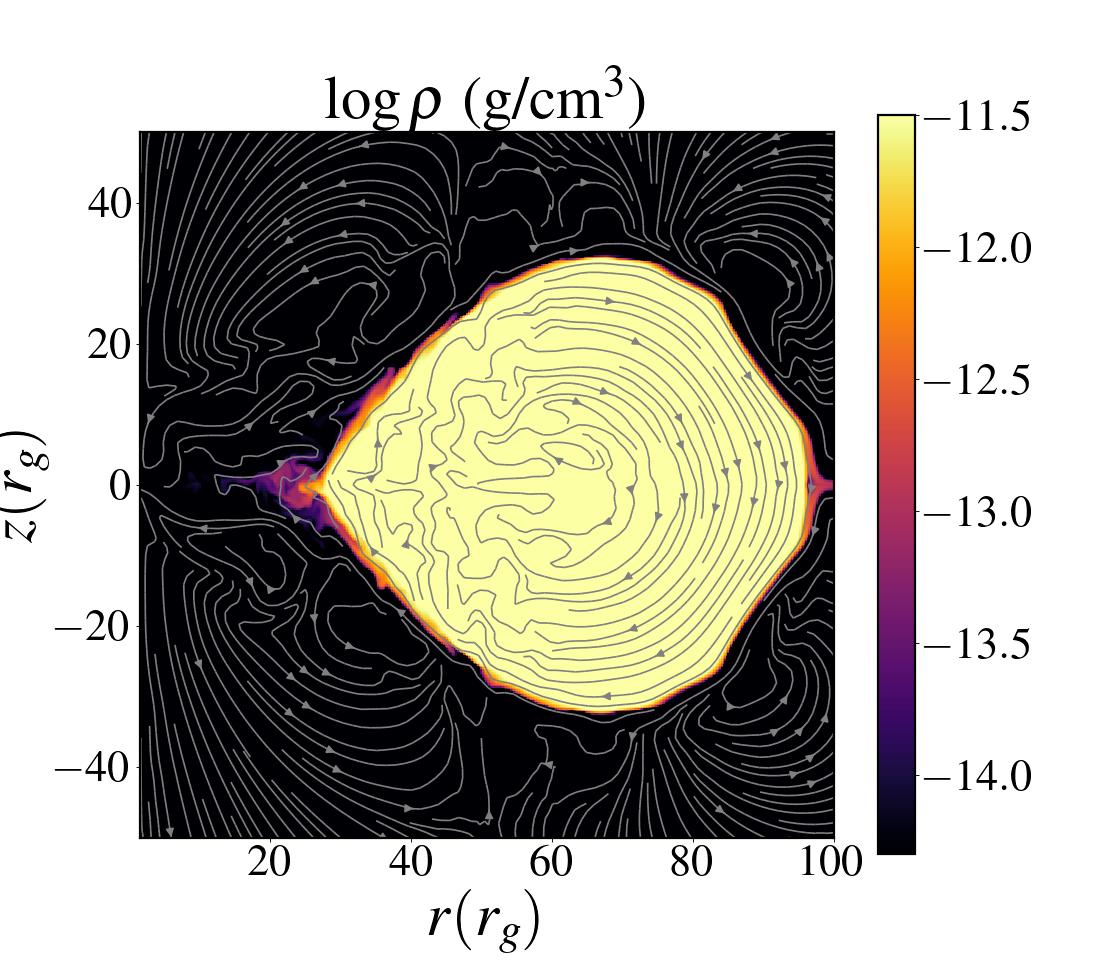} 
        \hskip -4 mm
        \includegraphics[width=0.26\textwidth]{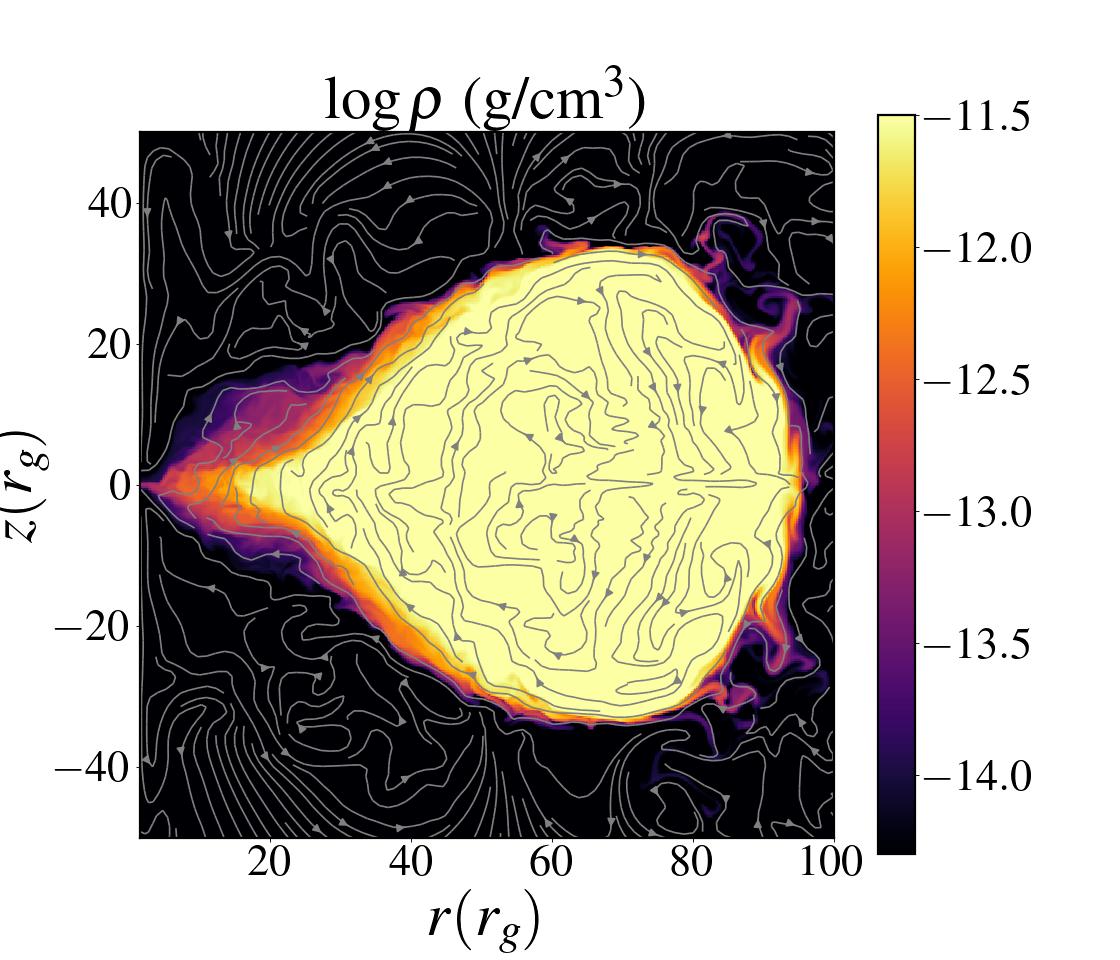} 
        \hskip -4 mm
        \includegraphics[width=0.26\textwidth]{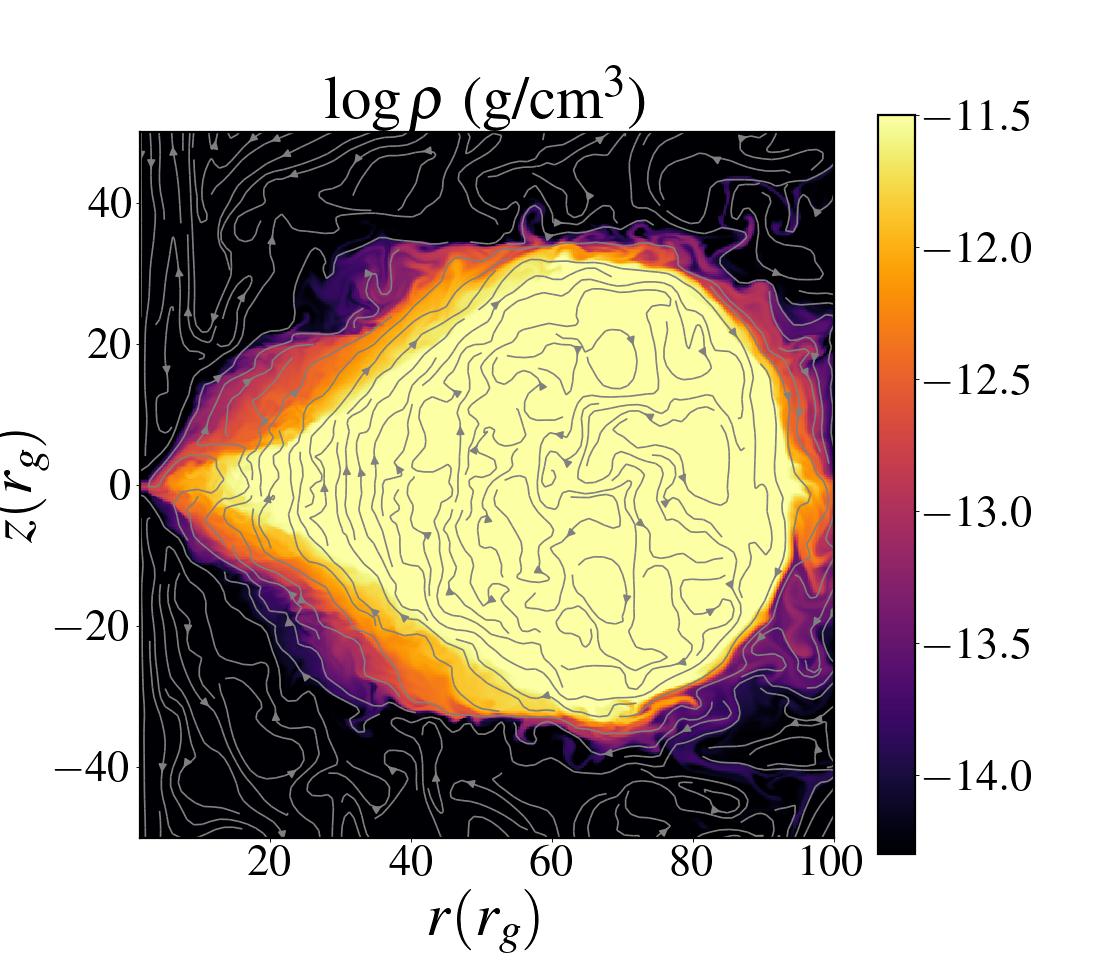} 
	\end{center}
	\caption{Distribution of gas density in $(r-z)$ plane for different initial plasma-$\beta$ at various times of evolution. The first, second, third, and fourth rows correspond to $\beta_0$ = 10, 50, 100, and 1000, respectively. Similarly, the first, second, third, and fourth columns represent at times $t_1 = 500 t_g$, $t_2 = 5000 t_g$, $t_3 = 10500 t_g$, and $t_4 = 17500 t_g$, respectively. The grey lines represent the magnetic field lines.}
	\label{Figure_2}
\end{figure*}

\begin{figure*}
	\begin{center}
		\includegraphics[width=0.26\textwidth]{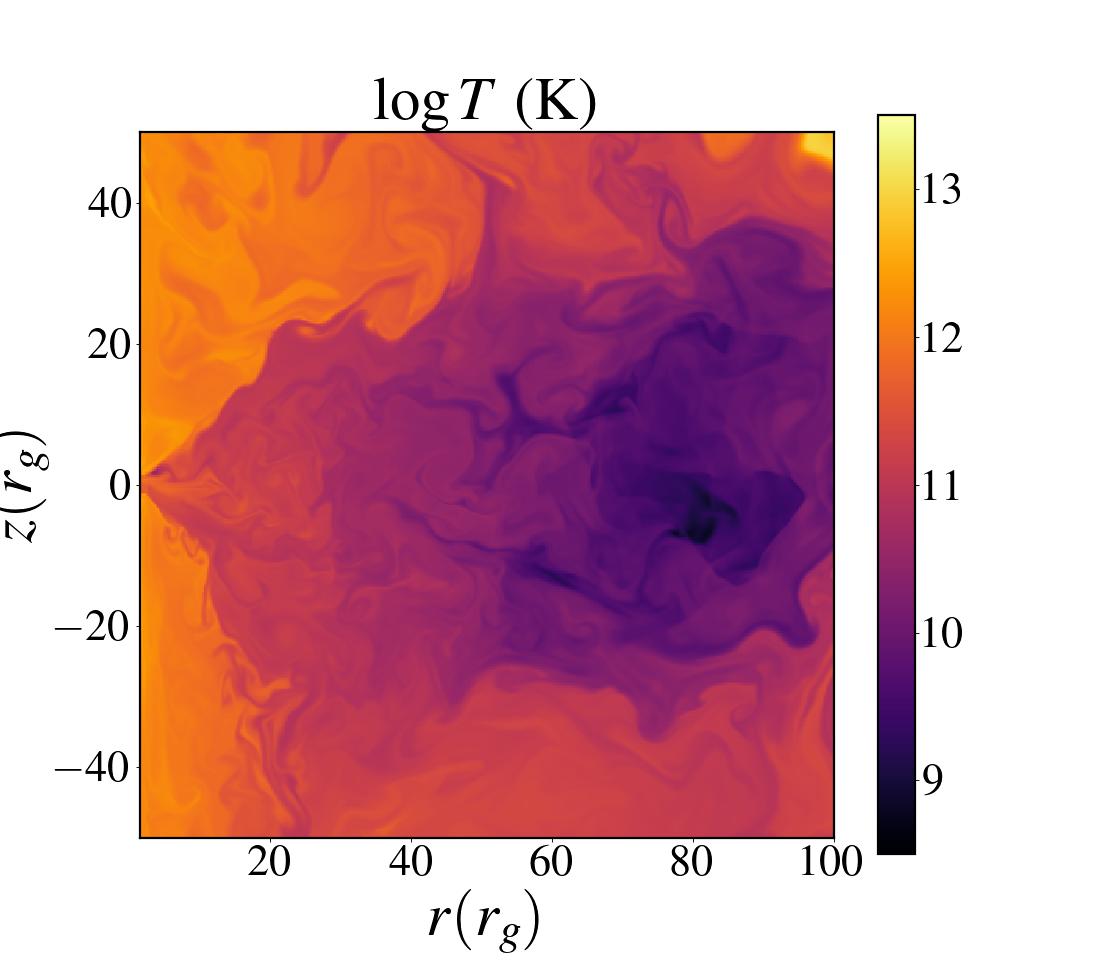} 
        \hskip -4 mm
        \includegraphics[width=0.26\textwidth]{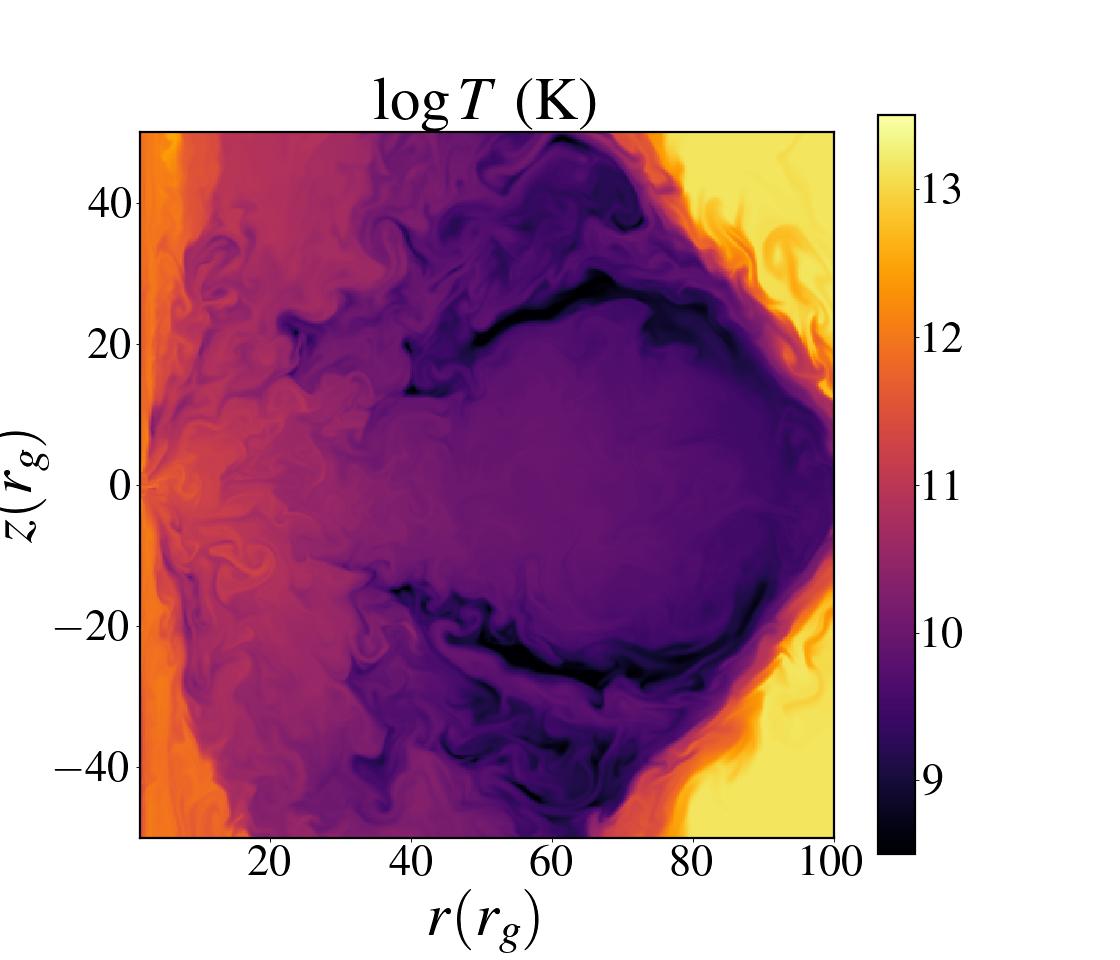} 
        \hskip -4 mm
        \includegraphics[width=0.26\textwidth]{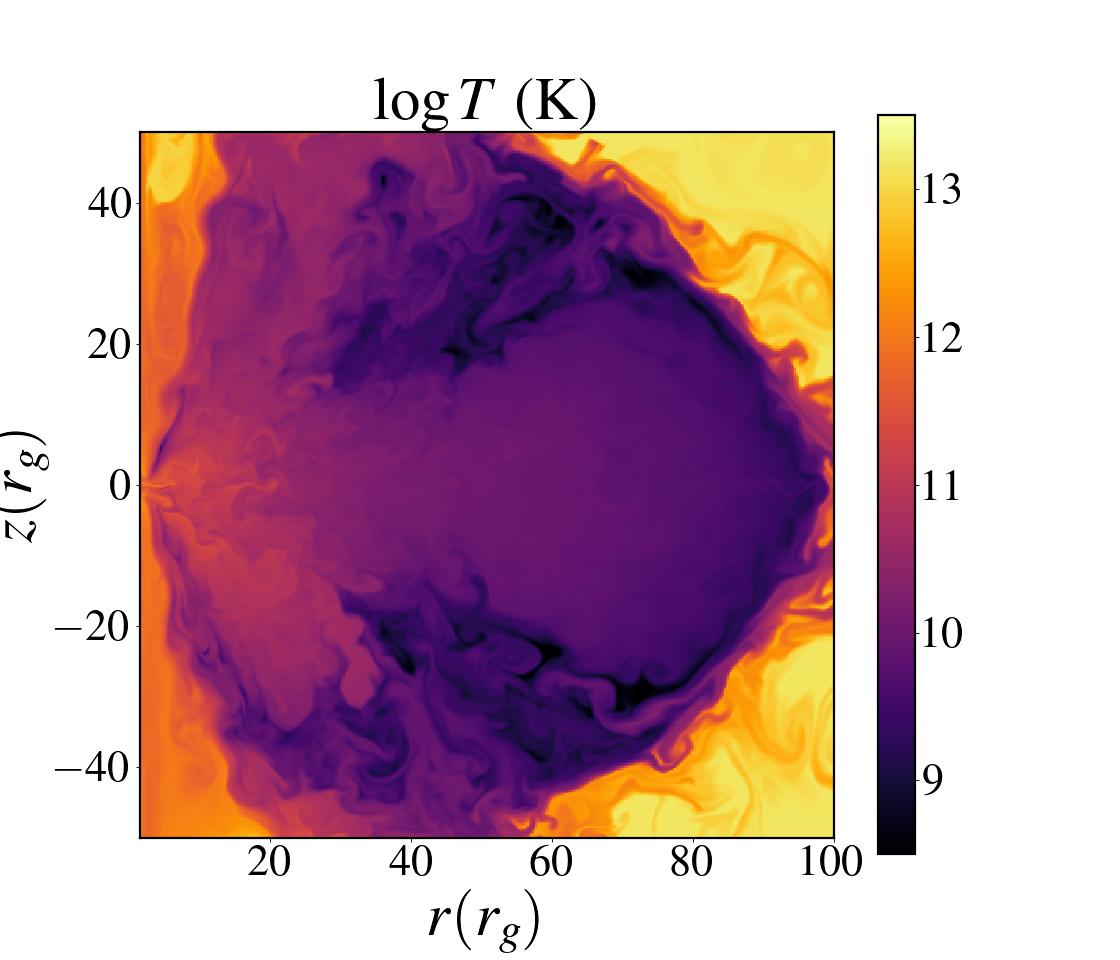} 
        \hskip -4 mm
        \includegraphics[width=0.26\textwidth]{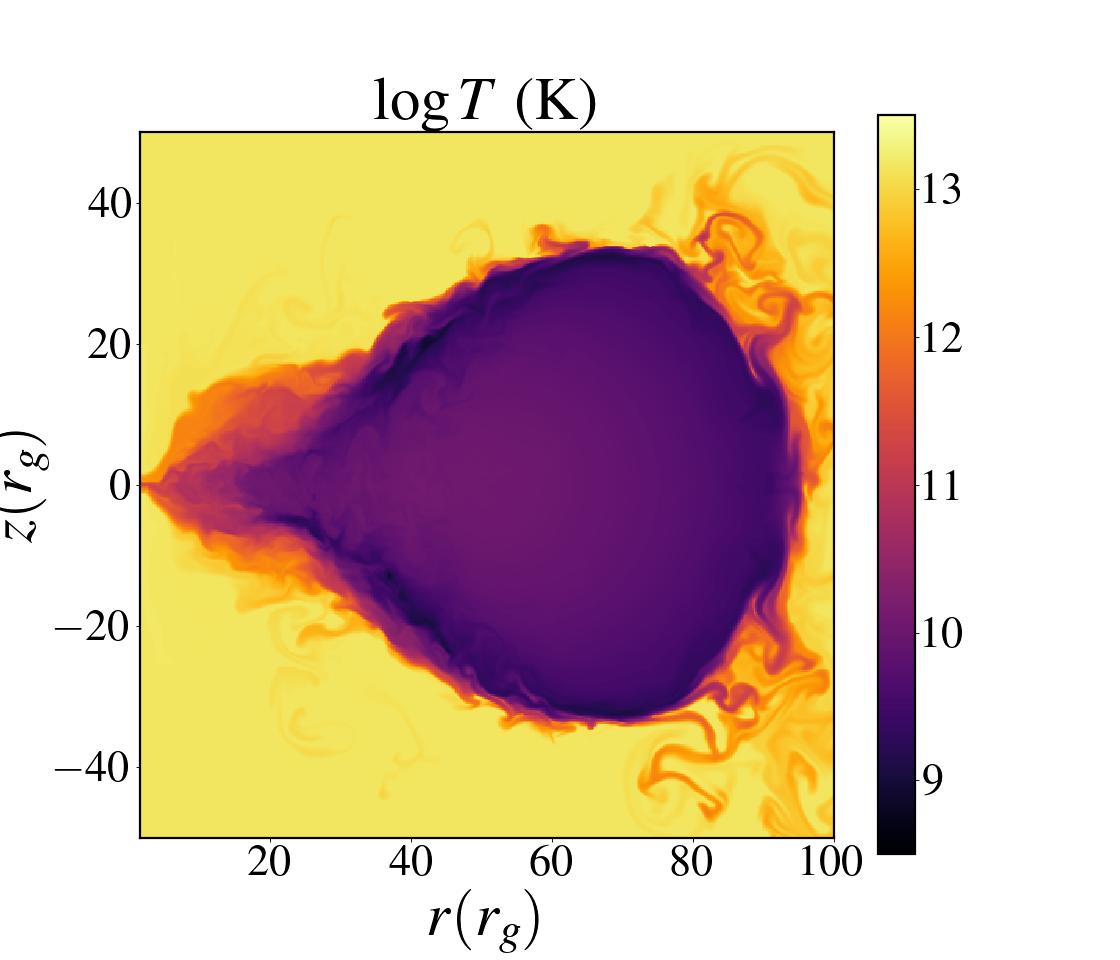} 
        
        \includegraphics[width=0.26\textwidth]{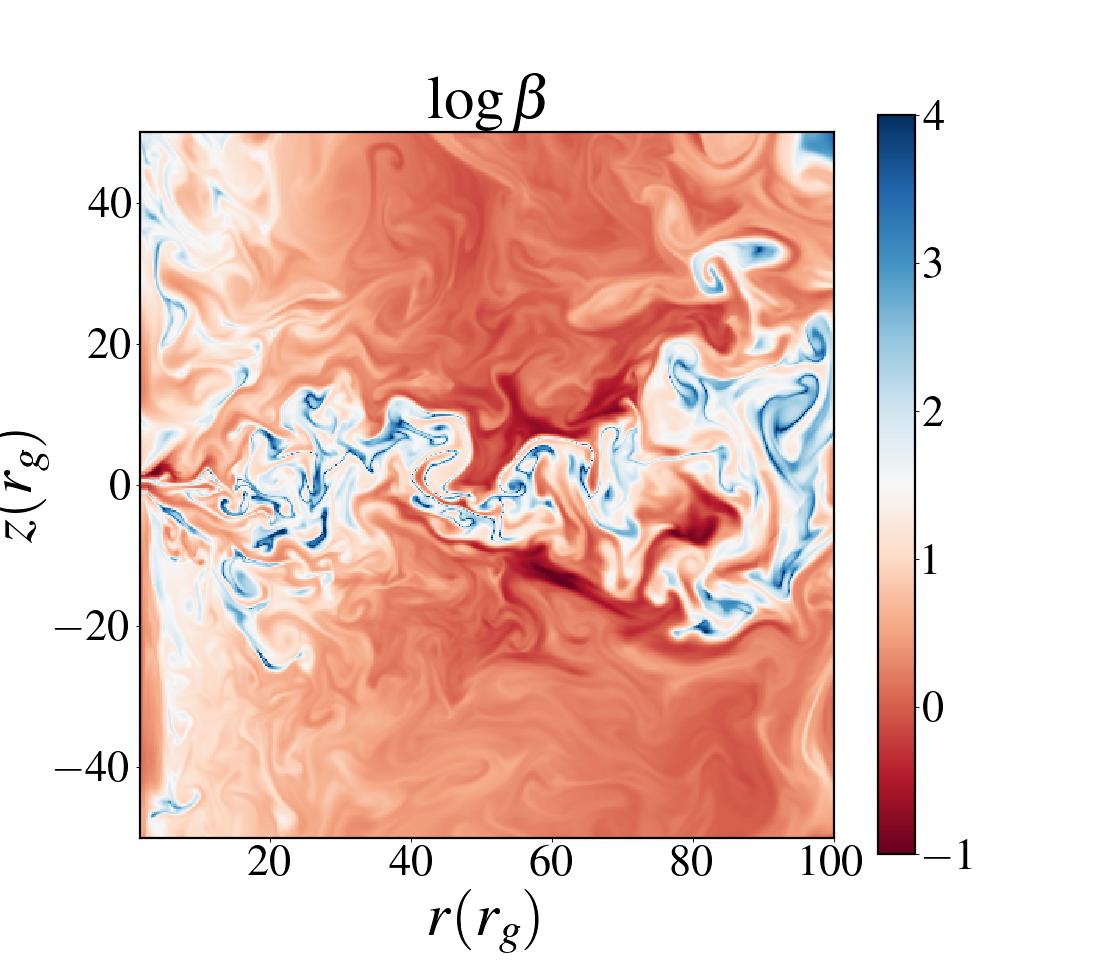} 
        \hskip -4 mm
        \includegraphics[width=0.26\textwidth]{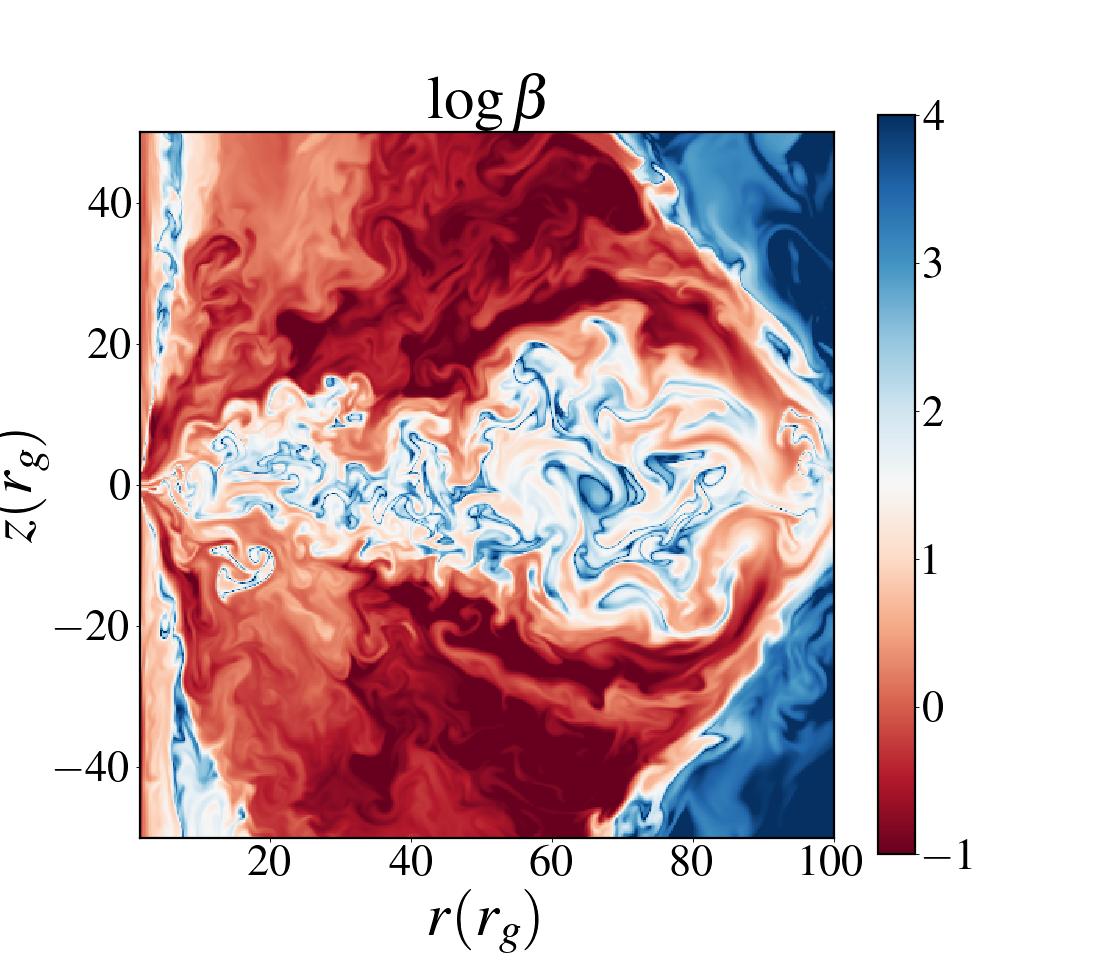} 
        \hskip -4 mm
        \includegraphics[width=0.26\textwidth]{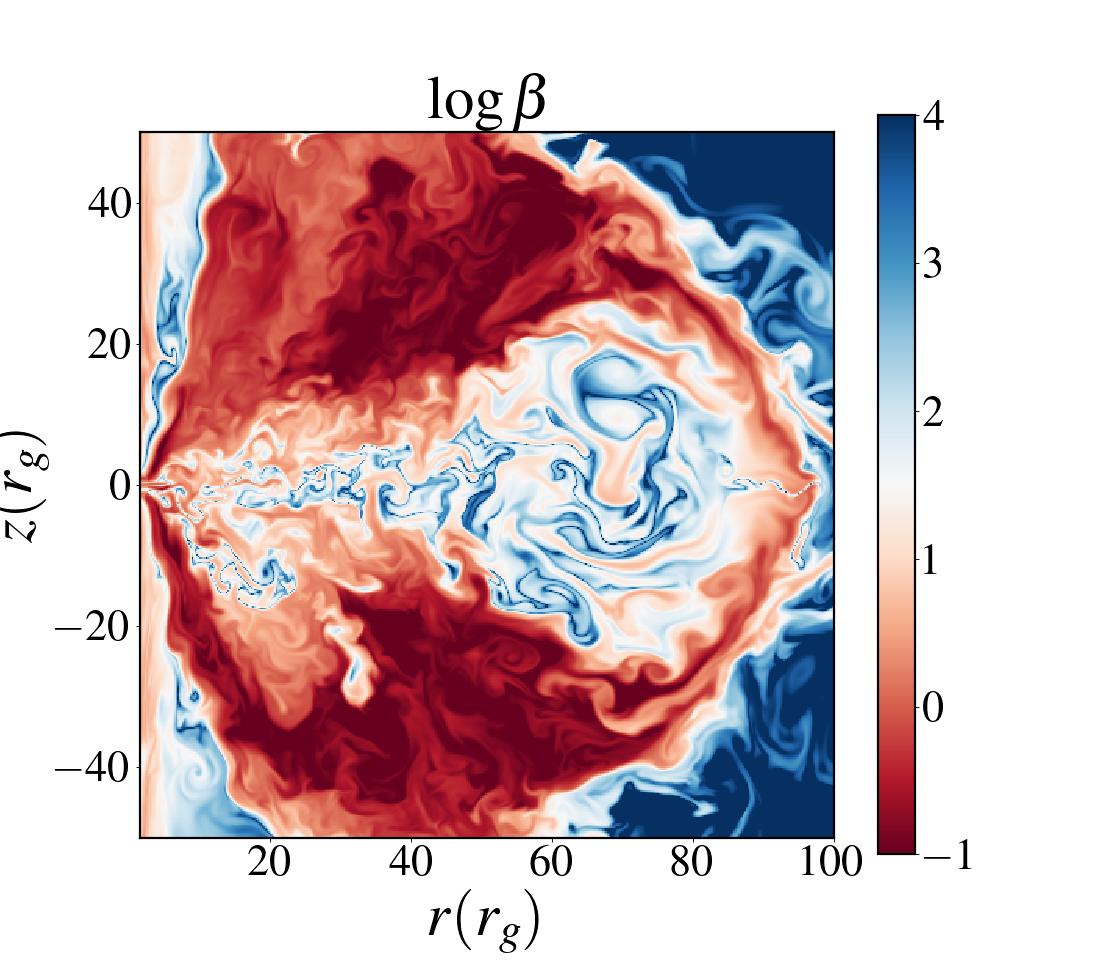} 
        \hskip -4 mm
        \includegraphics[width=0.26\textwidth]{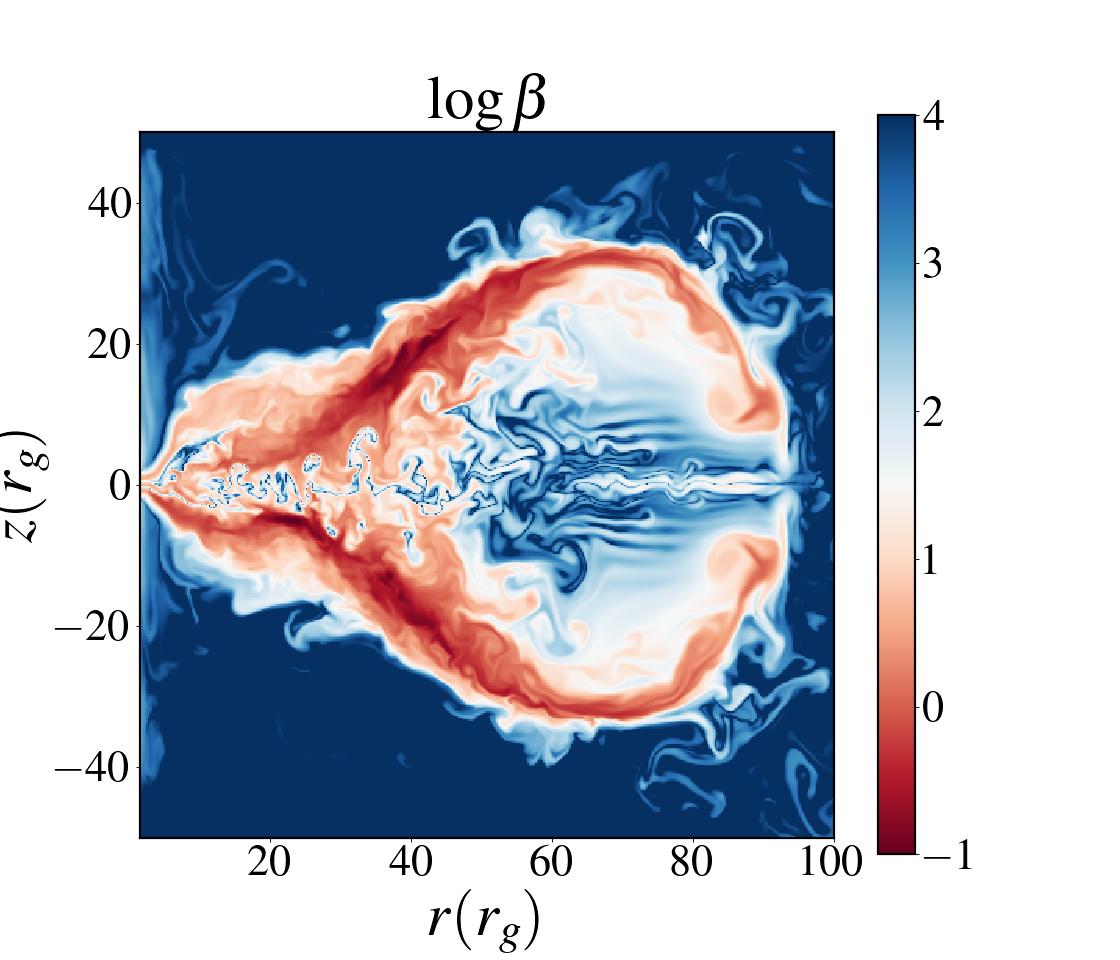} 
        
        \includegraphics[width=0.26\textwidth]{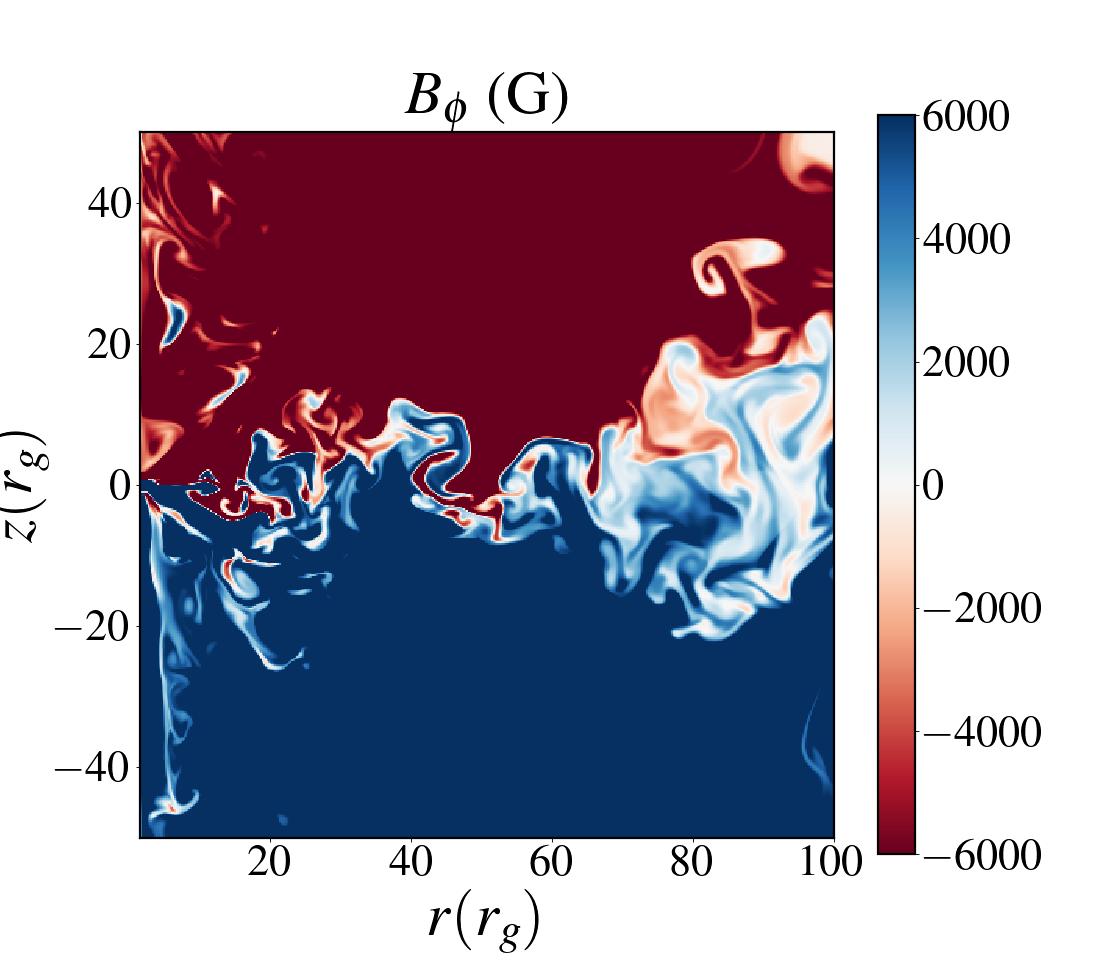} 
        \hskip -4 mm
        \includegraphics[width=0.26\textwidth]{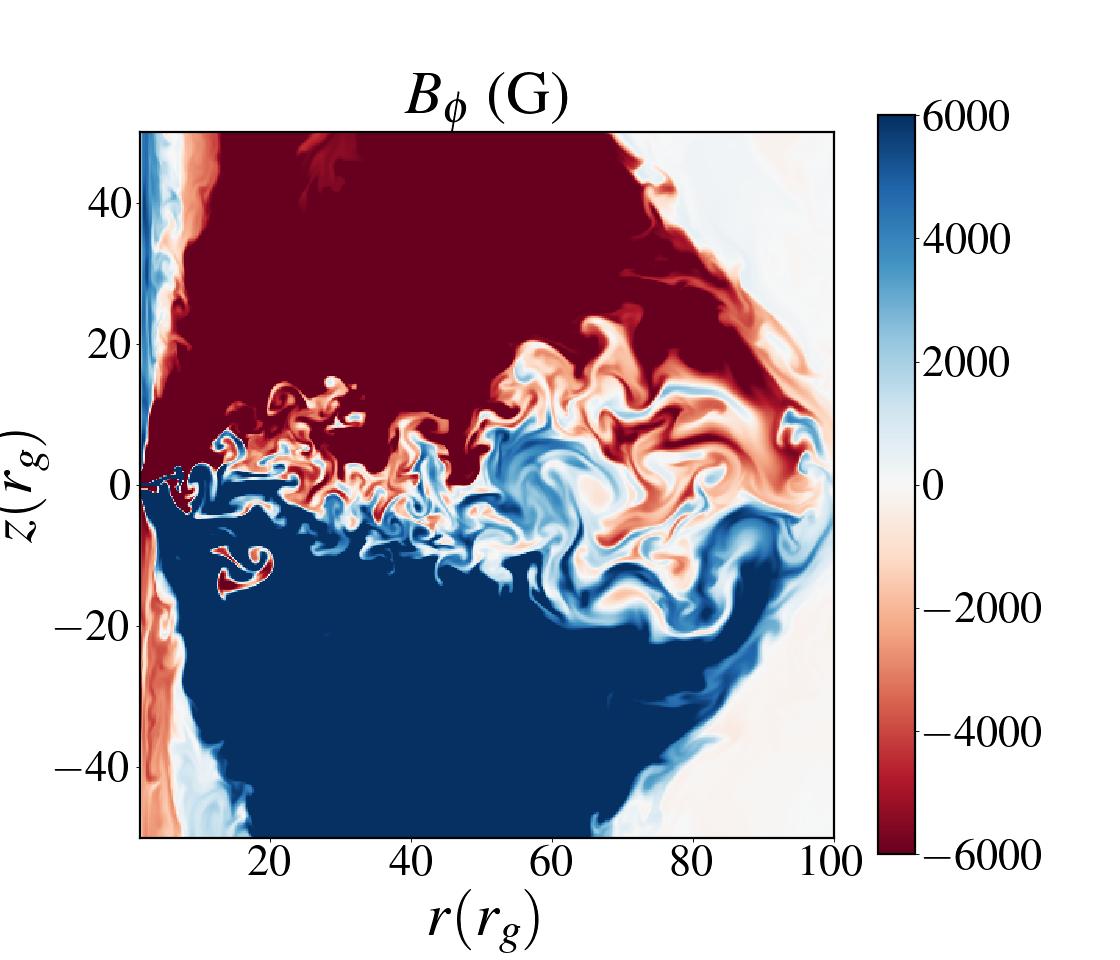} 
        \hskip -4 mm
        \includegraphics[width=0.26\textwidth]{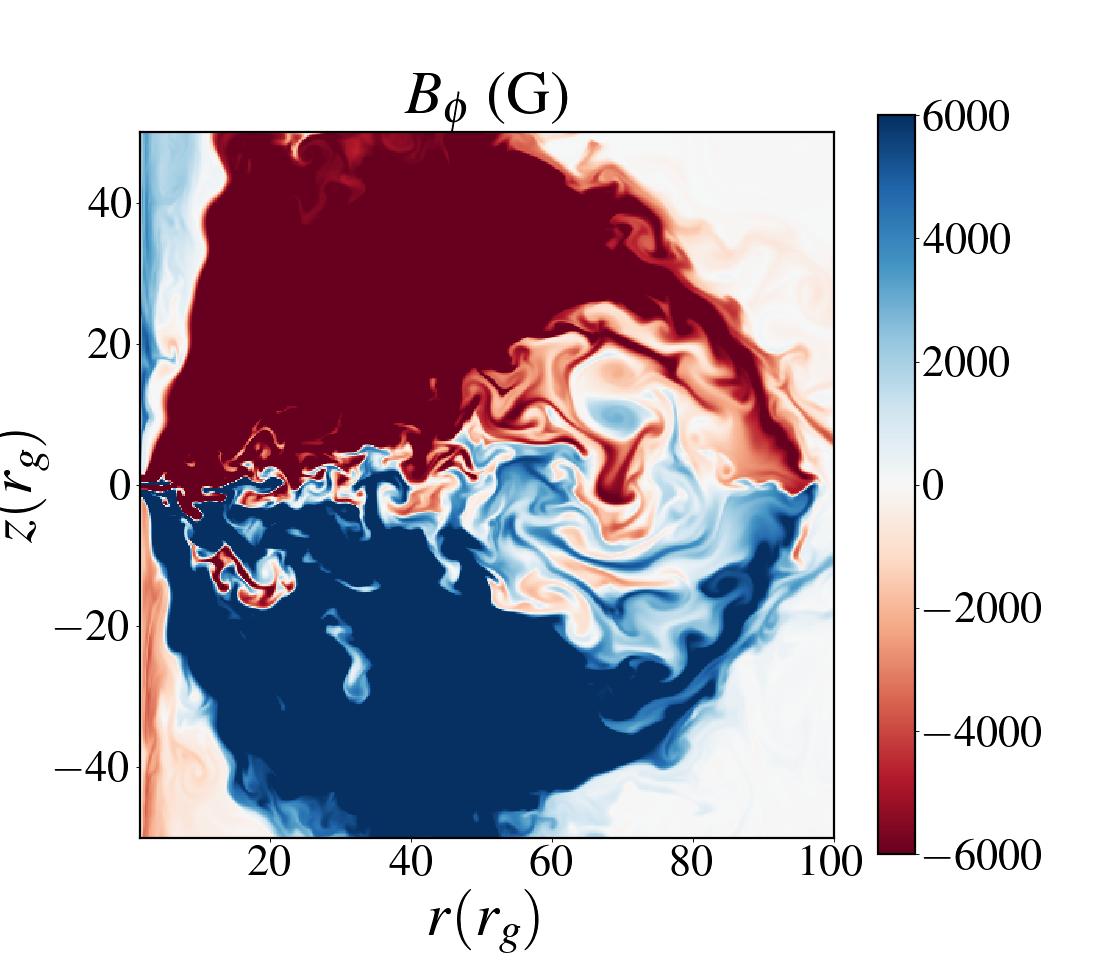} 
        \hskip -4 mm
        \includegraphics[width=0.26\textwidth]{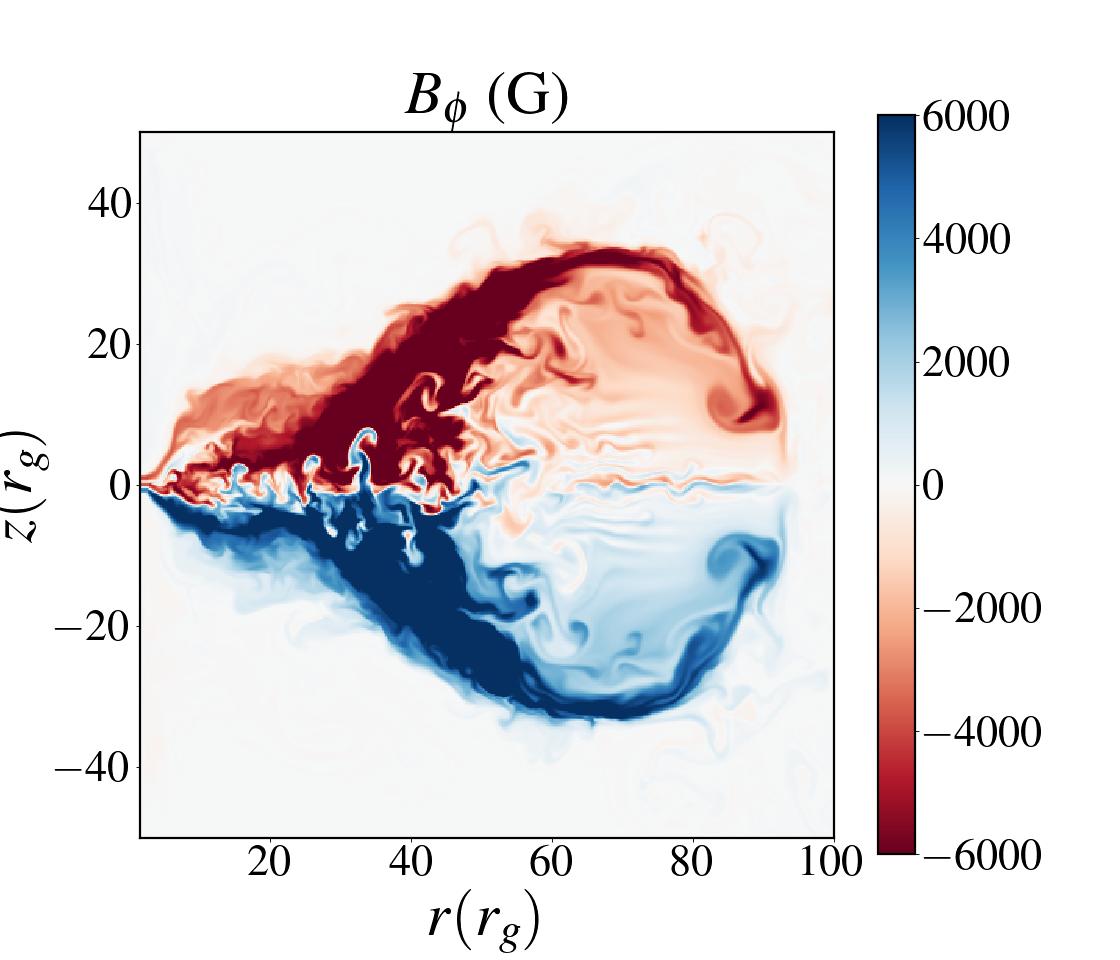} 

        \includegraphics[width=0.26\textwidth]{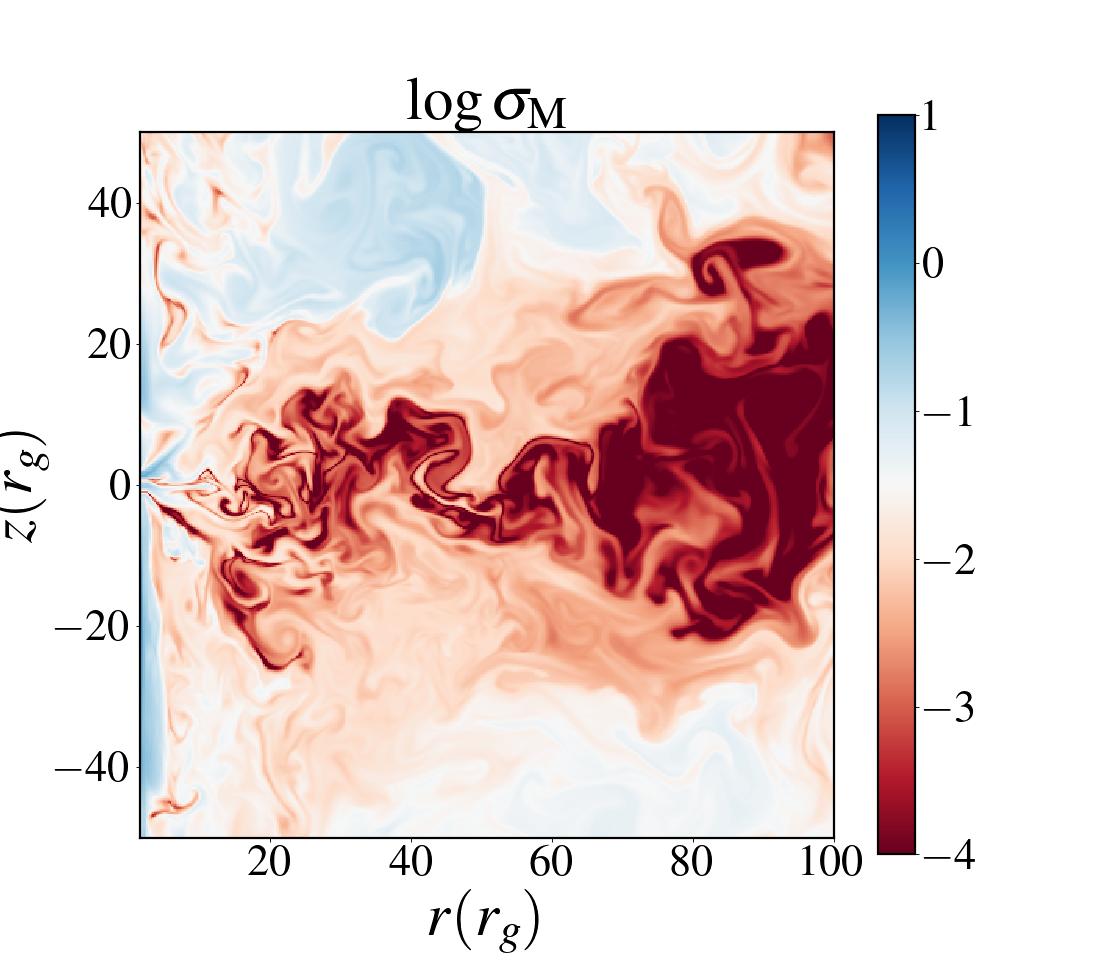} 
        \hskip -4 mm
        \includegraphics[width=0.26\textwidth]{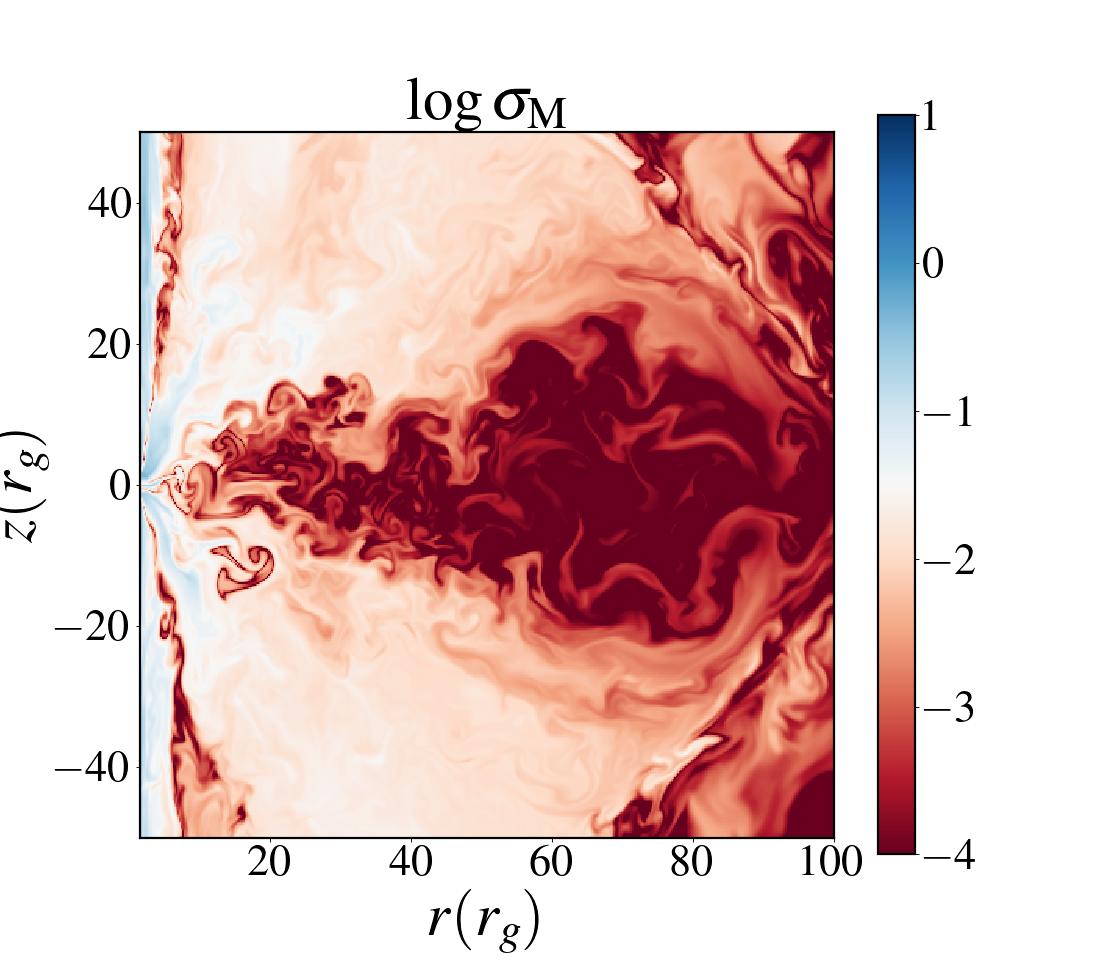} 
        \hskip -4 mm
        \includegraphics[width=0.26\textwidth]{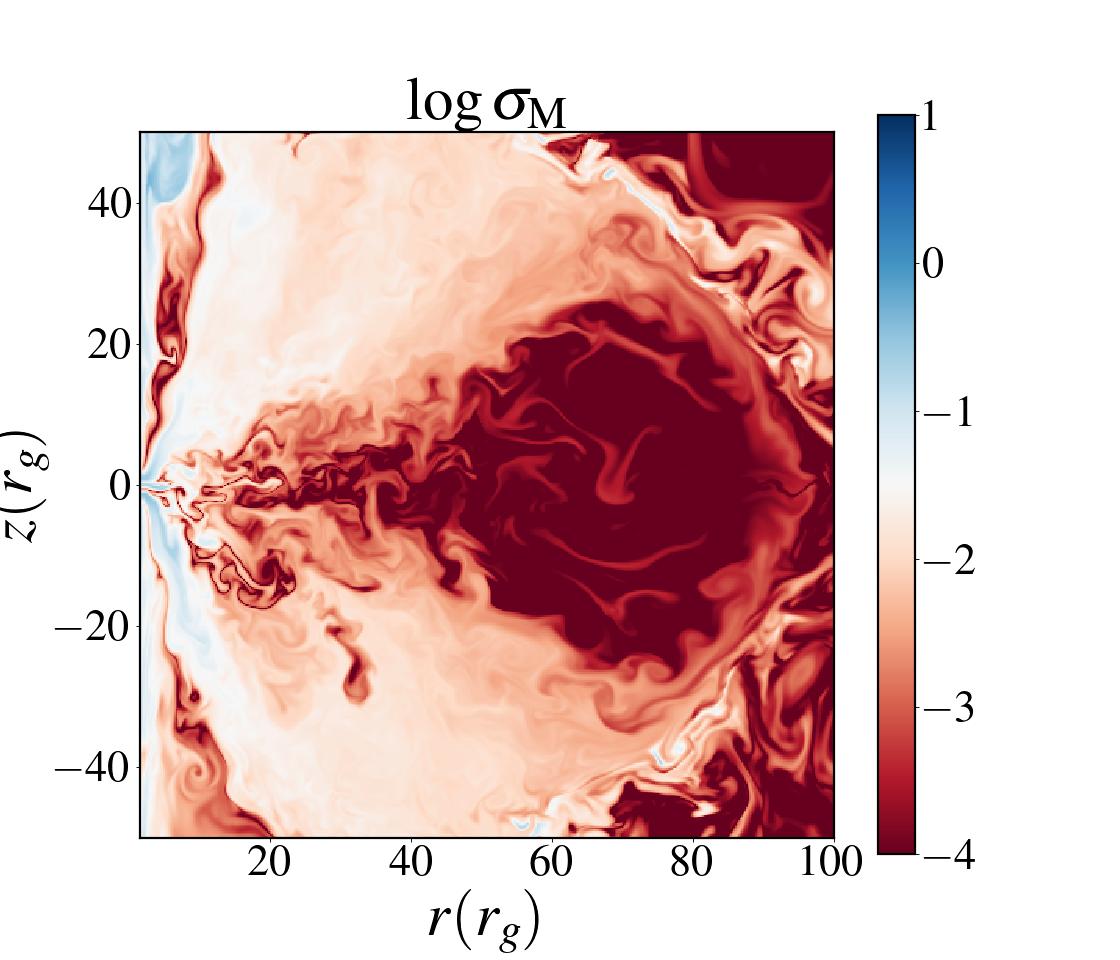} 
        \hskip -4 mm
        \includegraphics[width=0.26\textwidth]{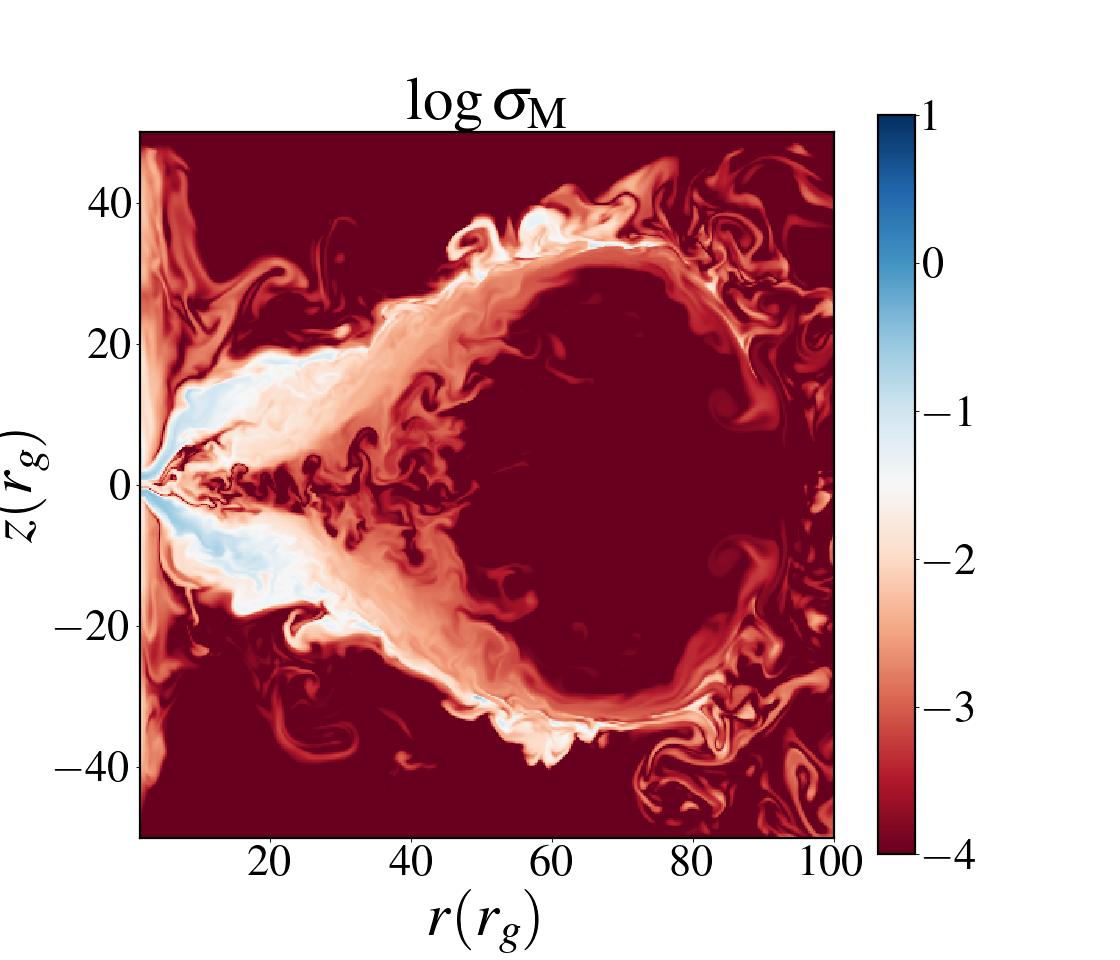} 
	\end{center}
	\caption{Distribution of Temperature ($T$), plasma-$\beta$, azimuthal magnetic field ($B_{\phi}$) and magnetization parameter $(\sigma_{\rm M})$ for different initial magnetic field ($\beta_0$). First, second, third, and fourth columns correspond to $\beta_0$ = 10, 50, 100, and 1000, respectively, at time $t_3 = 10500 t_g$. See the text for details.}
	\label{Figure_3}
\end{figure*}

\begin{figure}
	\begin{center}
		\includegraphics[width=0.5\textwidth]{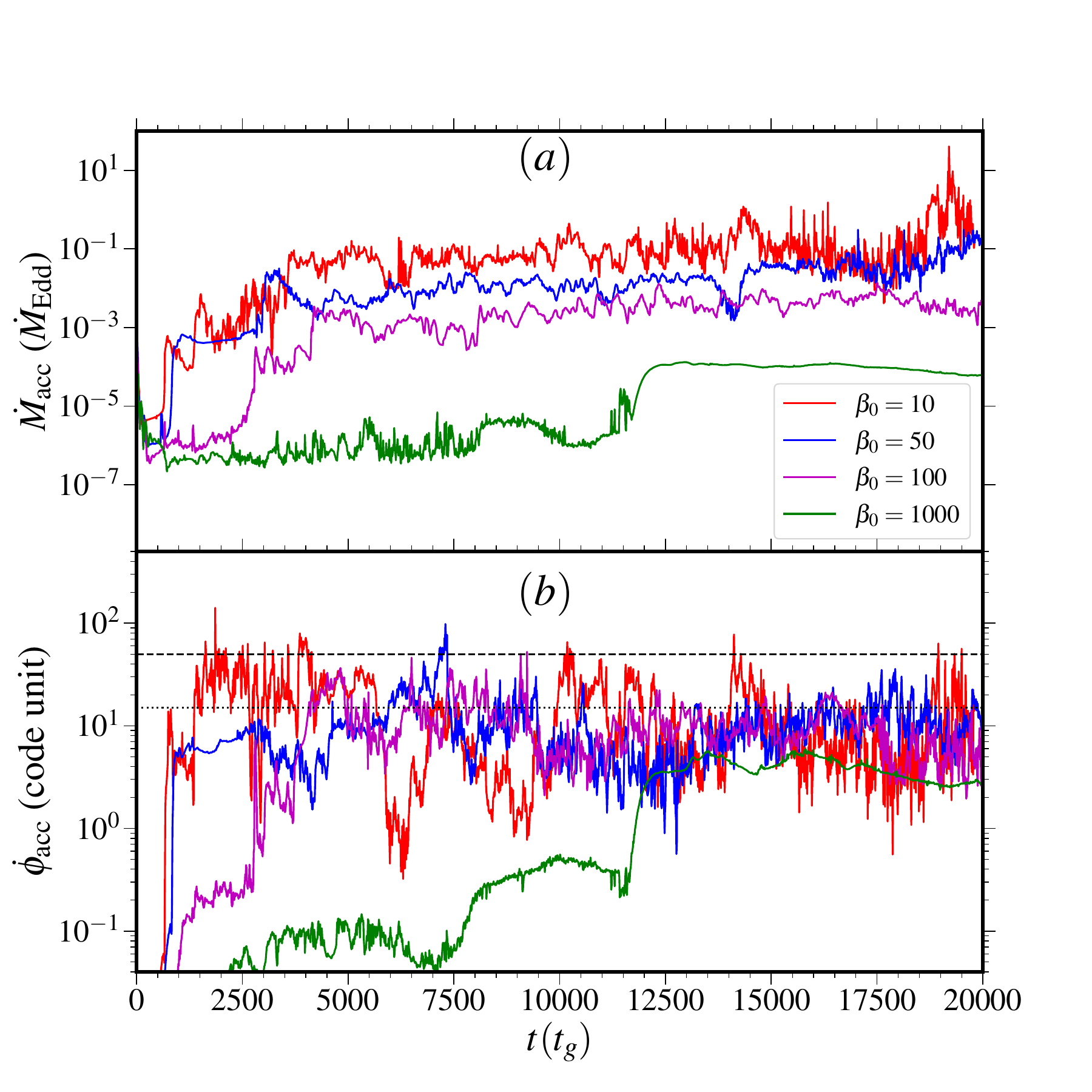} 
	\end{center}
	\caption{Temporal evolution of $(a)$ mass accretion rate $(\dot{M}_{\rm acc})$ in Eddington unit and $(b)$: normalized magnetic flux $(\dot{\phi}_{\rm acc})$ in code units accumulated at the black hole inner boundary with the simulation time for different initial plasma-$\beta$ parameter $\beta_0$ = 10 (red), 50 (blue), 100 (magenta) and 1000 (green), respectively. Dashed and dotted horizontal lines are for $\dot{\phi_{\rm acc}} = 15$ and $50$, respectively. See the text for details.}
	\label{Figure_4}
\end{figure}

\begin{figure*}
	\begin{center}
		\includegraphics[width=1.0\textwidth]{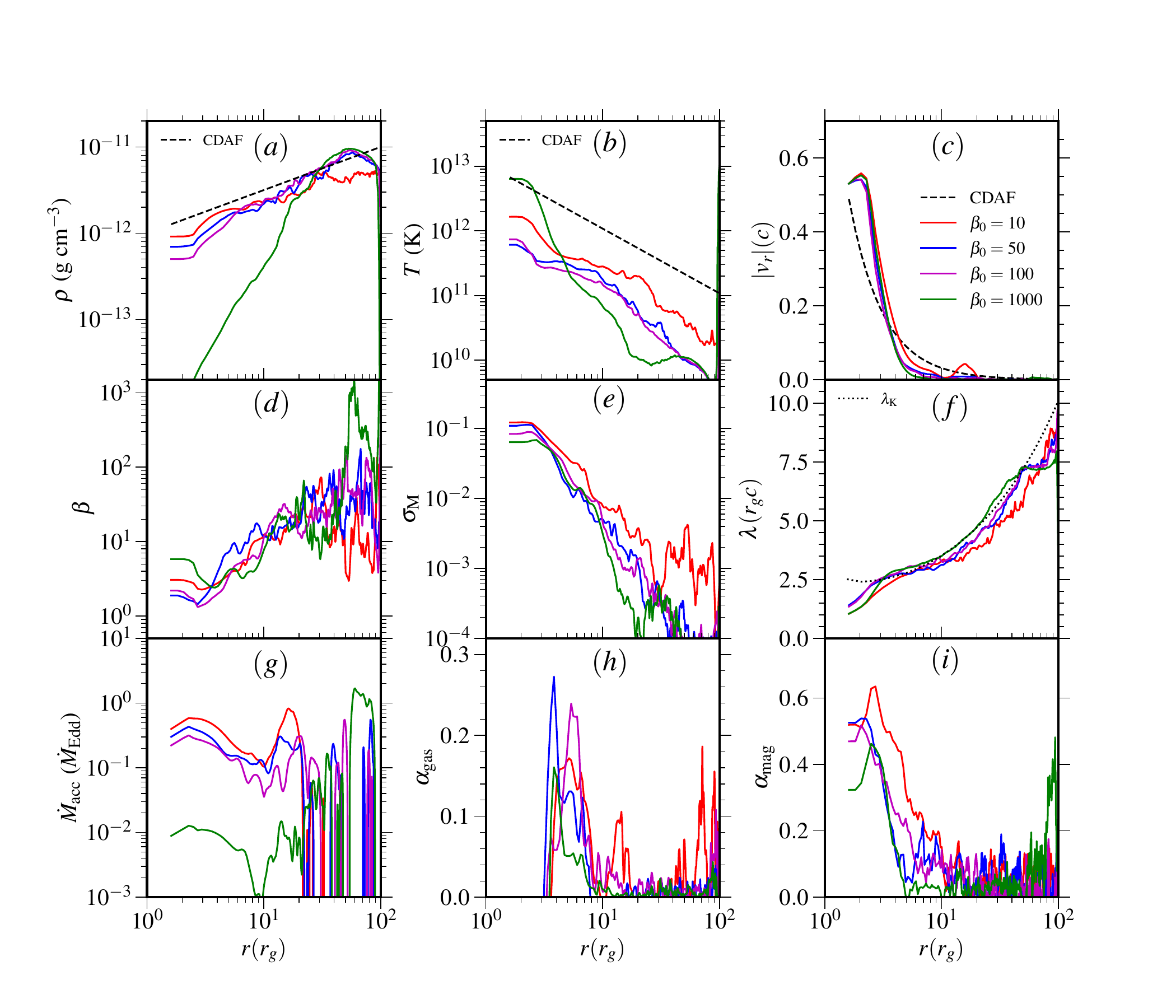} 
	\end{center}
	\caption{Radial variations of different flow variables: ($a$): density ($\rho$), ($b$): temperature ($T$), ($c$): radial velocity ($v_r$), $(d)$: plasma-beta ($\beta$), $(e)$:  magnetization parameter ($\sigma_{\rm M}$), $(f)$:  specific angular momentum ($\lambda$), $(g)$: mass accretion rate ($\dot{M}_{\rm acc}$), $(h)$: normalized Reynolds stress ($\alpha_{\rm gas}$), and $(i)$: normalized Maxwell stress ($\alpha_{\rm mag}$), respectively. Here, CDAF implies Convection Dominated Accretion Flow and $\lambda_{\rm K}$ is the Keplerian angular momentum. See the text for details.}
	\label{Figure_5}
\end{figure*}

\begin{figure}
	\begin{center}
		\includegraphics[width=0.5\textwidth]{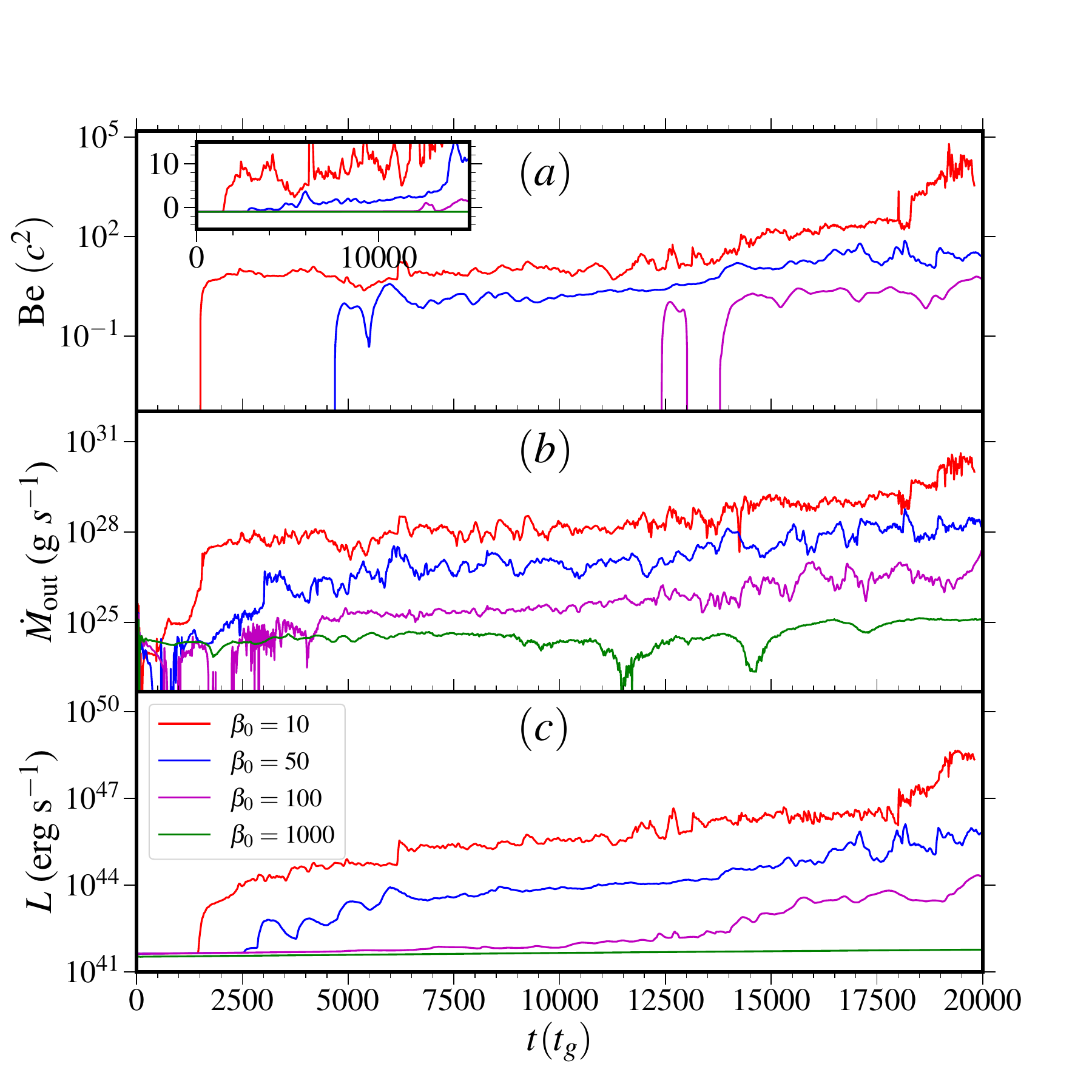} 
	\end{center}
	\caption{Variation of $(a)$: Bernoulli parameter ($Be$) $(b)$: mass outflow rates $(\dot{M}_{\rm out})$ and $(c)$: luminosity $(L)$ with time for different $\beta_0$ =10 (red), 50 (blue), 100 (magenta) and 1000 (green), respectively. See the text for details.} 
	\label{Figure_6}
\end{figure}

\begin{figure*}
	\begin{center}
		\includegraphics[width=0.26\textwidth]{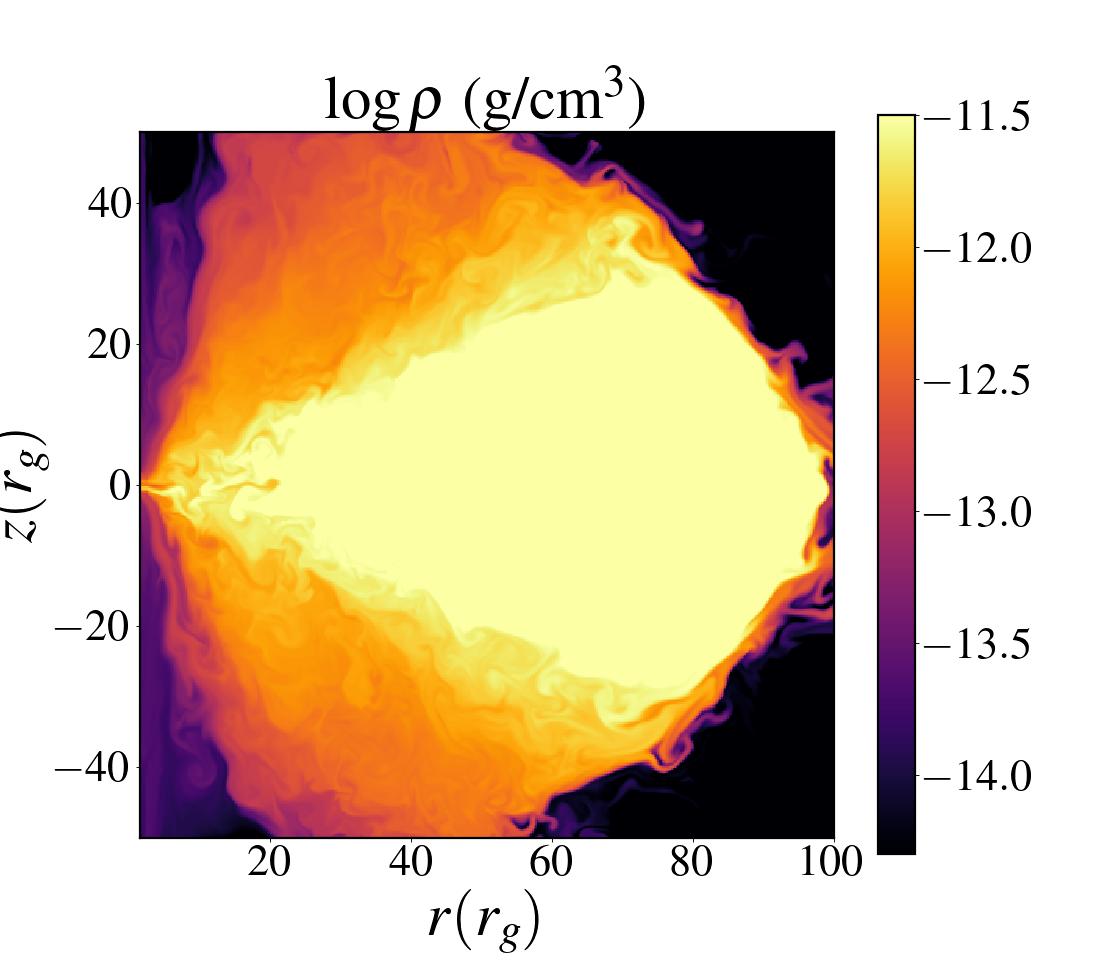} 
        \hskip -4 mm
        \includegraphics[width=0.26\textwidth]{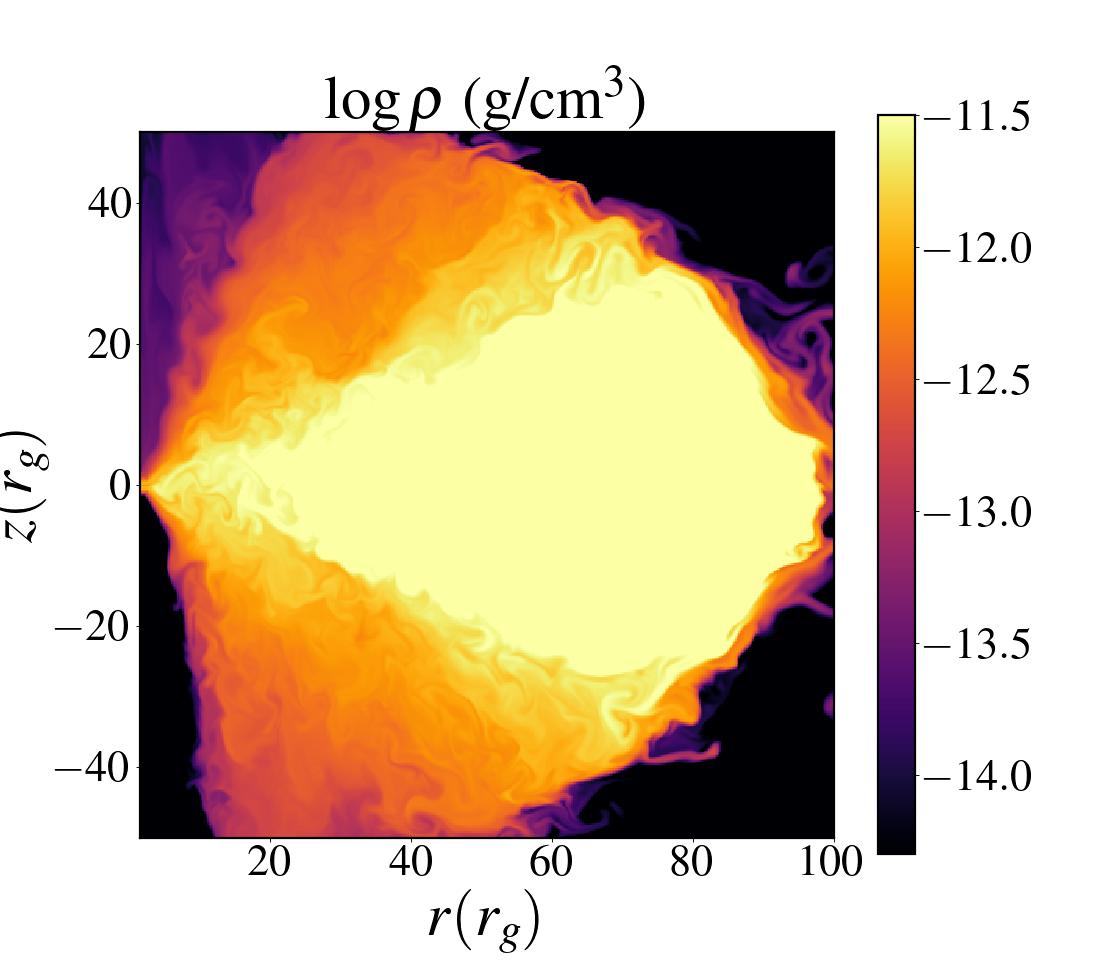} 
        \hskip -4 mm
        \includegraphics[width=0.26\textwidth]{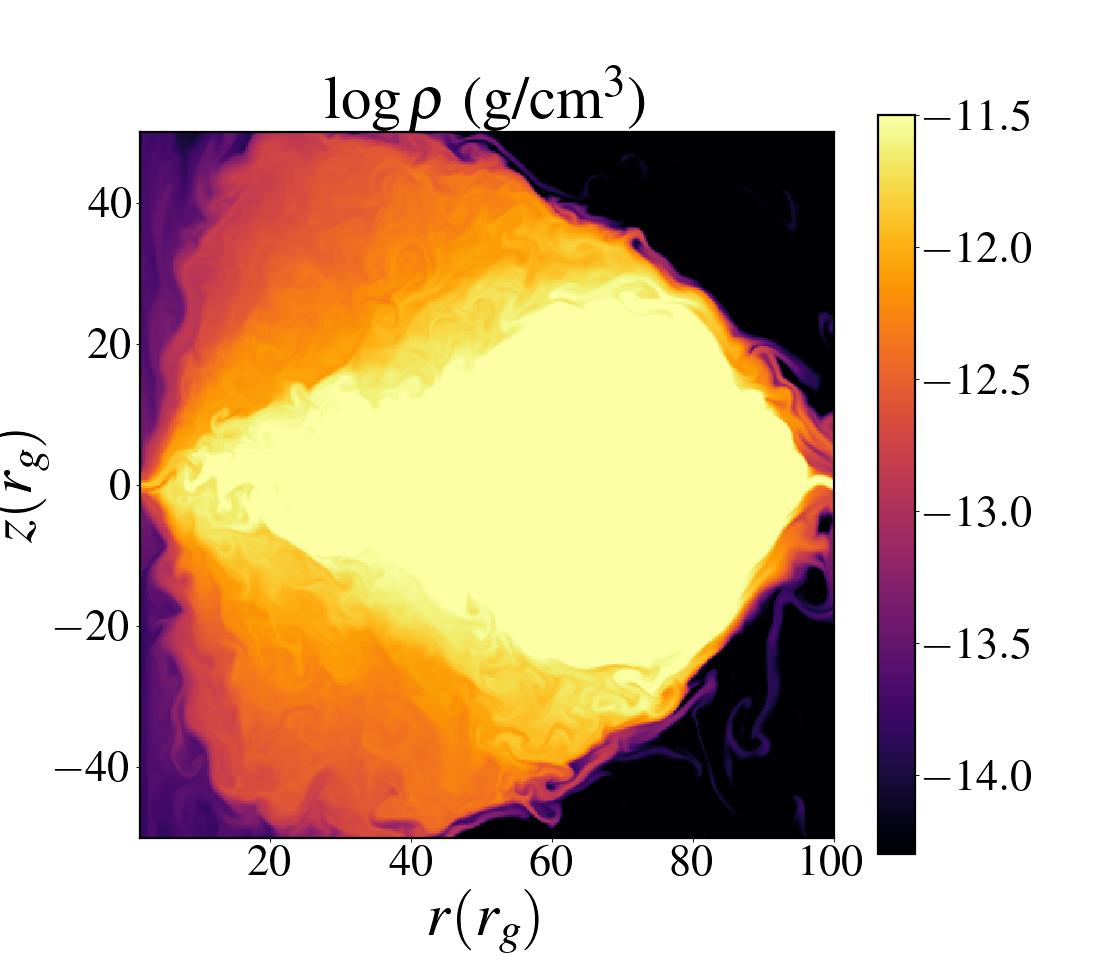} 
        \hskip -4 mm
        \includegraphics[width=0.26\textwidth]{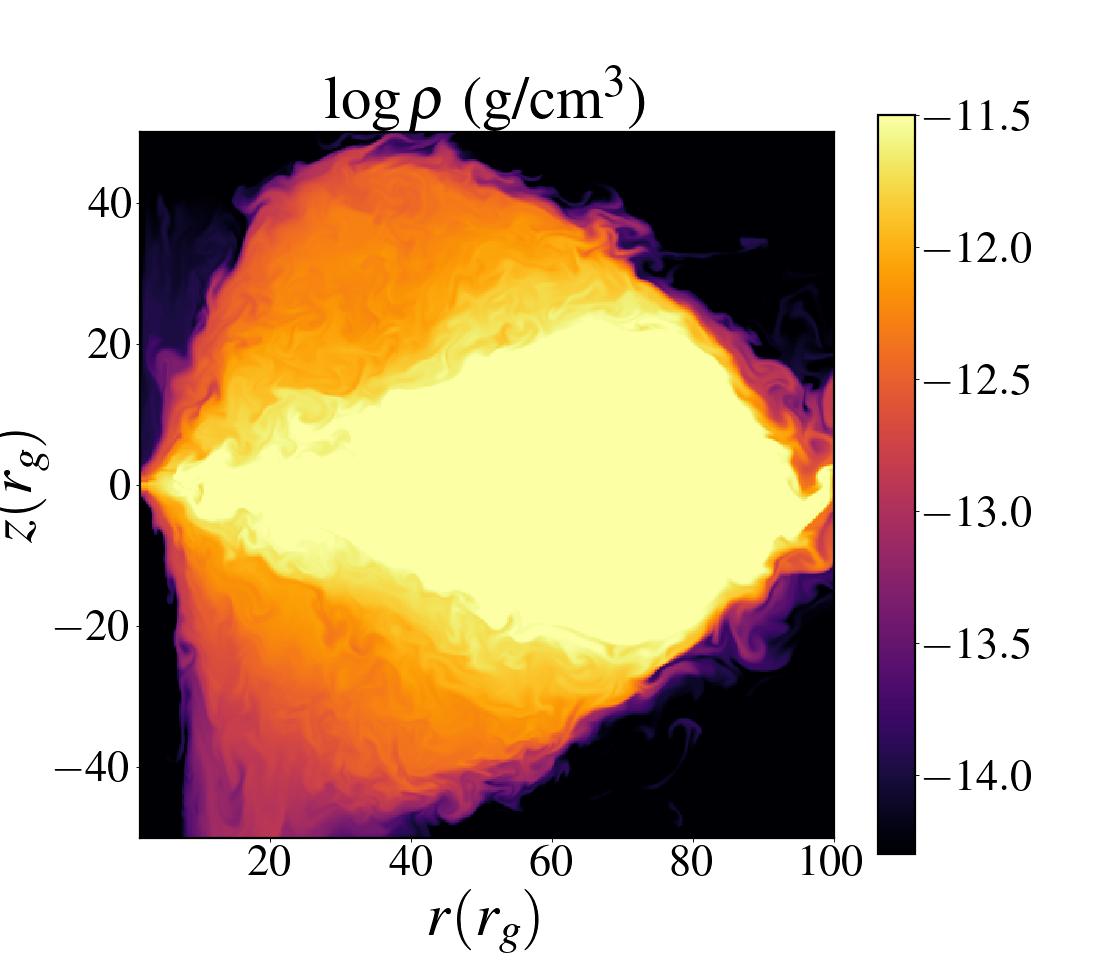} 
        
        \includegraphics[width=0.26\textwidth]{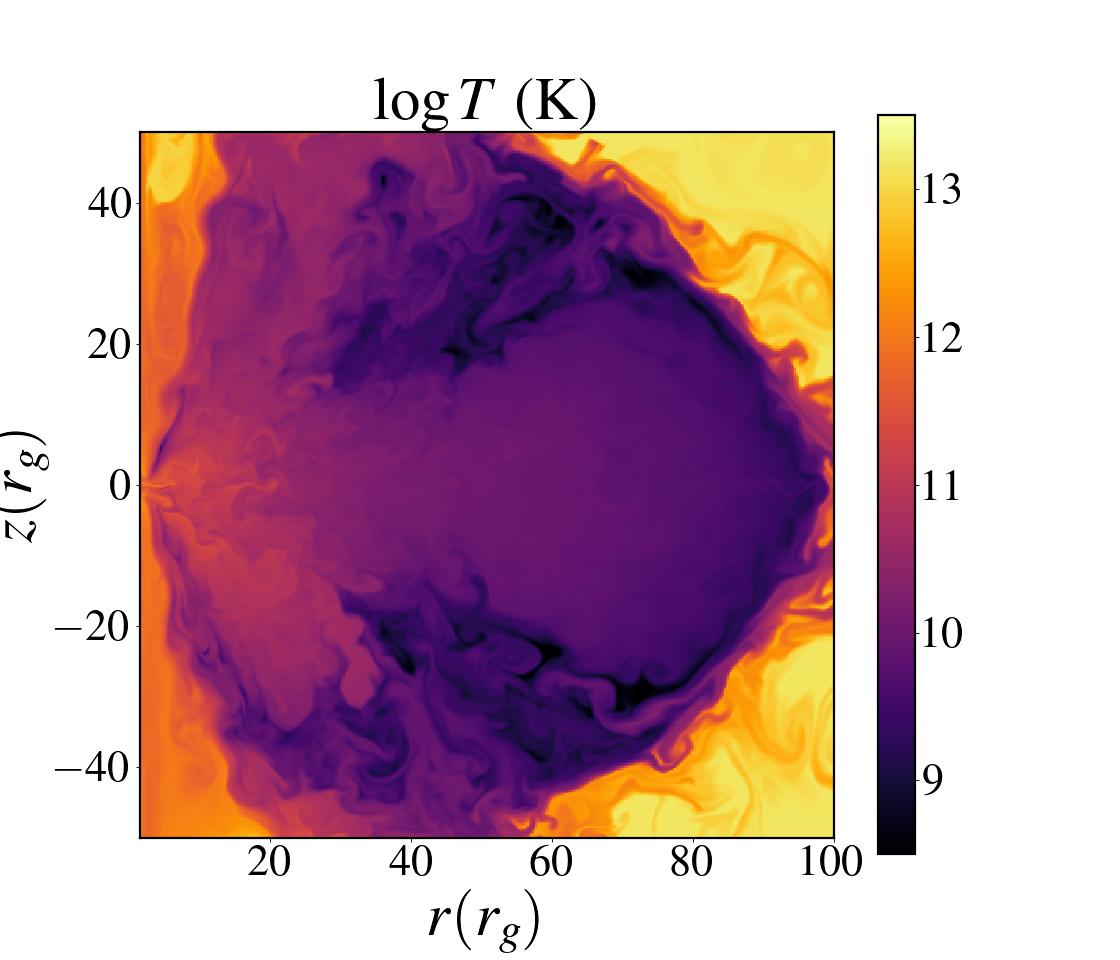} 
        \hskip -4 mm
        \includegraphics[width=0.26\textwidth]{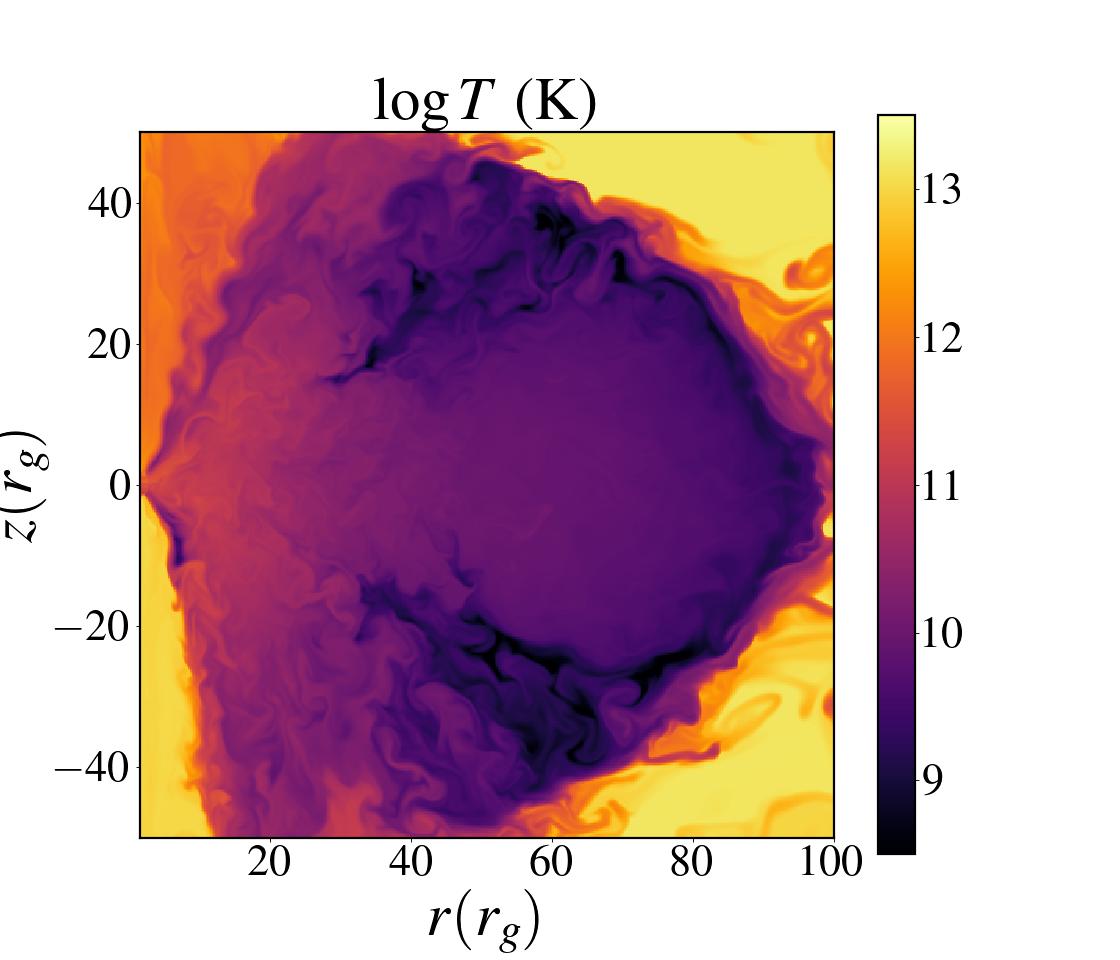} 
        \hskip -4 mm
        \includegraphics[width=0.26\textwidth]{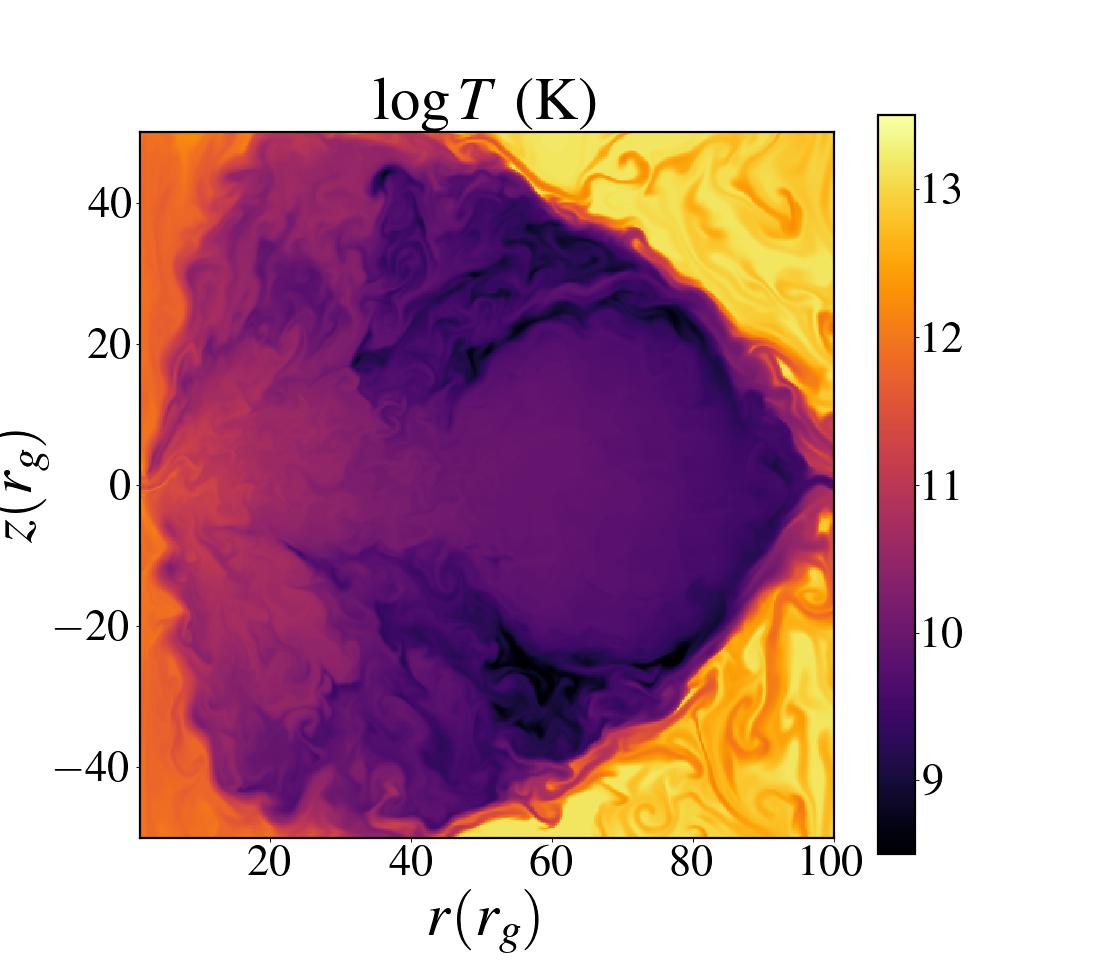} 
        \hskip -4 mm
        \includegraphics[width=0.26\textwidth]{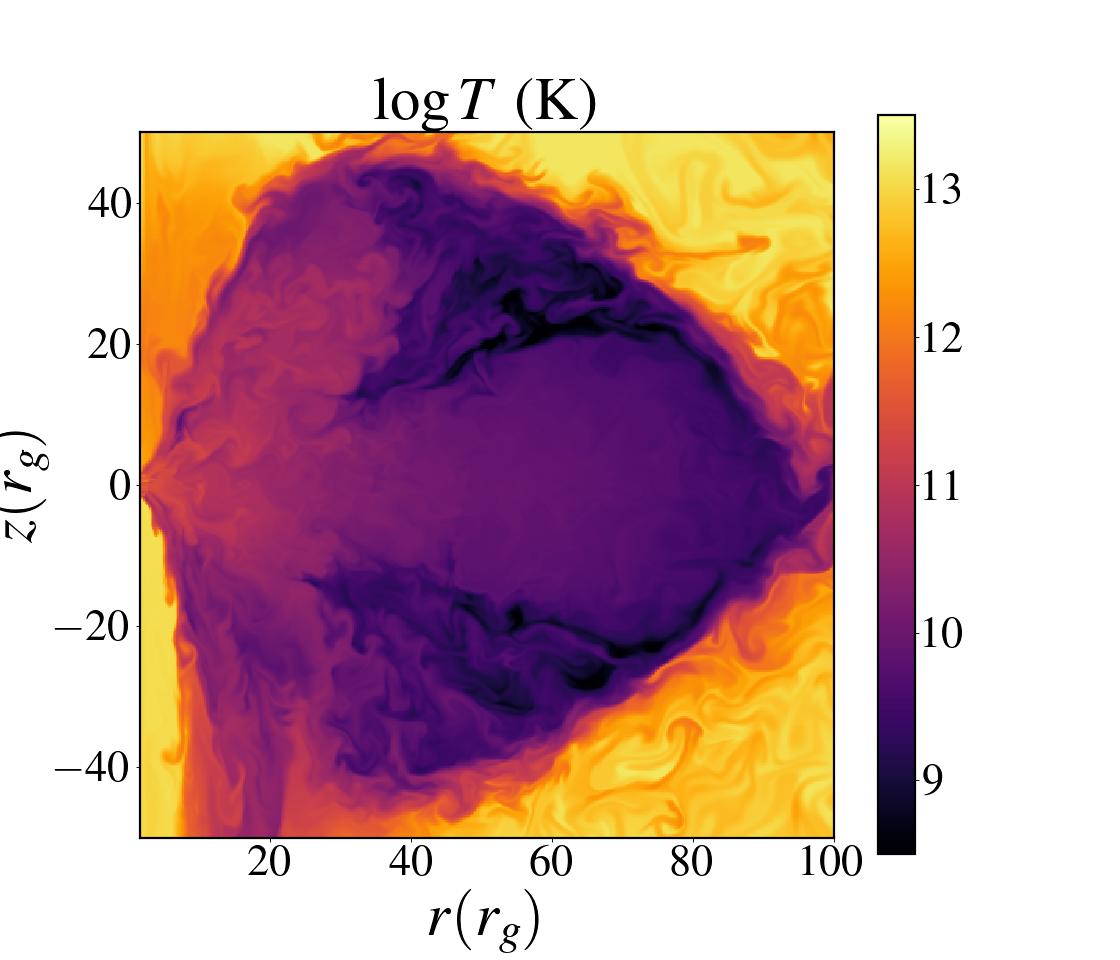} 
       
        \includegraphics[width=0.26\textwidth]{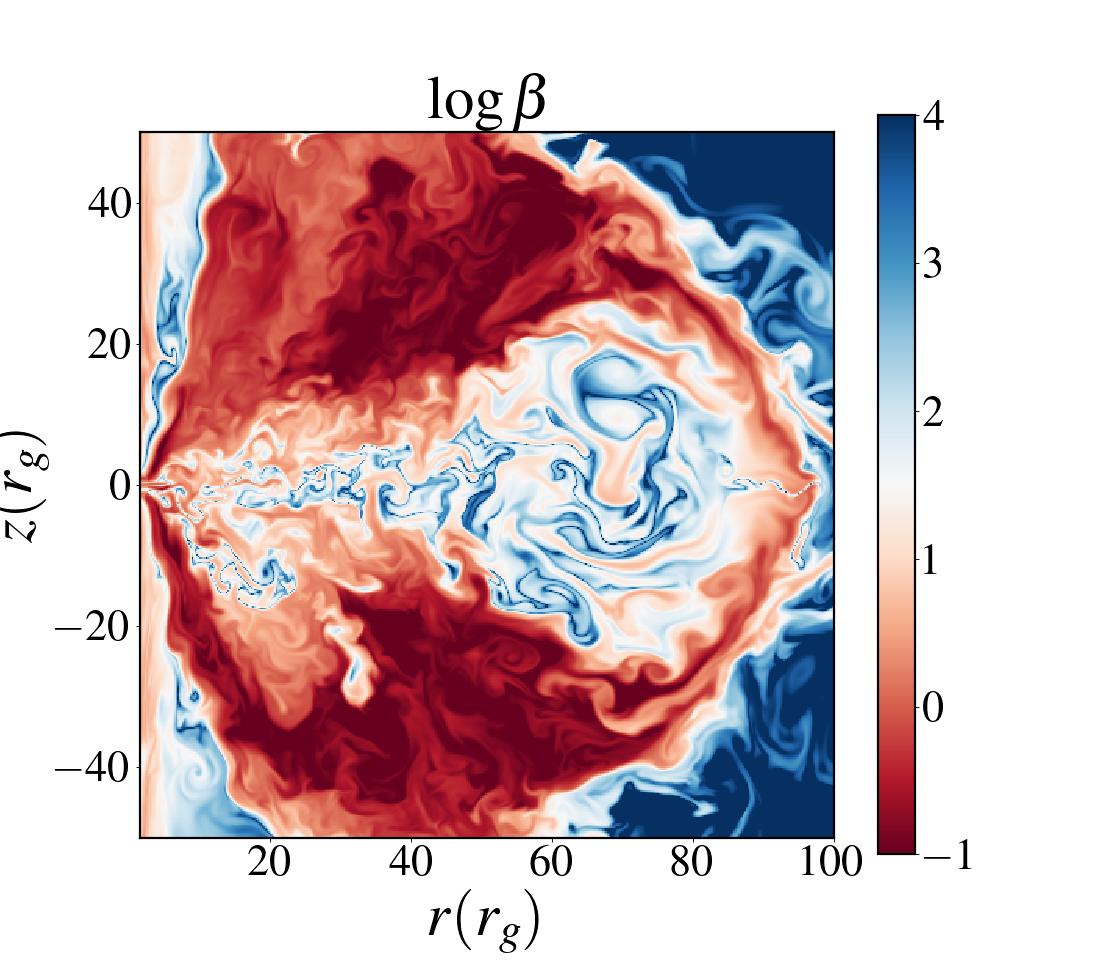} 
        \hskip -4 mm
        \includegraphics[width=0.26\textwidth]{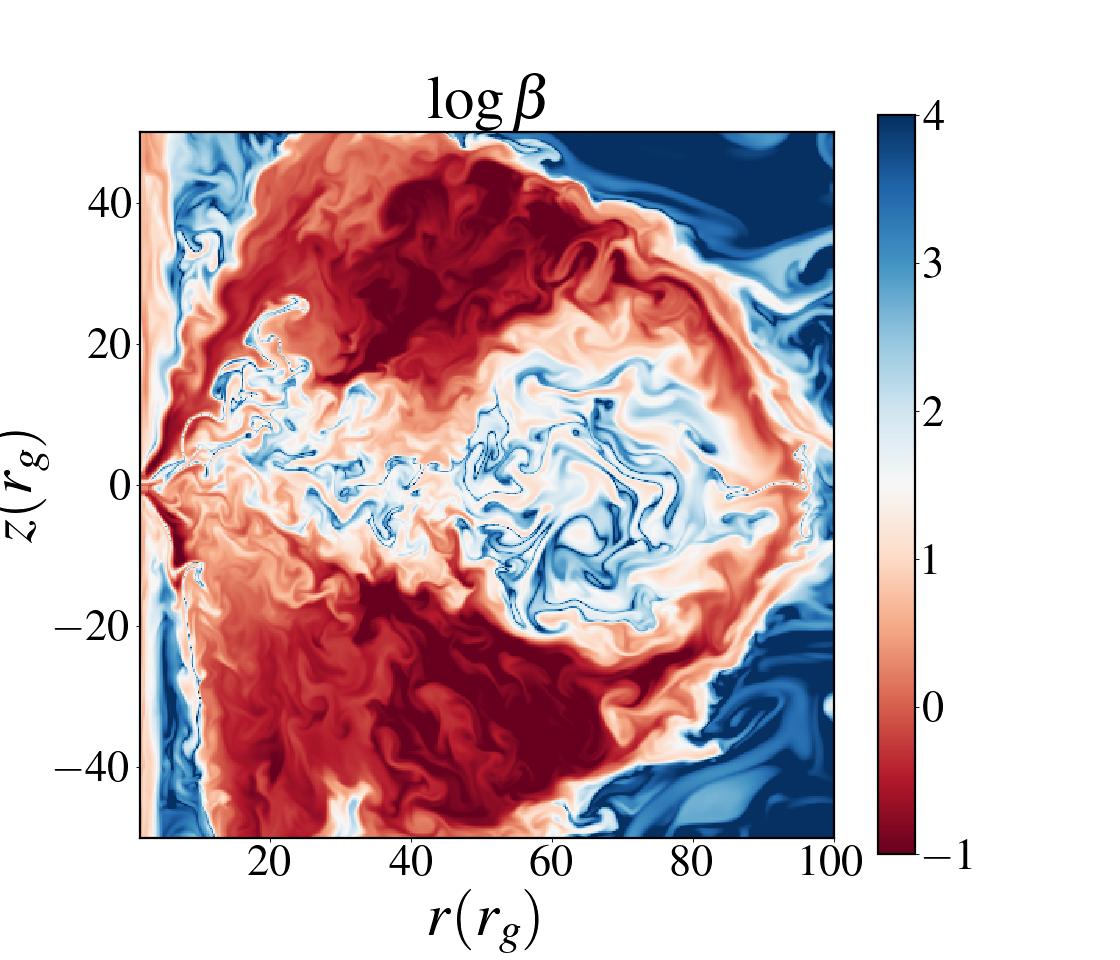} 
        \hskip -4 mm
        \includegraphics[width=0.26\textwidth]{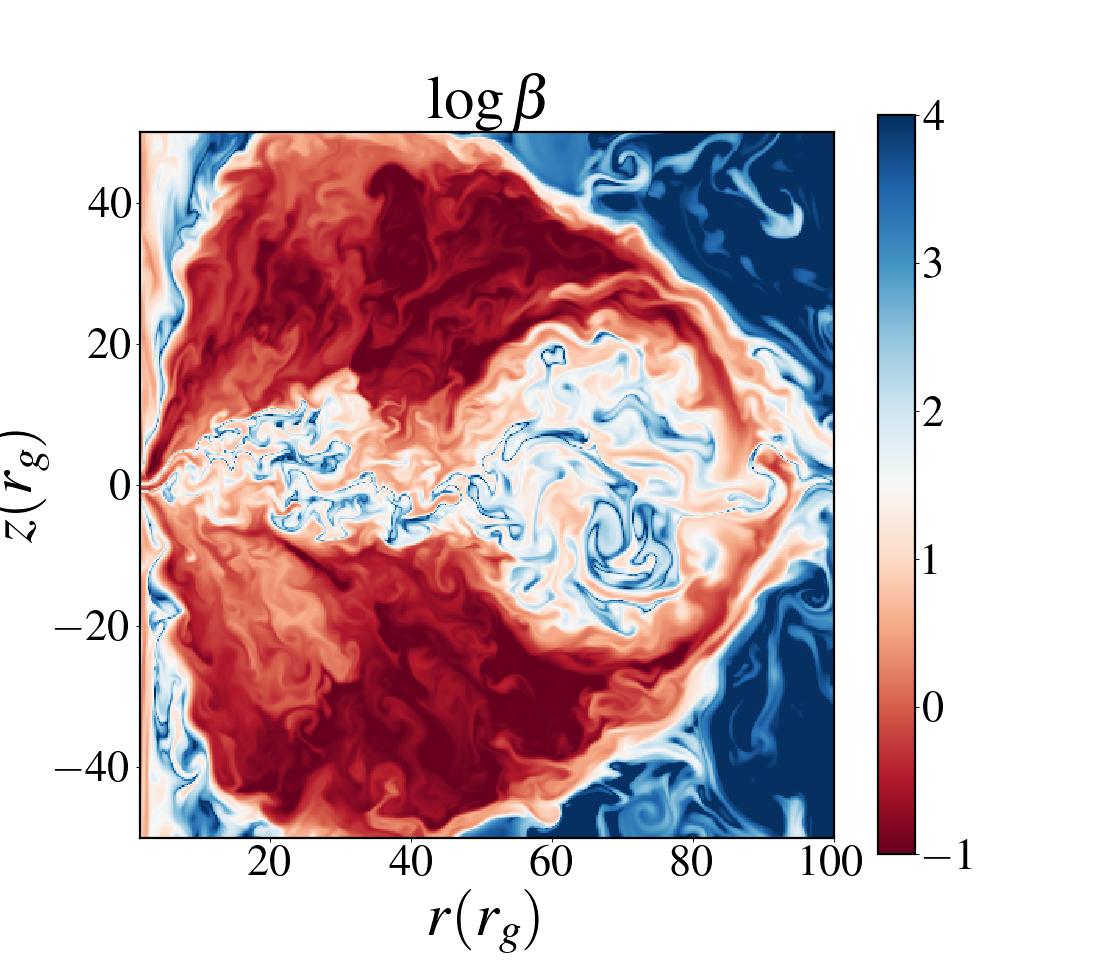} 
        \hskip -4 mm
        \includegraphics[width=0.26\textwidth]{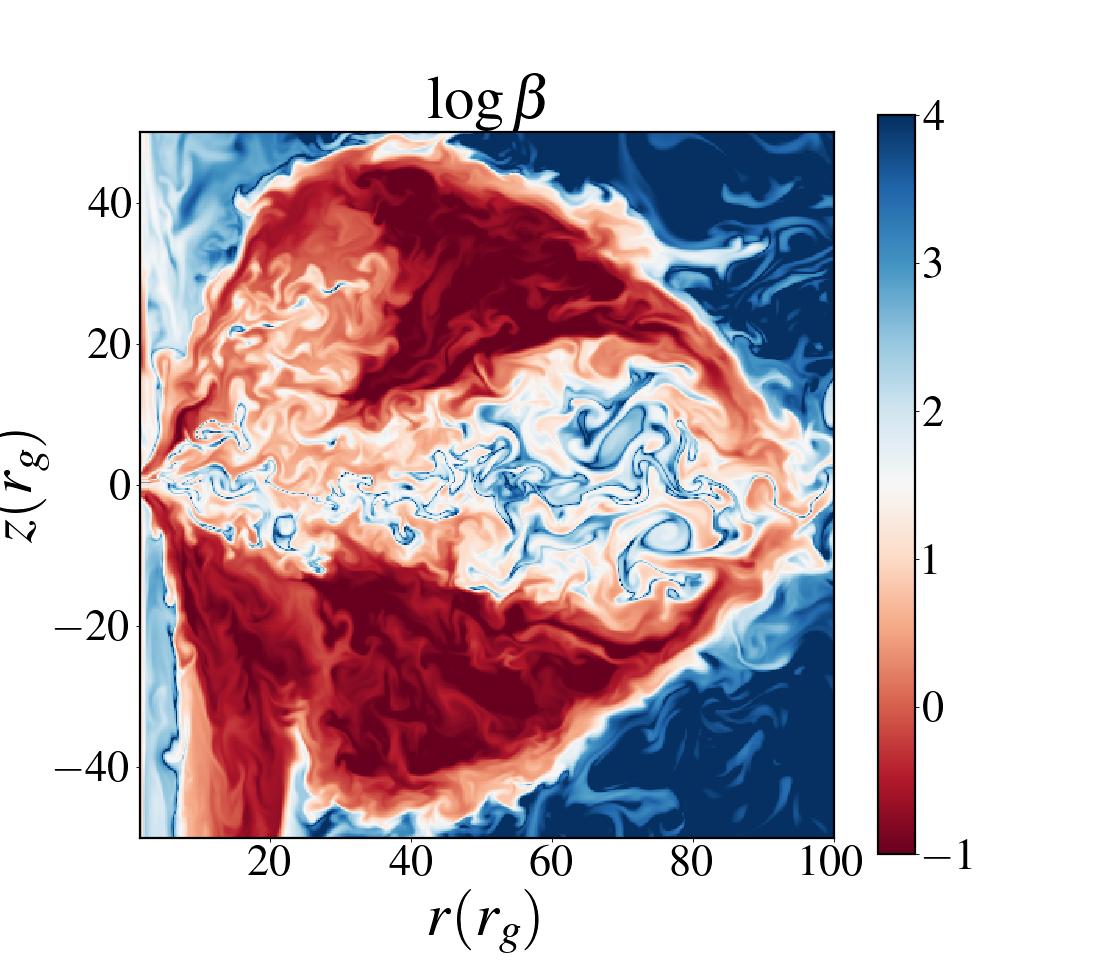} 

        \includegraphics[width=0.26\textwidth]{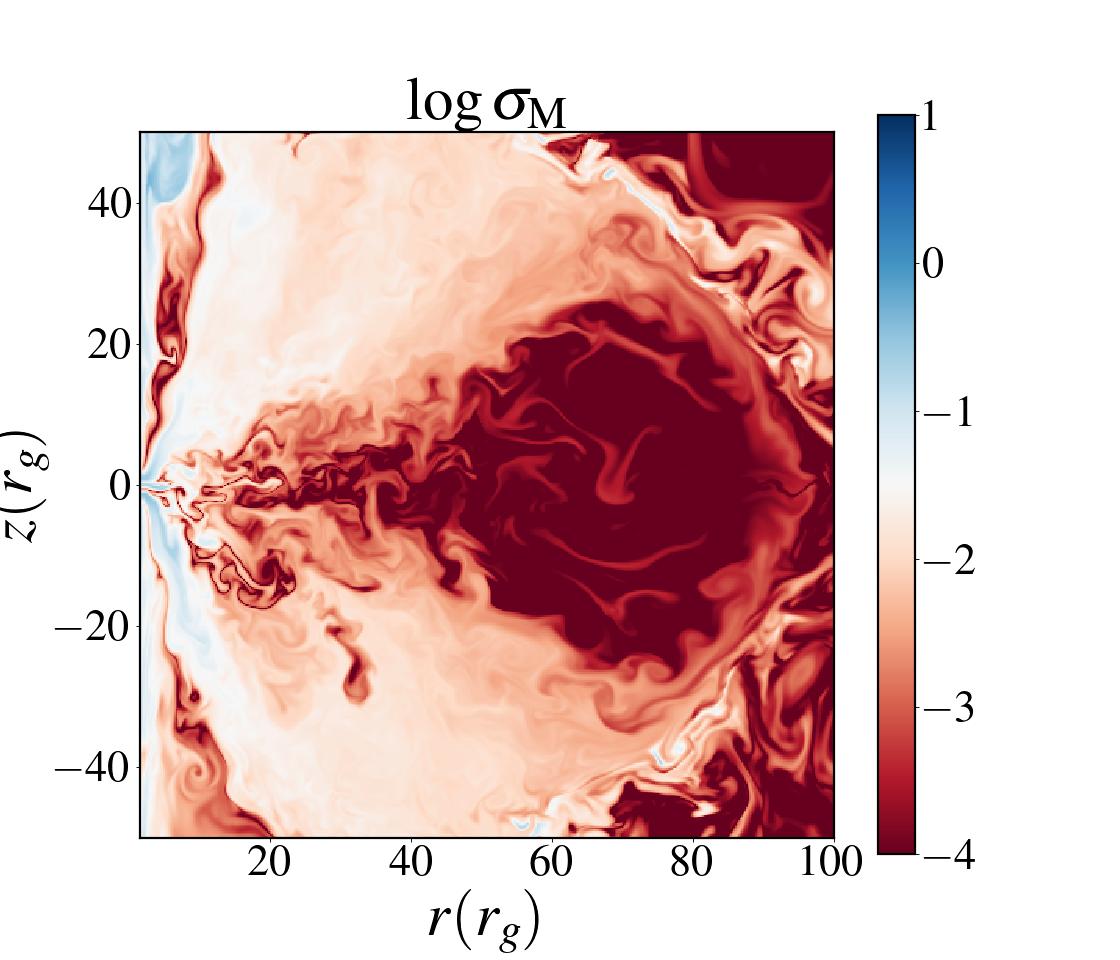} 
        \hskip -4 mm
        \includegraphics[width=0.26\textwidth]{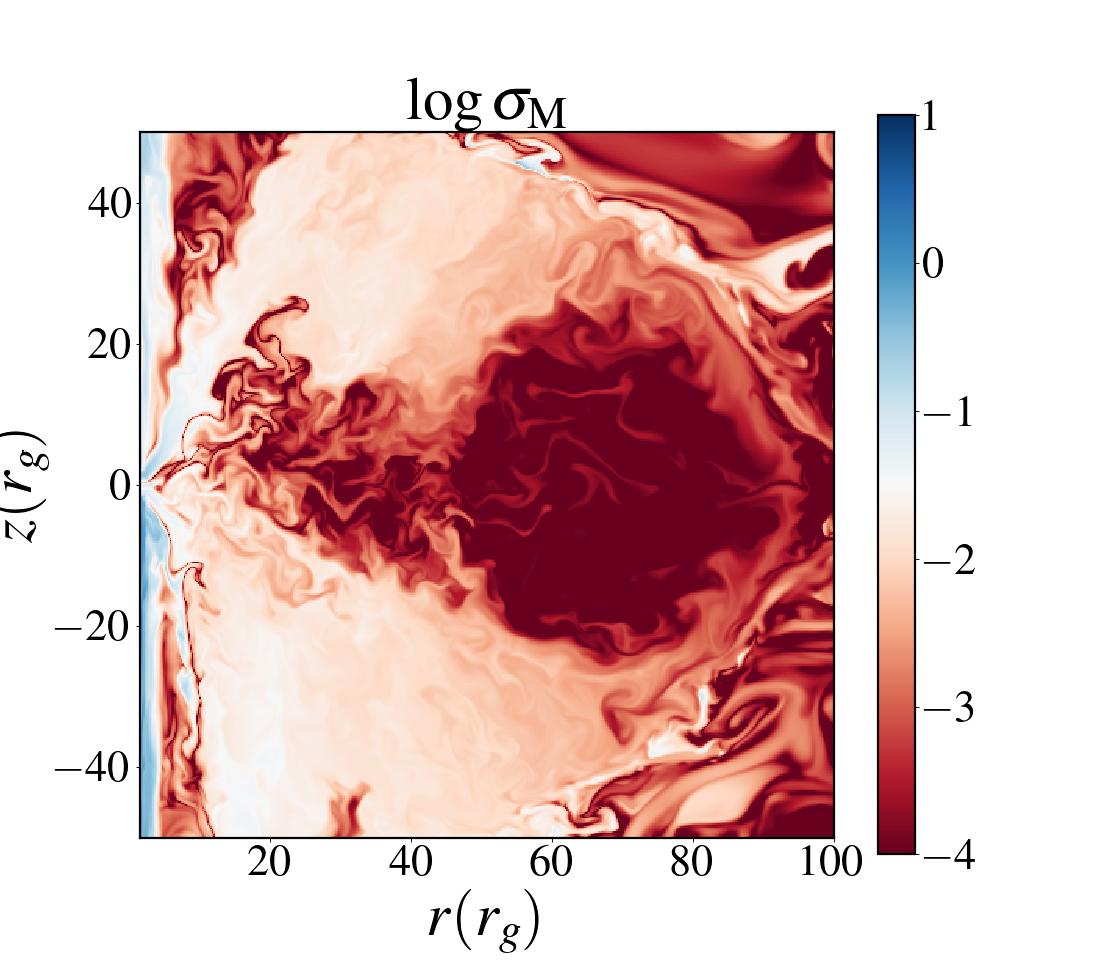} 
        \hskip -4 mm
        \includegraphics[width=0.26\textwidth]{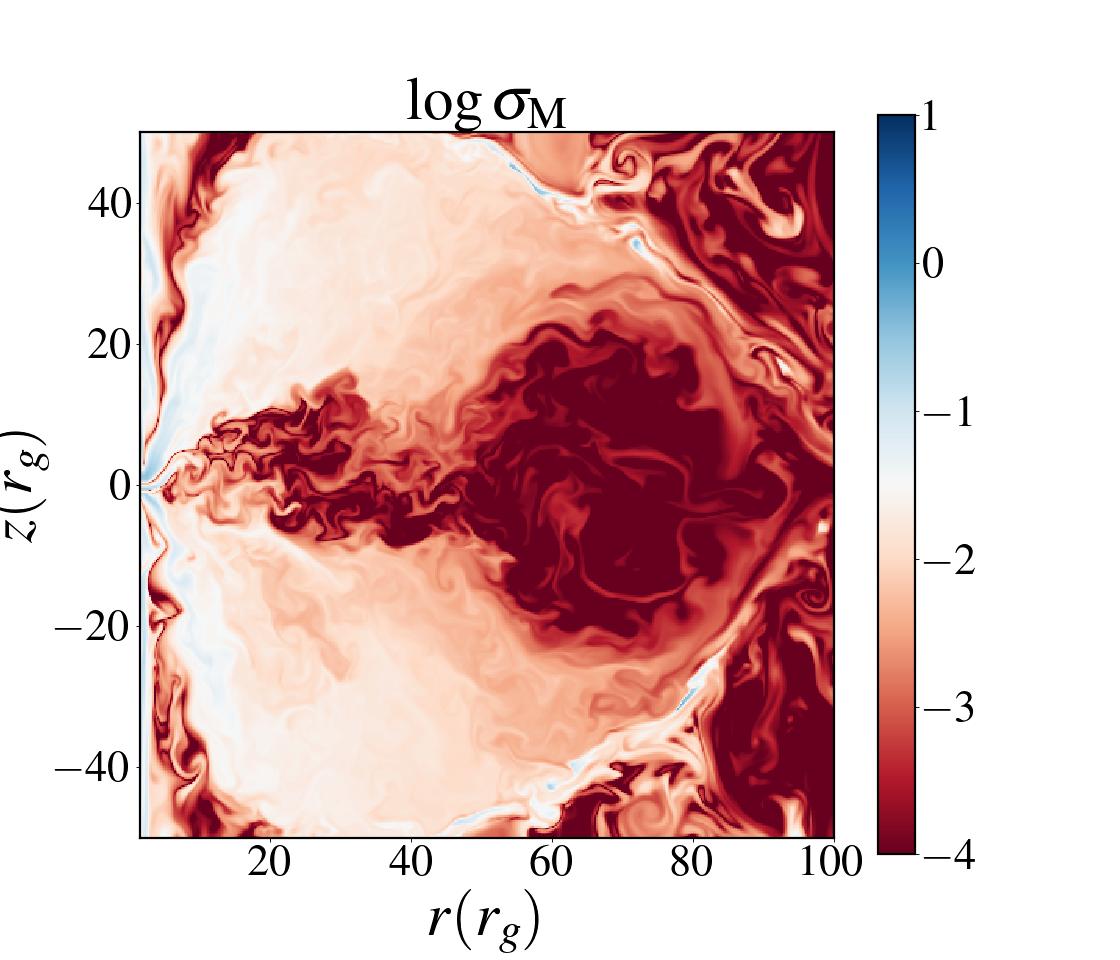} 
        \hskip -4 mm
        \includegraphics[width=0.26\textwidth]{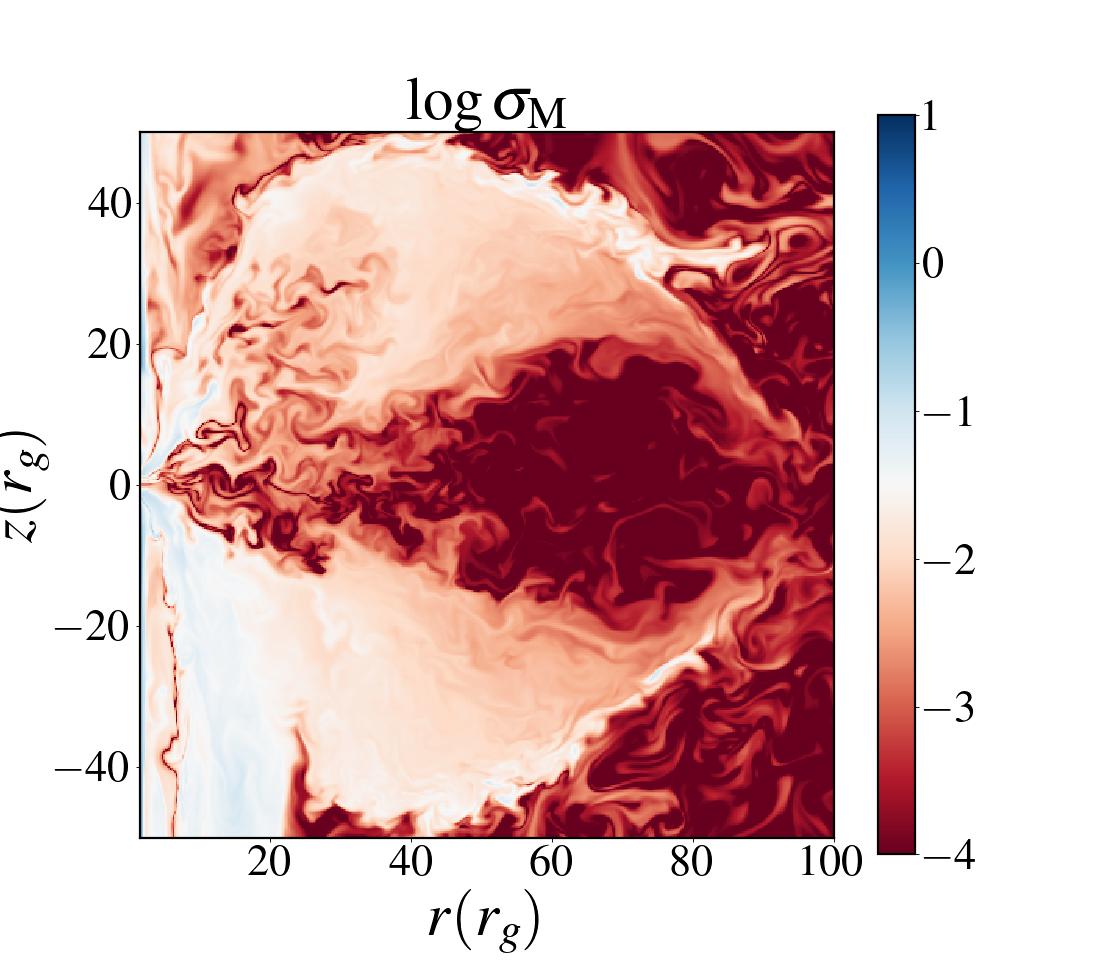} 
	\end{center}
	\caption{Distribution of density $(\rho)$, Temperature ($T$), plasma-$\beta$ ($\beta$) and magnetization parameter $(\sigma_{\rm M})$ for various spin ($a_k$) values. The first, second, third and fourth column are for $a_k = 0.99, 0.8, 0.5$, and 0.0, respectively at time $t_3 = 10500 t_g$.}
	\label{Figure_7}
\end{figure*}


\begin{figure}
	\begin{center}
		\includegraphics[width=0.5\textwidth]{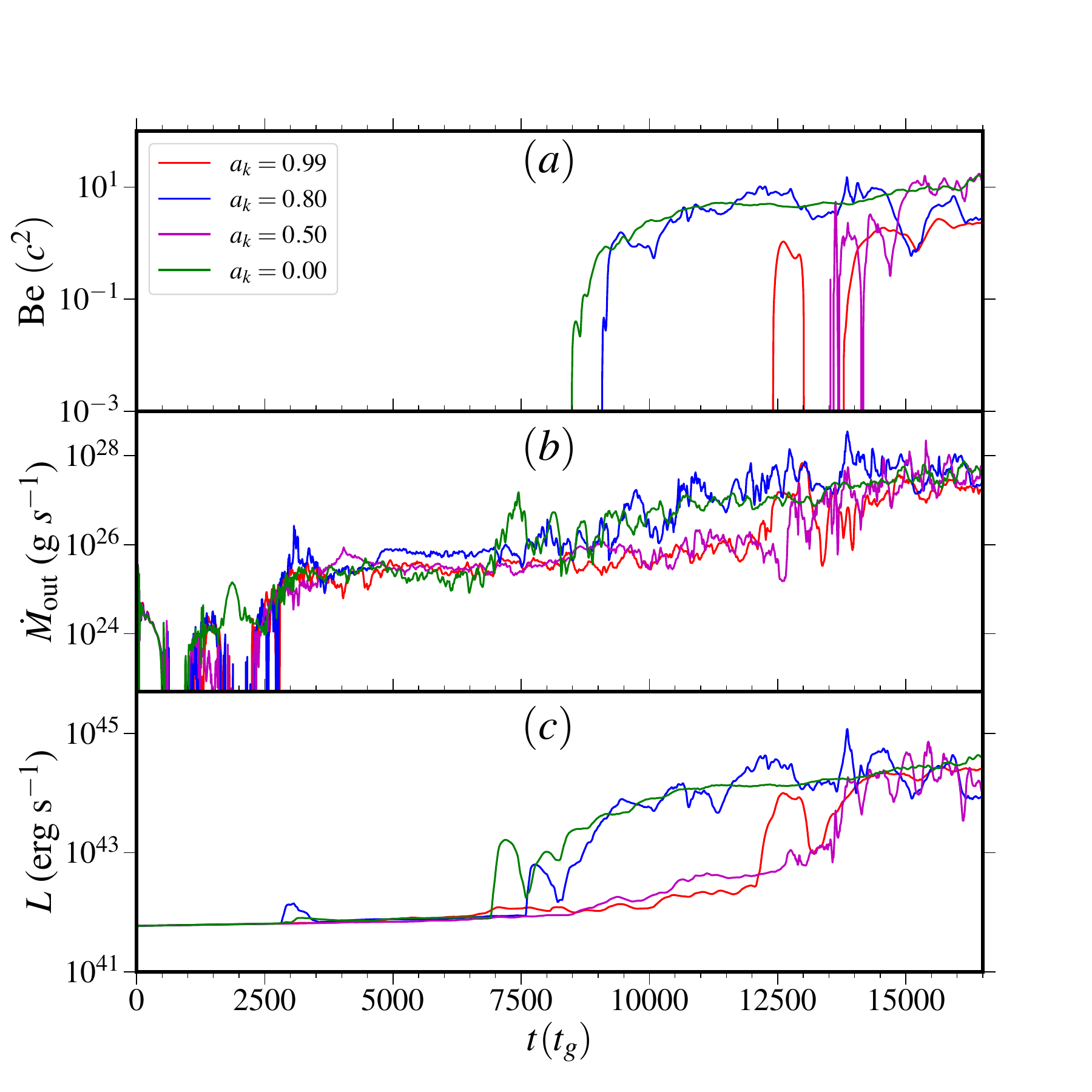} 
	\end{center}
	\caption{Variation of $(a)$: Bernoulli parameter ($\rm Be$) $(b)$: mass outflow rates $(\dot{M}_{\rm out})$ and $(c)$: luminosity $(L)$ with time for different spin values as $a_k$ = 0.99 (red), 0.80 (blue), 0.50 (magenta) and 0.0 (green), respectively.} 
	\label{Figure_8}
\end{figure}

\begin{figure*}
	\begin{center}
		\includegraphics[width=0.26\textwidth]{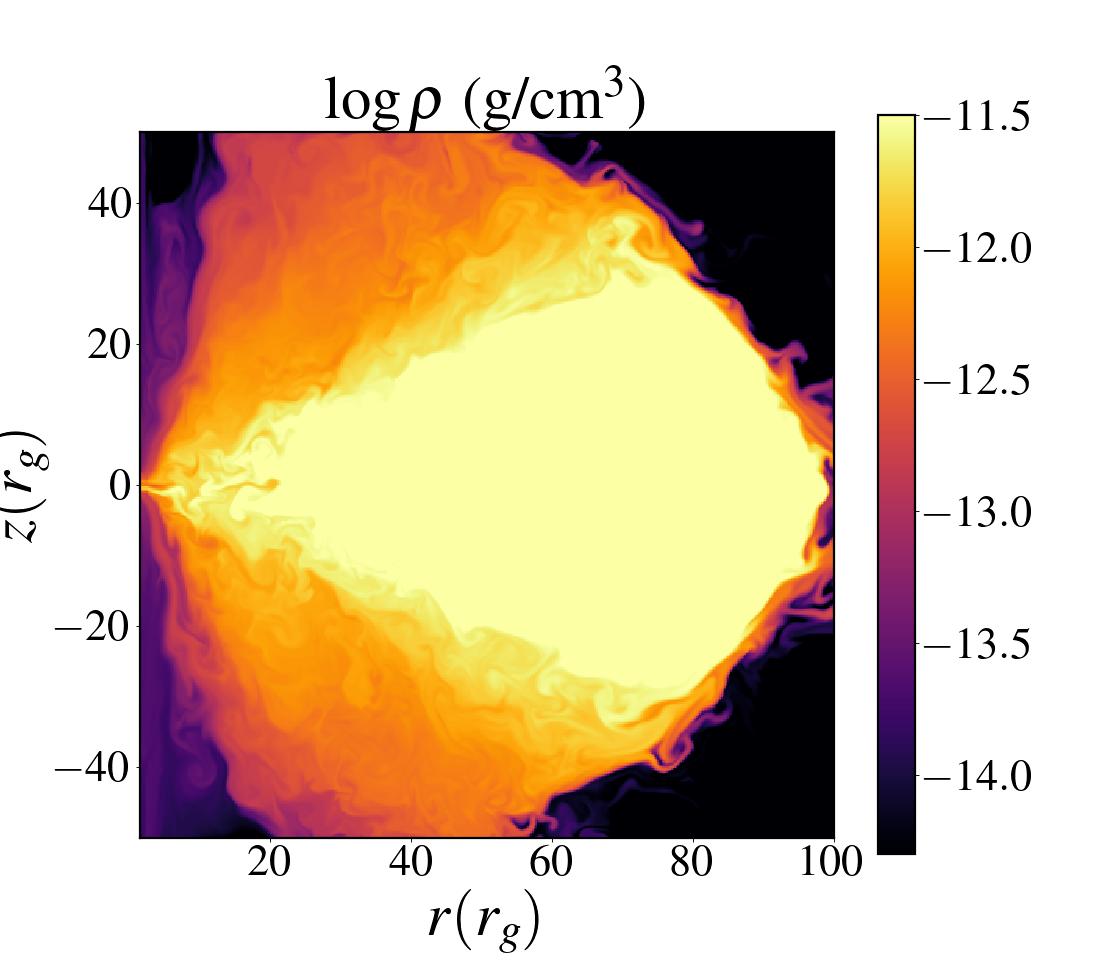} 
        \hskip -4 mm
        \includegraphics[width=0.26\textwidth]{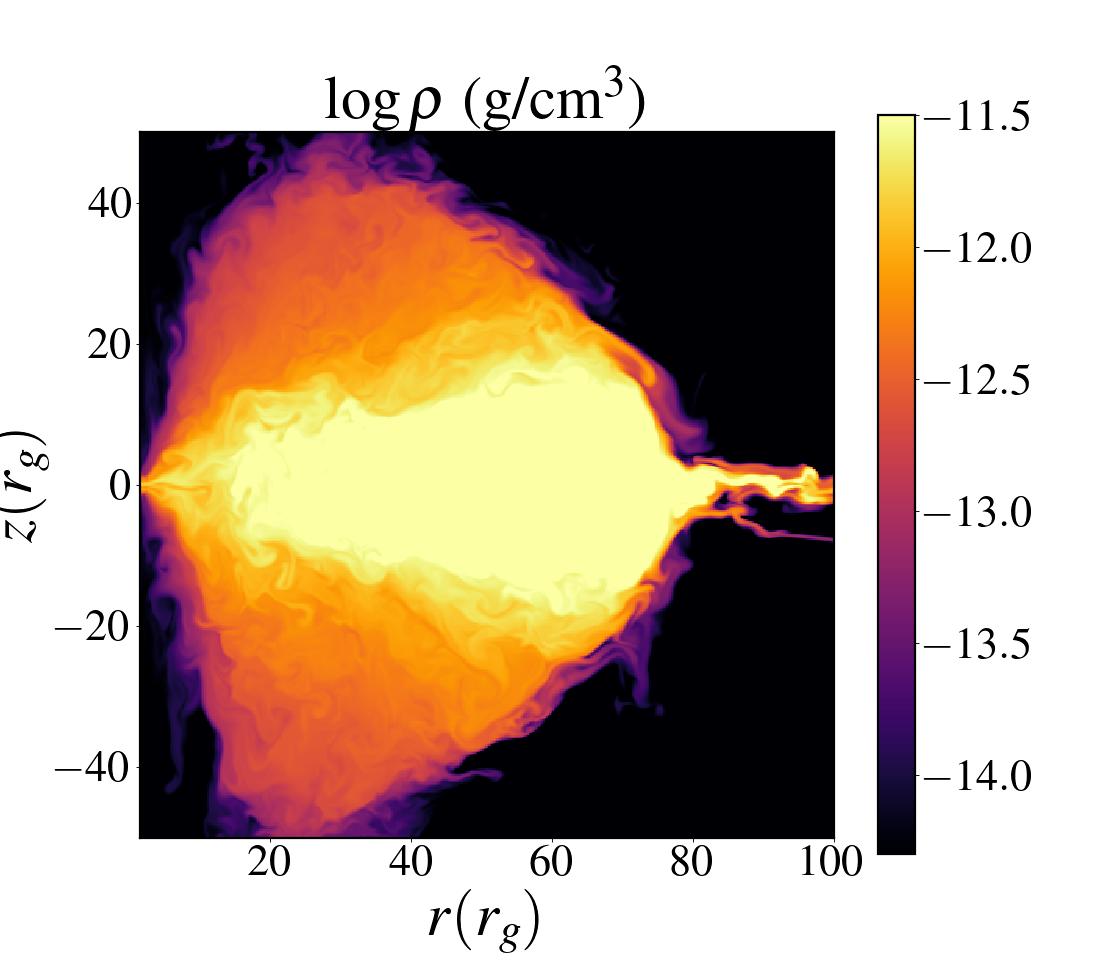} 
        \hskip -4 mm
        \includegraphics[width=0.26\textwidth]{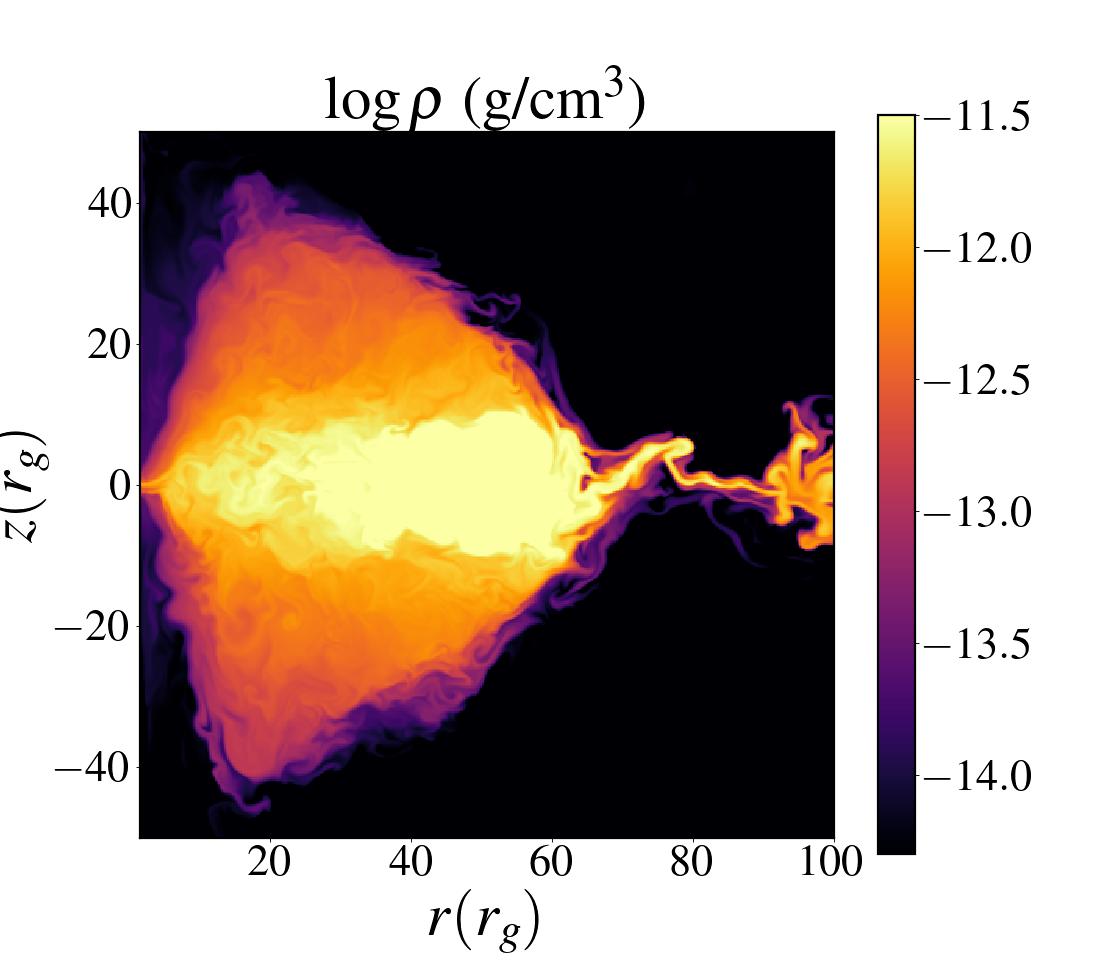} 
        \hskip -4 mm
        \includegraphics[width=0.26\textwidth]{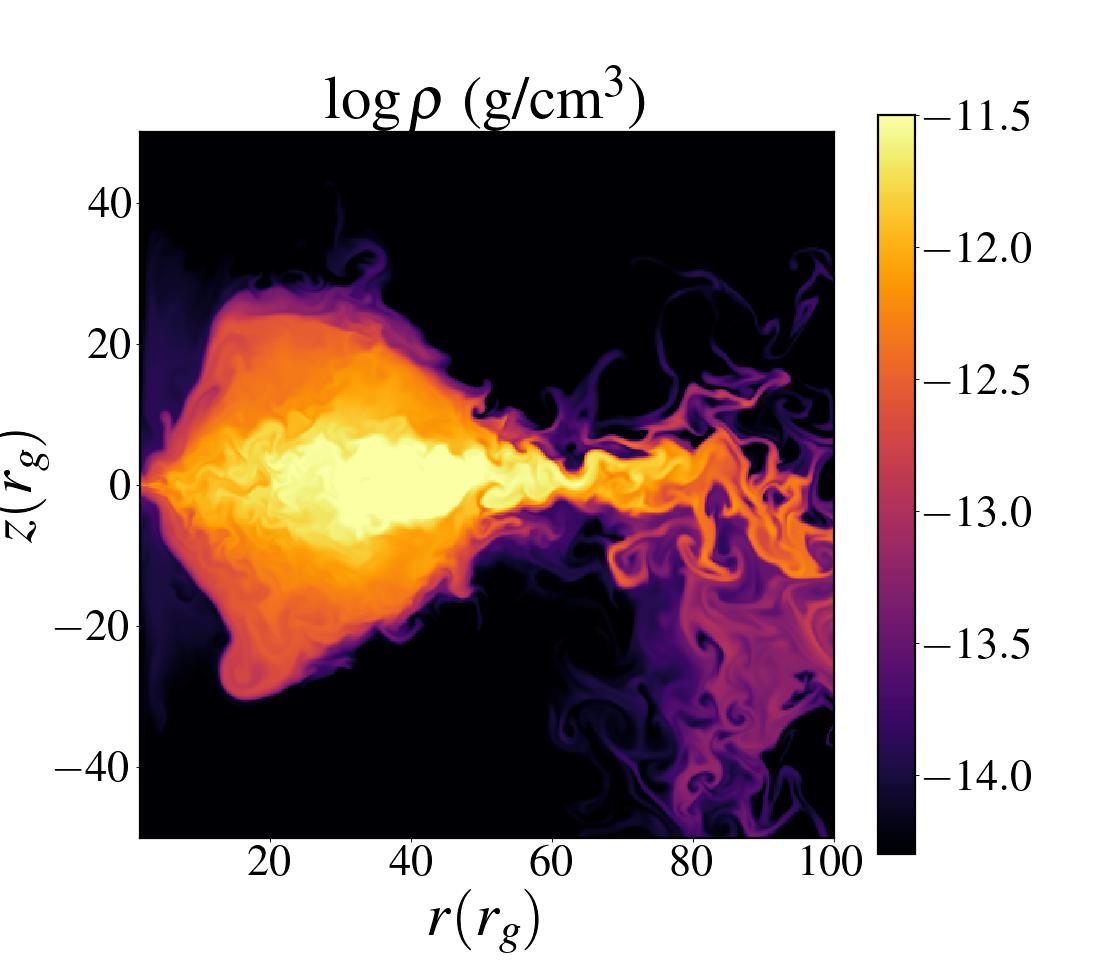} 
        
        \includegraphics[width=0.26\textwidth]{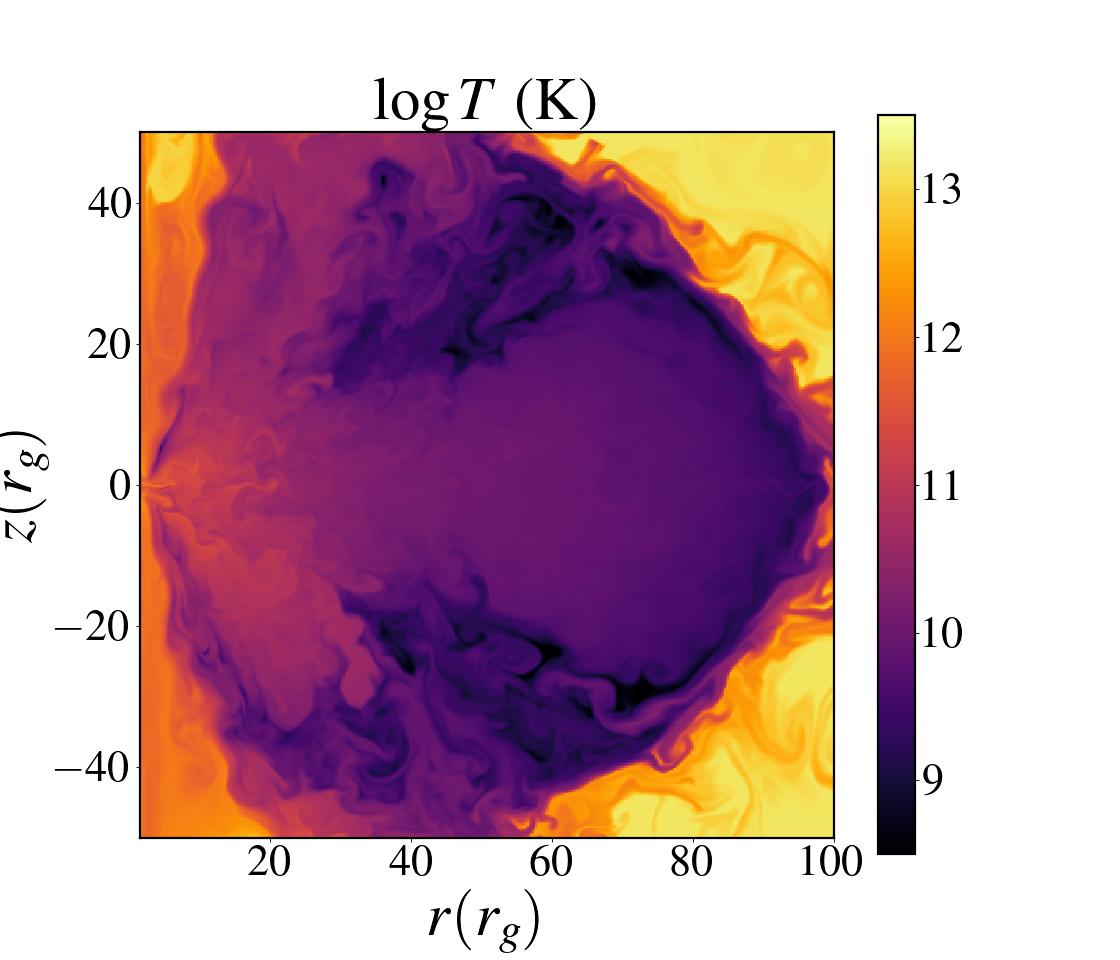} 
        \hskip -4 mm
        \includegraphics[width=0.26\textwidth]{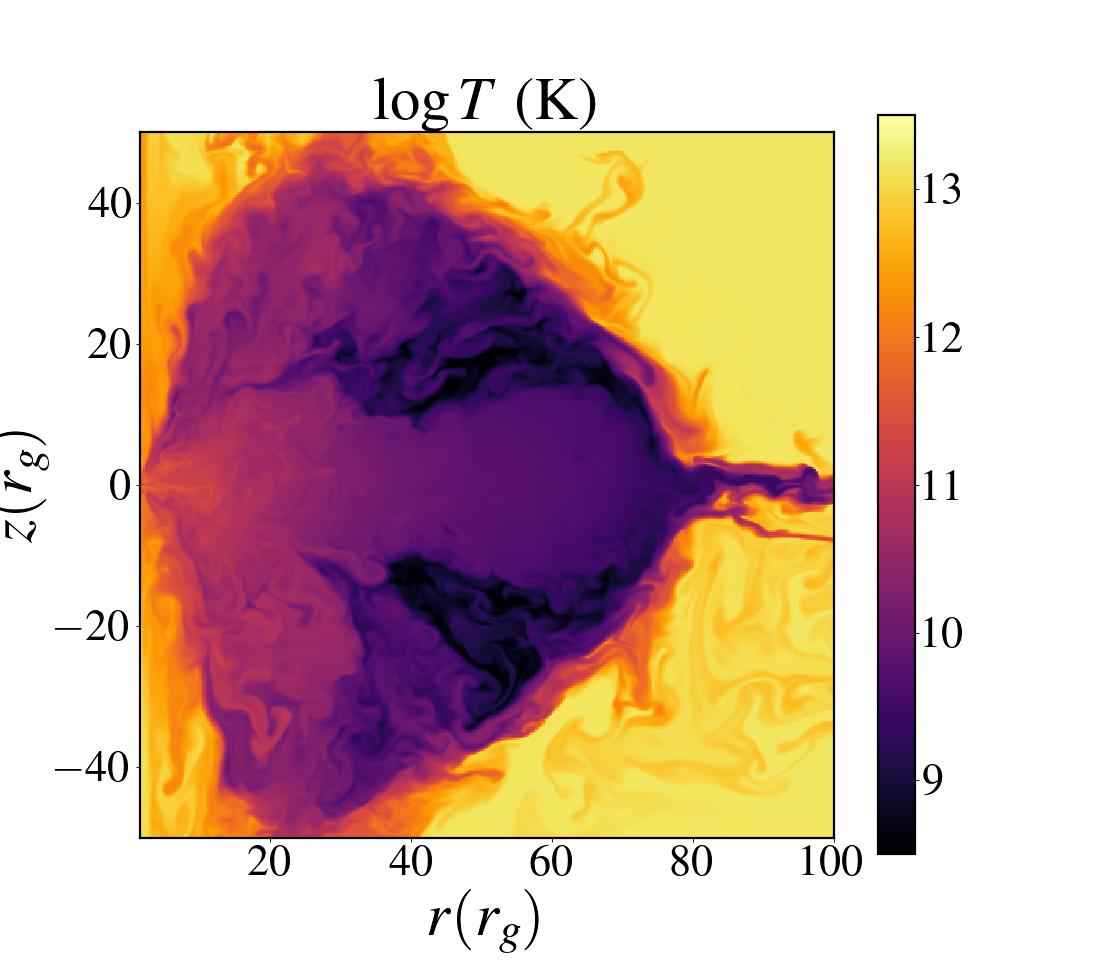} 
        \hskip -4 mm
        \includegraphics[width=0.26\textwidth]{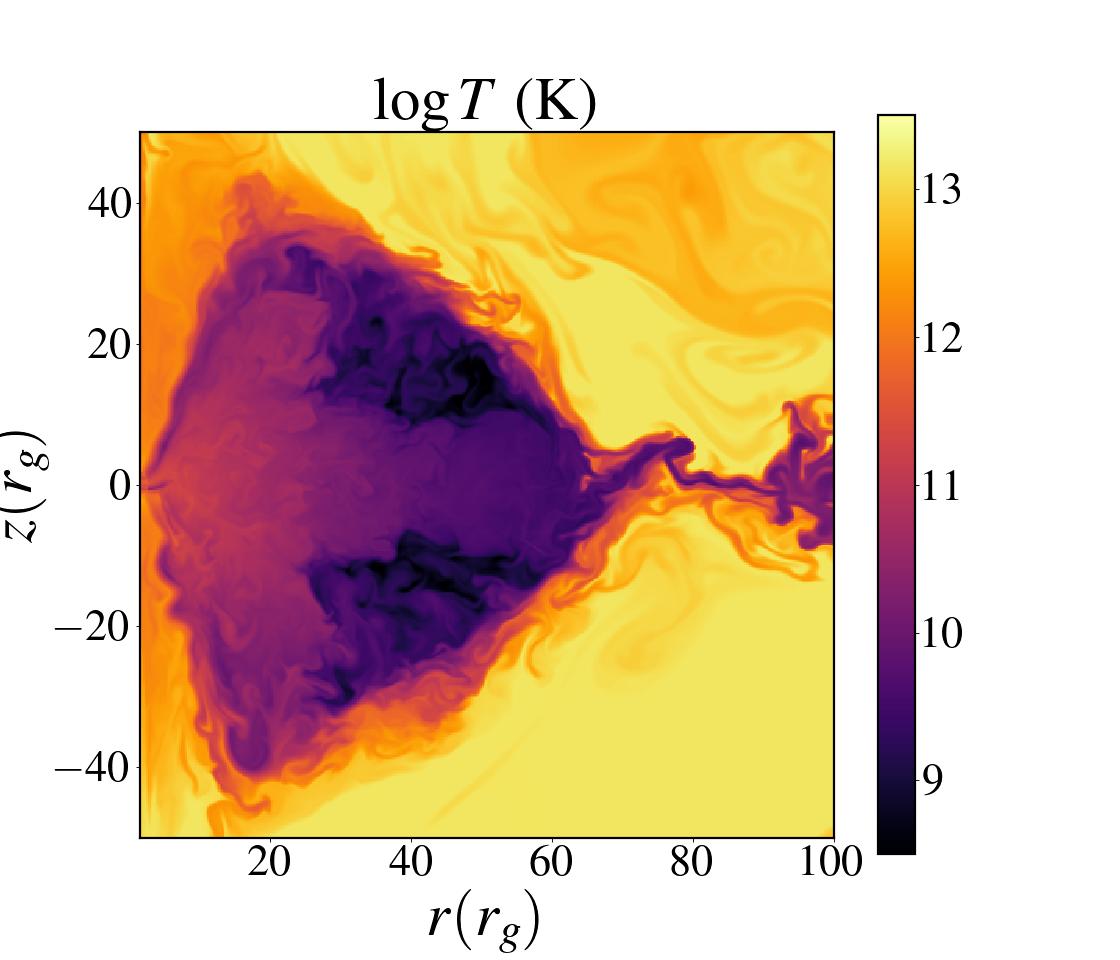} 
        \hskip -4 mm
        \includegraphics[width=0.26\textwidth]{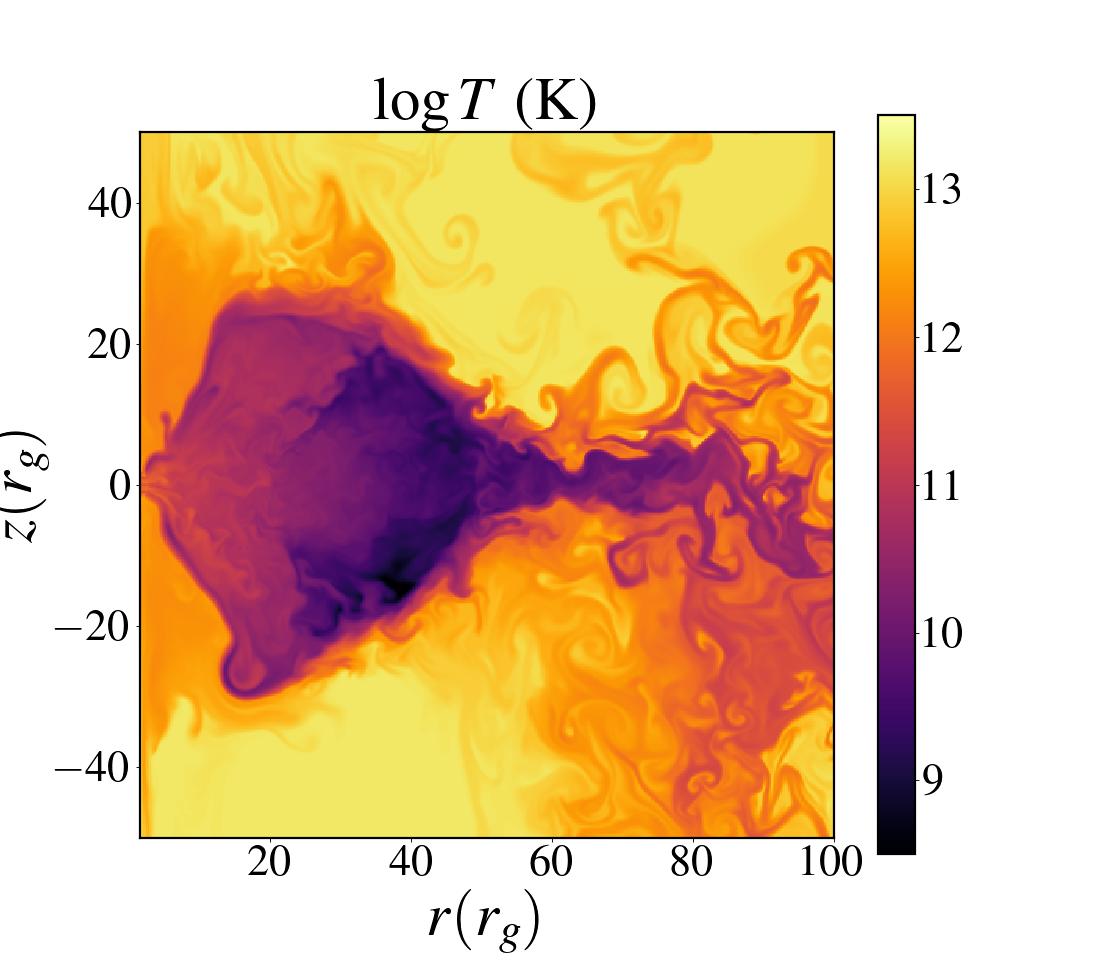} 
        
        \includegraphics[width=0.26\textwidth]{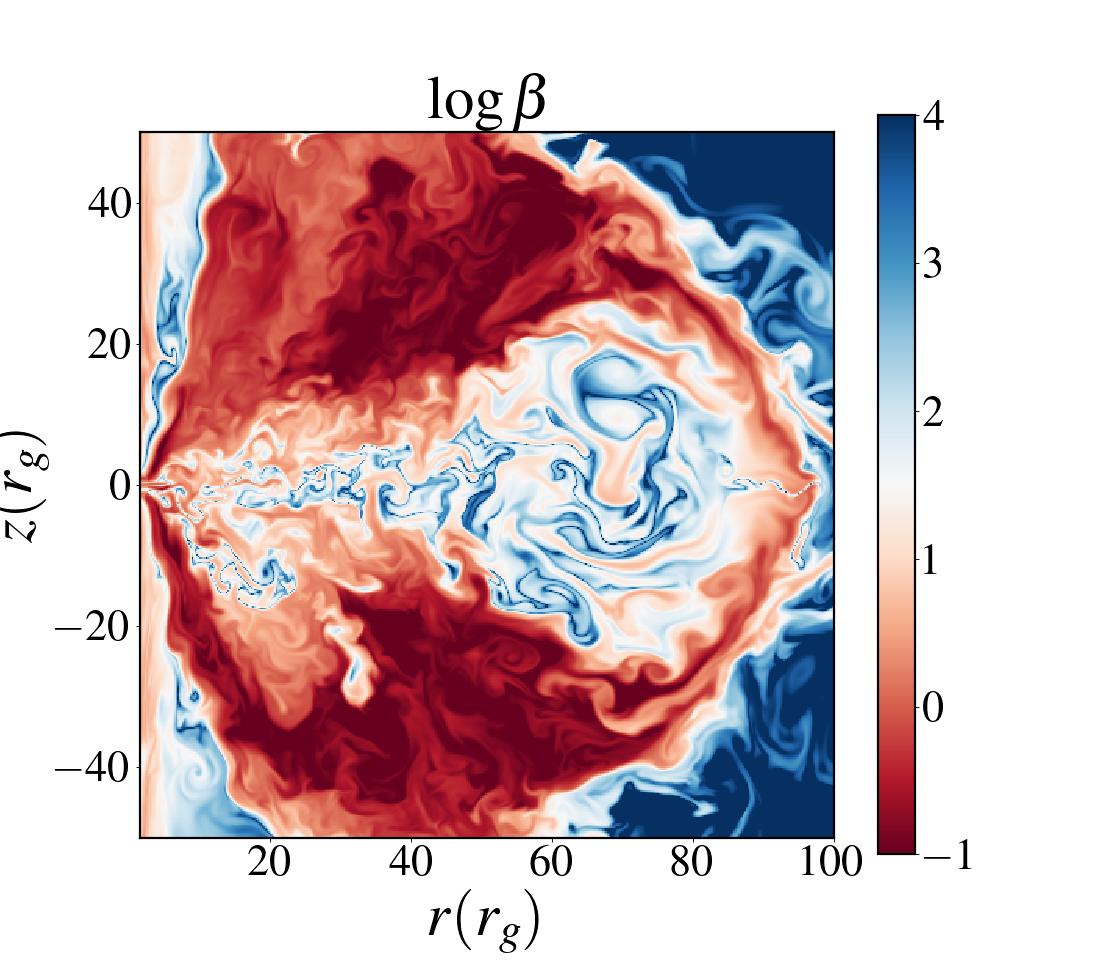} 
        \hskip -4 mm
        \includegraphics[width=0.26\textwidth]{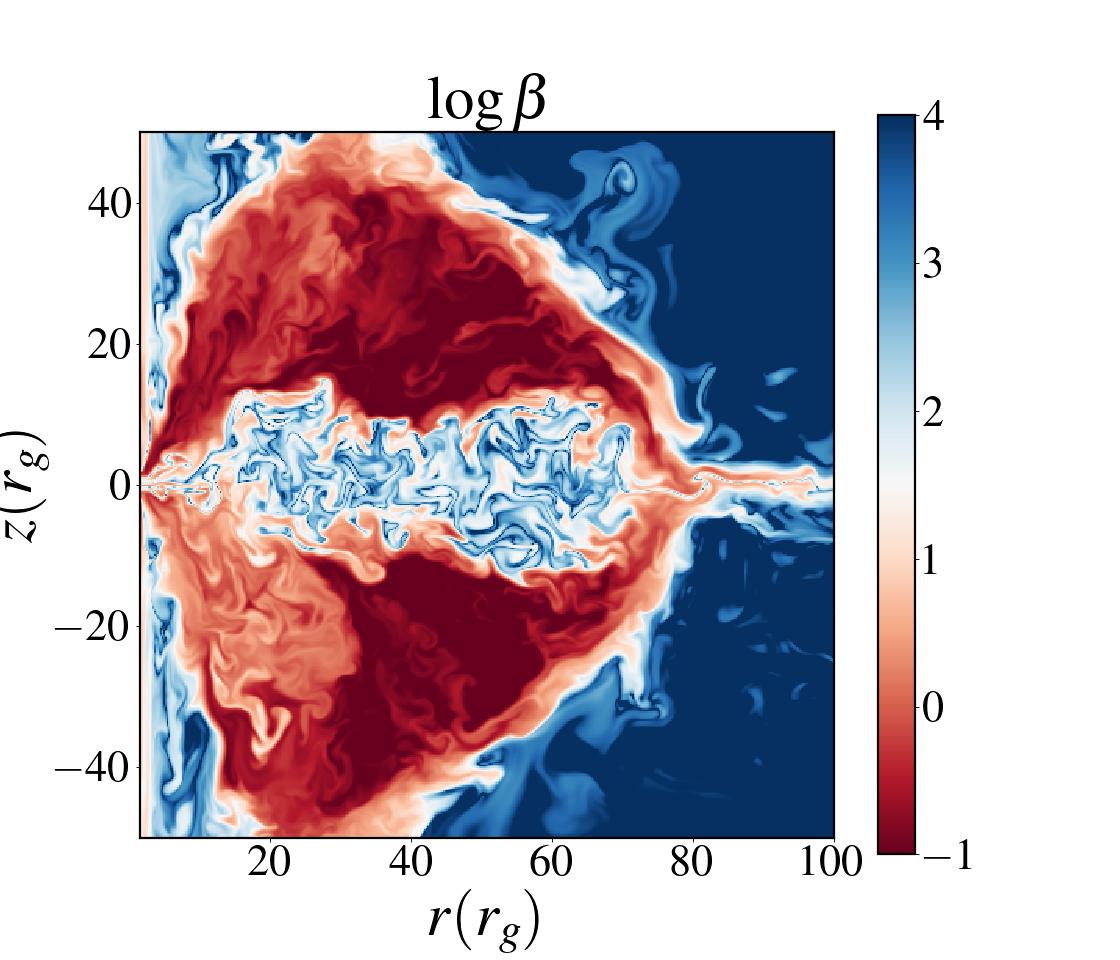} 
        \hskip -4 mm
        \includegraphics[width=0.26\textwidth]{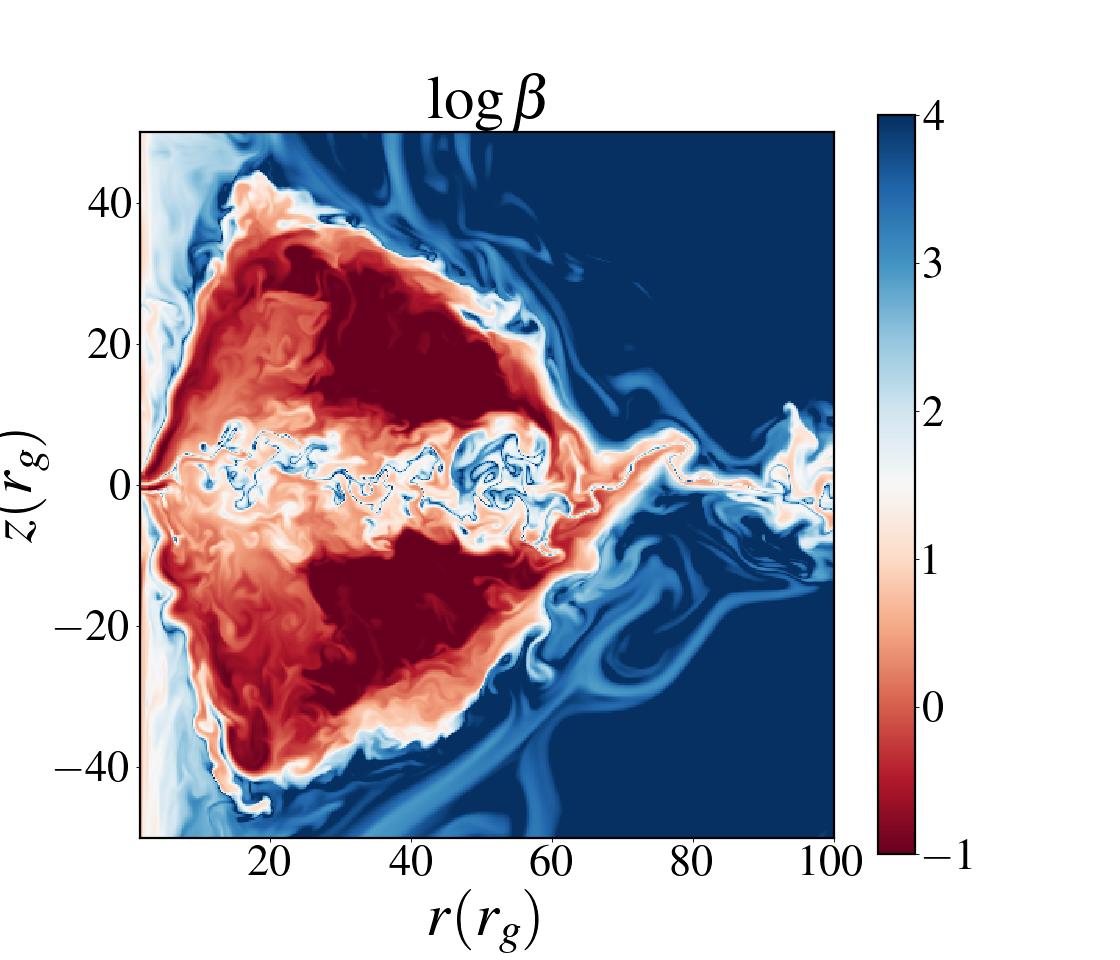} 
        \hskip -4 mm
        \includegraphics[width=0.26\textwidth]{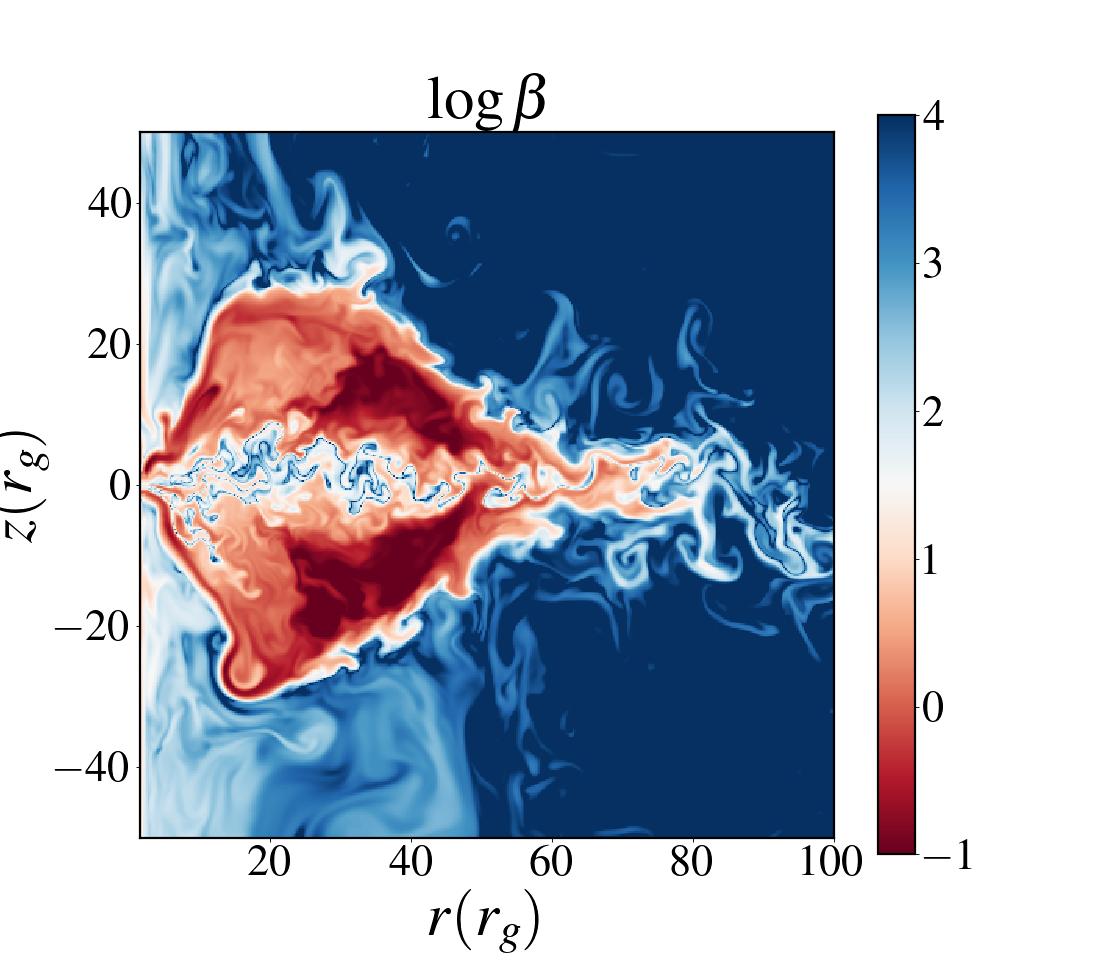} 

        \includegraphics[width=0.26\textwidth]{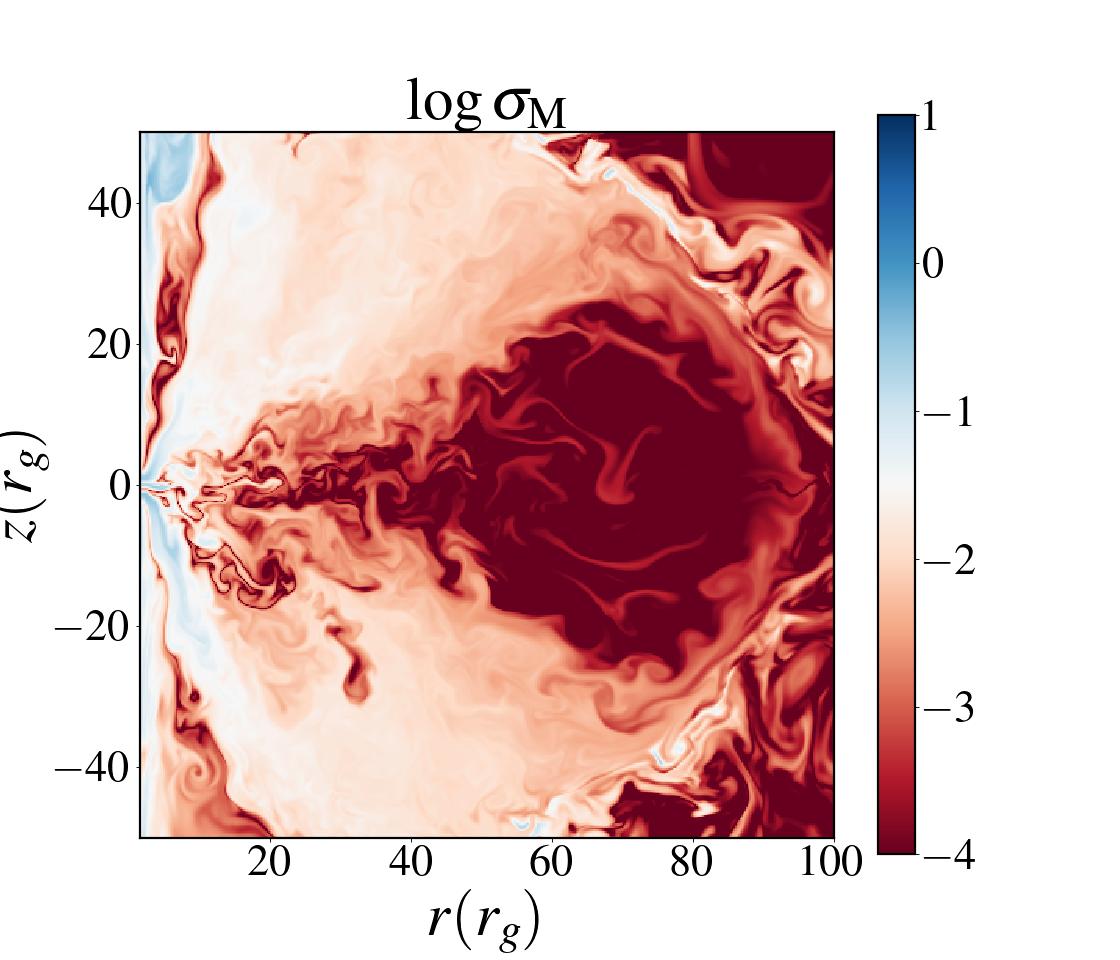} 
        \hskip -4 mm
        \includegraphics[width=0.26\textwidth]{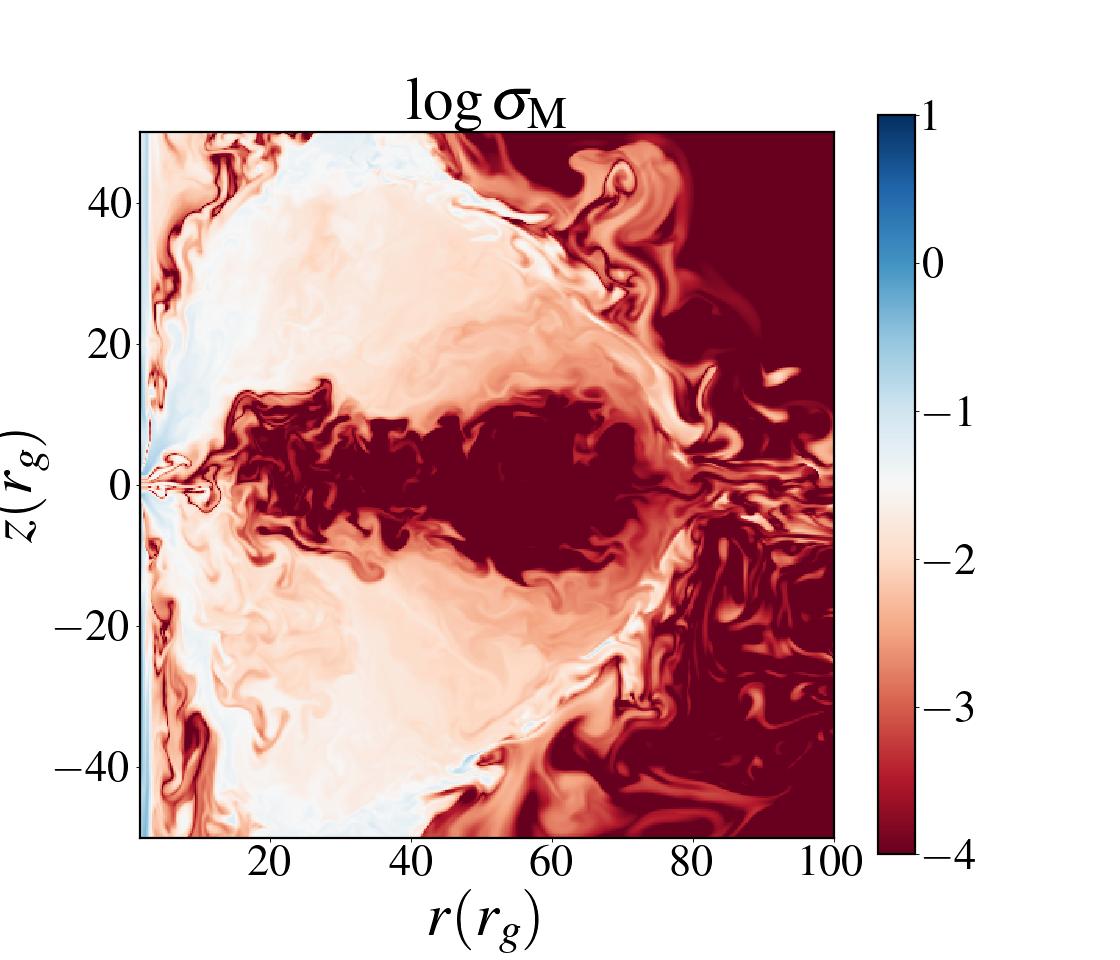} 
        \hskip -4 mm
        \includegraphics[width=0.26\textwidth]{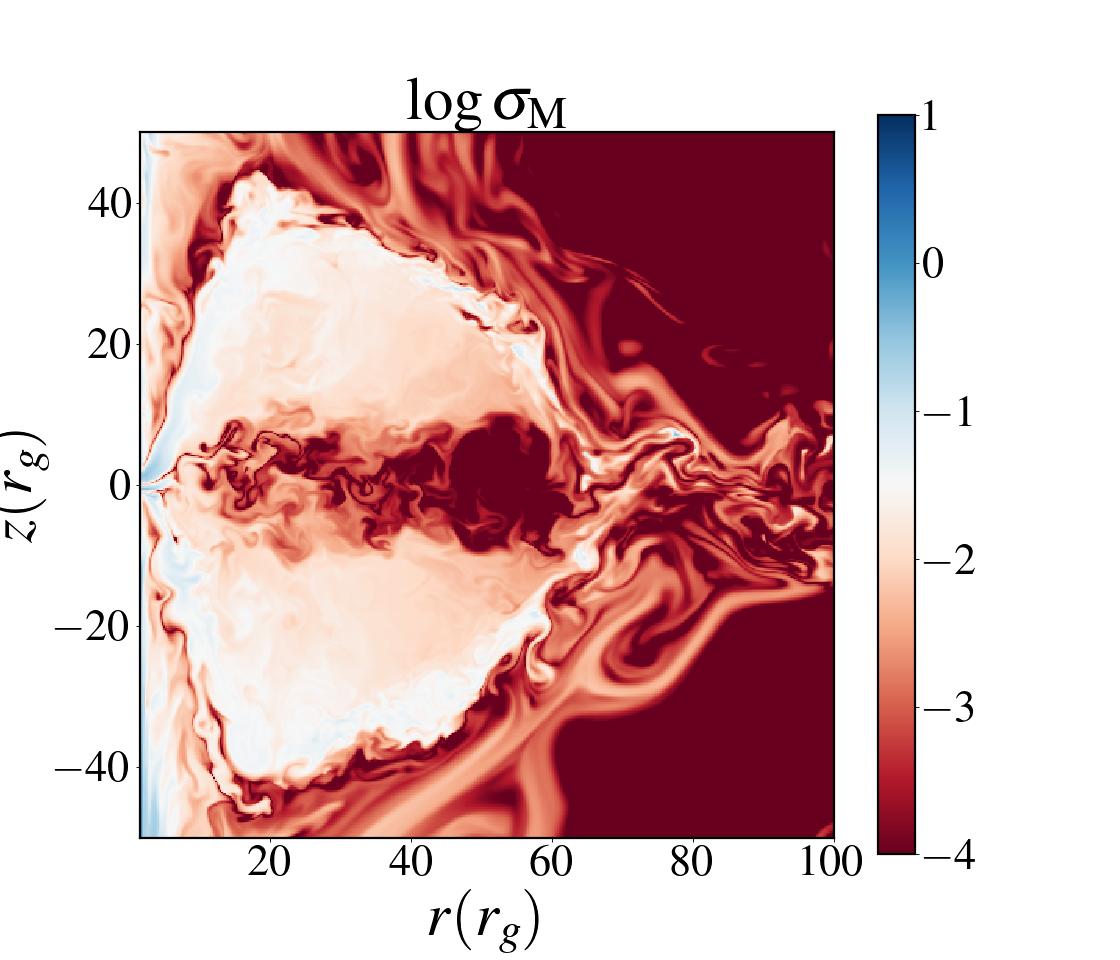} 
        \hskip -4 mm
        \includegraphics[width=0.26\textwidth]{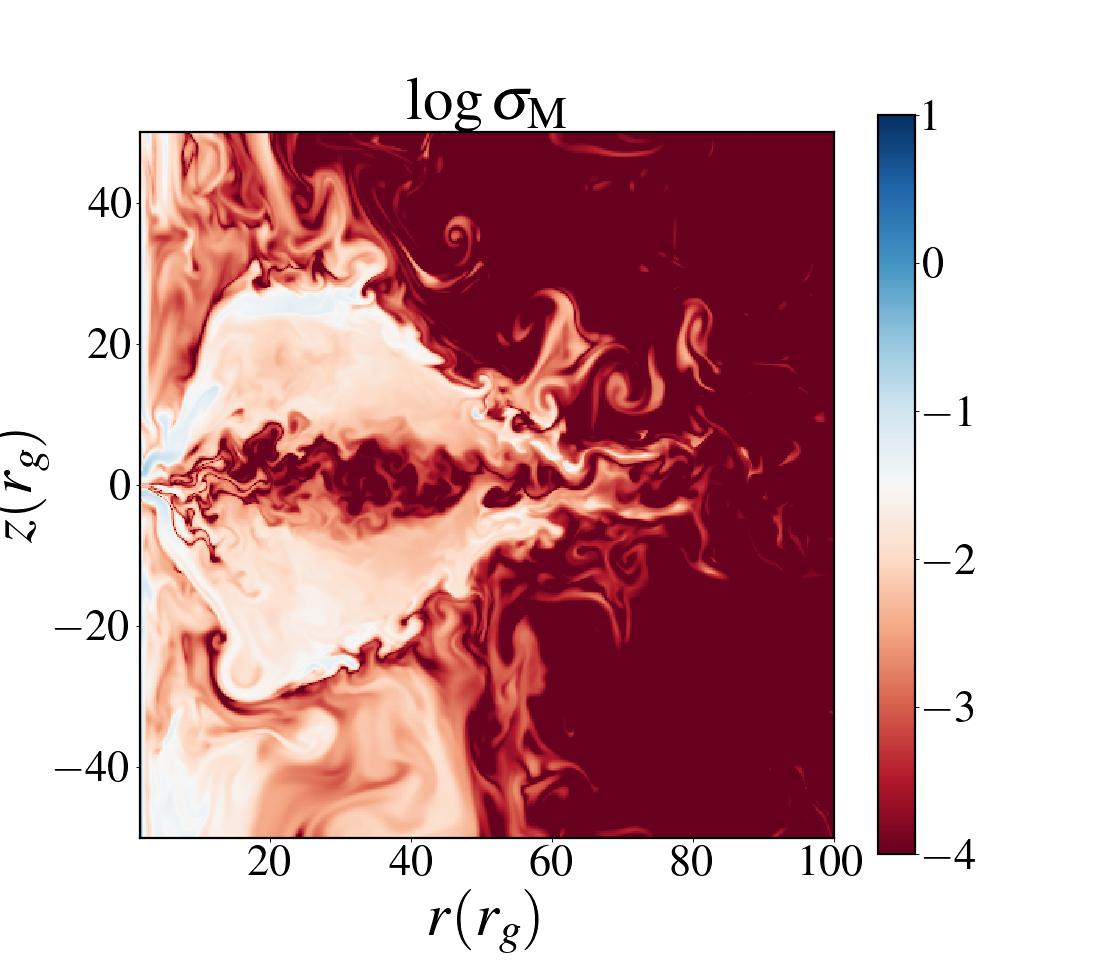} 
	\end{center}
	\caption{Distribution of density $(\rho)$ Temperature ($T$), plasma-$\beta$ ($\beta$) and magnetization parameter $(\sigma_{\rm M})$ for various angular momentum. The first, second, third and fourth columns are for $\lambda = \lambda_{\rm K}, 7.00, 6.80$ and 6.60, respectively at time $t_3 = 10500 t_g$}.
	\label{Figure_9}
\end{figure*}

\begin{figure}
	\begin{center}
		\includegraphics[width=0.5\textwidth]{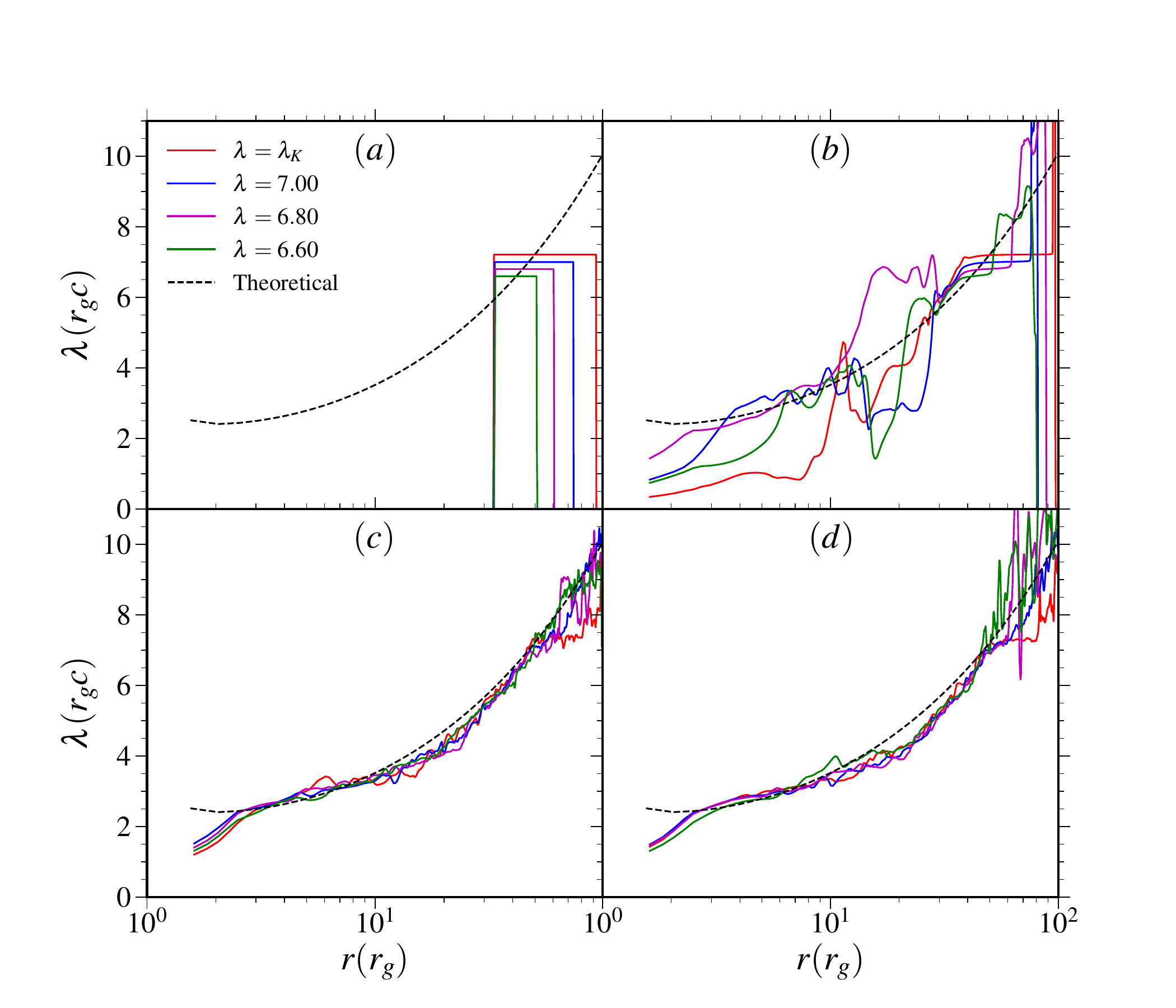} 
	\end{center}
	\caption{Variation of specific angular momentum $(\lambda)$ for different time of evolution $(a)$: $t= 0$ s, $(b)$: $t = 7.42\times 10^5$s, $(c)$: $t = 5.2 \times 10^6$ s and $(d)$: $t = 7.42 \times 10^6$ s, respectively. }
	\label{Figure_10}
\end{figure}

\begin{figure}
	\begin{center}
		\includegraphics[width=0.5\textwidth]{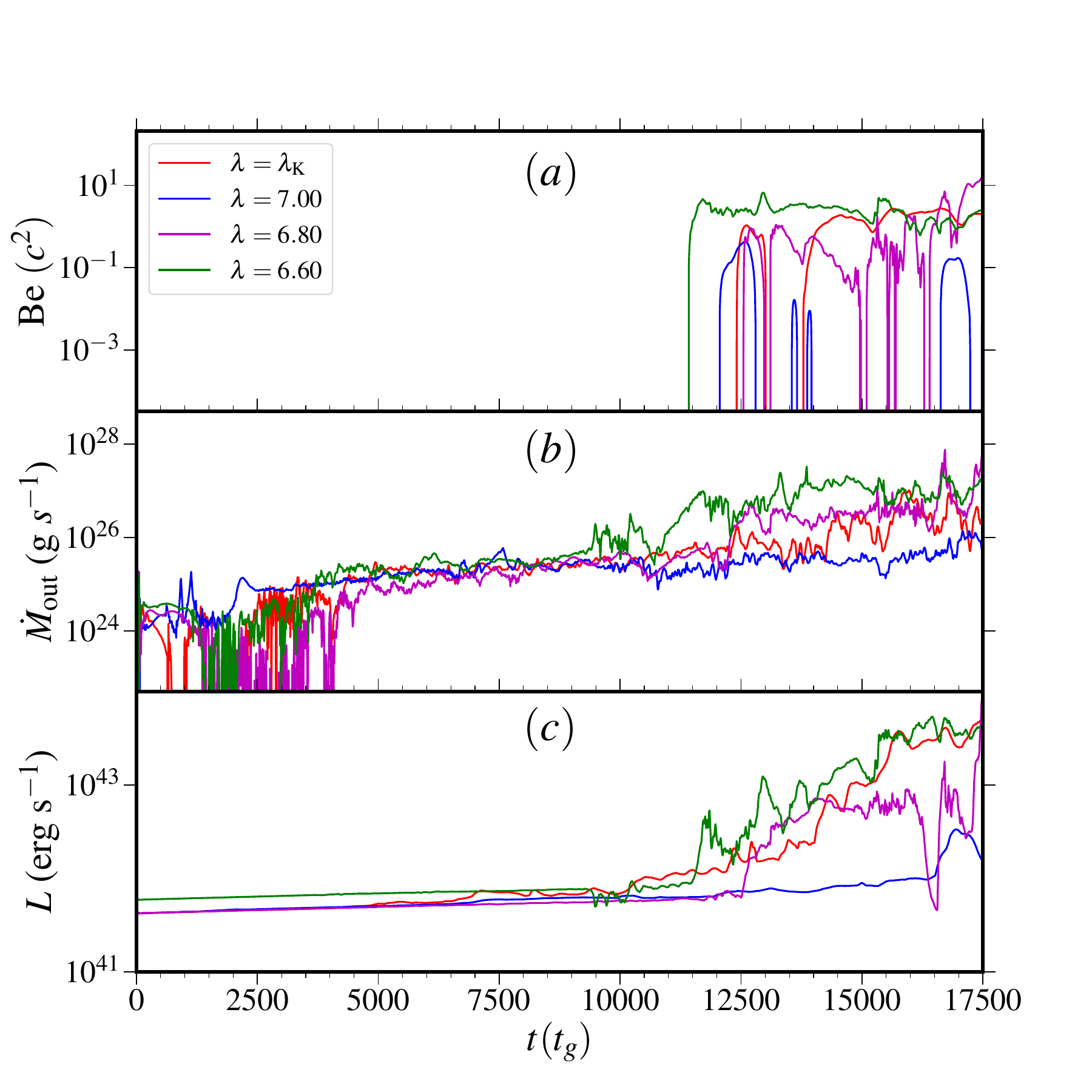} 
	\end{center}
	\caption{Variation of $(a)$: Bernoulli parameter $(\sigma)$, $(b)$: mass outflow rates $(\dot{M}_{\rm out})$ and $(c)$: luminosity $(L)$ with time for different specific angular momentum as $\lambda = \lambda_{\rm K}, 7.00, 6.80$ and 6.60, respectively.} 
	\label{Figure_11}
\end{figure}

\subsection{Magnetic field configuration}
\label{Mag_config}

In this work, we consider a poloidal magnetic field configuration following \citet{Hawley-etal02}. The initial magnetic field configuration is set by considering a toroidal component of vector potential as 
\begin{align}
 A_\phi = B_0 [\rho(r,z) - \rho_{\rm min}],   
\end{align}
where, $\rho_{\rm min}$ is the minimum density in the torus. $B_0$ is the normalized initial magnetic field strength. The magnetic field strength is parameterized by the ratio of the gas pressure to the magnetic pressure, and it is constant in the initial disc. The initial magnetic field strength is set by using the initial plasma-$\beta$ parameter as $\beta_0 = \frac{2 P_{\rm gas}}{B_0^2}$. The poloidal magnetic field configuration is obtained using $\bm{B} = \nabla \times \bm{A}$. Moreover, in MHD simulation, it is usually convenient to define a parameter known as magnetization parameter $\sigma_{\rm M} = \frac{B^2}{\rho}$. The magnetization parameter represents the ratio of magnetic energy to rest mass energy in the flow \citep{Dihingia-etal21, Dihingia-etal22, Dhang-etal23, Curd-Narayan23}.

\subsection{Initial and boundary conditions}

Here, we consider two-dimensional MHD flows in cylindrical coordinate $(r, \phi, z)$ in PLUTO code. To simulate the MHD accretion flows, we employ HLLD Riemann solver, second-order-in-space linear interpolation, and the second-order-in-time Runge–Kutta algorithm. We also enforce the hyperbolic divergence cleaning method for solving the induction equation $\nabla \cdot \bm{B} = 0$ \citep{Migone-etal07}. In this work, we consider the computational domain is $1.5 r_g \leq r \leq 200 r_g$ in the radial direction and $-100 r_g \leq z \leq 100 r_g$ in the vertical direction. The number of grid is $(n_r, n_z) = (896, 896)$. We perform a convergence test of our simulation model, described in Appendix \ref{convergence_test}. We use uniform grid spacing in both the radial and vertical directions. We set the inner boundary at $r_{\rm in} = 2.5 r_g$. The inner boundary of the black hole is set to be absorbing boundary conditions \citep{Okuda-etal19, Okuda-etal22, Okuda-etal23}. Moreover, the axisymmetric boundary condition is set at the origin, and all the rest are set to outflow boundary conditions. To set up the initial torus around the black hole, we consider the inner edge of the torus at $r_{\rm min} = 32 r_g$. The maximum pressure surface is at $r_{\rm max} = 50 r_g$. The minimum density in the torus is set to be $\rho_{\rm min} = 0.5 \rho_{\rm max}$. The maximum density is choosen as $\rho_{\rm max} = 10 \rho_0$ (see the Table \ref{Table-1}). We initially set the magnetic field is zero at $\rho \leq \rho_{\rm min}$ .i.e., the initial magnetic field is embedded only inside the torus; otherwise, it is zero. Here, the mass of the AGN is chosen as $10^8 M_{\odot}$, and the adiabatic index is $\gamma = 4/3$. Also, the density fraction between the maximum density of the torus to the density of the halo $\eta$ is set $10^{-4}$ (see equation \ref{eqn_density_halo}). The C.G.S. and code unit system used in this work is summarized in Table \ref{Table-1}. In the PLUTO code, we are required to specify three fundamental units, such as unit density $(\rho_0)$, unit length $(L_0)$, and unit velocity $(v_0)$. Therefore, we first specify the physical unit (C.G.S.) system for our model and then convert the unit system into code units for the simulation run, as depicted in Table \ref{Table-1}.

In numerical simulation, especially in MHD simulation, sometimes it can be encountered that density and pressure become negative due to evacuated matter in the supersonic and highly magnetized flows. In general, fixing pressure and density to a particular lower value as floor values is recommended to avoid such a situation throughout the simulation run. Additionally, in the PLUTO code, there is an efficient way to handle this situation by implementing two flags turned on, i.e., (i) SHOCK  FLATTENING to MULTID and (ii) FAILSAFE to YES. In this FAILSAFE scheme, PLUTO first saves the solution arrays before attempting the step before encountering a negative density. Then, it retries the step again while tagging the critical zone(s) to be updated using FLAT reconstruction and the HLL Riemann solver. Therefore, we fix the floor density and pressure in our simulation as $\rho_{\rm floor} = 10^{-5}$ and $P_{\rm floor} = 10^{-8}$, respectively. Moreover, we turn on the FAILSAFE scheme in PLUTO, as mentioned above. Moreover, our choice for the magnetic field strength is confined by the requirement that the flow is super-Alfv\'enic, i.e., $|v_p|>|v_{Ap}|$ which implies $\sigma_{\rm M}< 4 \pi v_p^2$, where $v_p$ is the poloidal fluid velocity and $v_{Ap}=[B^2/4 \pi \rho]^{1/2}$ is the poloidal Alfv\'en velocity, respectively \citep{Proga-Begelman-03}. Therefore, we impose the conditions $\sigma_{\rm M}< 4 \pi v_p^2$ for all the radii for the entire computational domain in our model.

\begin{table}
\caption{Units used in this paper 
\label{Table-1}}
 \begin{tabular}{@{}c c c   } 
 \hline
 Units &   C.G.S values &   code units   \\ 
 \hline  
 Density  &  $\rho_0 = 1 \times 10^{-12}$ g cm$^{-3}$ &  $\rho_0 = 1 \times 10^{-12}$\\
 Length   &  $r_g = 1.485 \times 10^{13}$ cm &    $ r_g = GM_{\rm BH}/c^2$\\
 Velocity &  $c = 2.998 \times 10^{10}$ cm s$^{-1}$ &  $c$ \\
 Time     &  $t_g = 4.953 \times 10^{2}$ s  &  $t_g = GM_{\rm BH}/c^3$\\
 Magnetic field &   $B_0 = 1.063 \times 10^5$  G & $B = \frac{B_0}{c \sqrt{4 \pi \rho_0}}$\\
 Mass     &   $M_{\rm BH} = 10^8 M_{\odot}^*$ &    \\
 \hline
$*$ $M_{\odot}$ is the mass of the sun. \\

 \end{tabular}
\end{table}


\section{Simulation Results}
\label{results}

First, to set up the initial torus around the black hole, we fix the black hole spin ($a_k$) and supply Keplerian angular momentum ($\lambda_{\rm K}$) (equation \ref{kep_ang}) at the pressure maximum surface. The formation of the initial torus is mainly dependent on the three parameters such as flow angular momentum ($\lambda$), spin ($a_k$) of the black hole, and location of the inner edge of the torus ($r_{\rm min}$). The initial magnetic field is embedded inside the torus (see sub-section \ref{Mag_config}), and with time evolving, MRI grows in the disc \citep{Balbus-Hawley91}. We represent the initial equilibrium torus of the density profile with magnetic field lines in Figure \ref{Figure_1}. Here, we consider the spin, angular momentum, and plasma-$\beta$ are $a_k =0.00$, $\lambda=\lambda_{\rm K}$ and $\beta_0 = 10$, respectively. MRI transports angular momentum outwards and gradually increases the accretion of matter toward the black hole. As a result, the edge of the disc drifts towards the black hole horizon. The accretion flow becomes turbulence via MRI. In this turbulent state, the gas spreads in the vertical direction. With time evolve, the poloidal magnetic field enhances in the disc, and the magnetic pressure drives the matter outward from the disc as outflow \citep{Machida-etal00, Hawley-Krolik01, Hawley-Balbus02, De-Villiers-etal03a}.

\subsection{Effect of magnetic field}
\label{effect_mag}

In this section, we first examine the effect of magnetic field on the evolution of torus around spinning AGN. For the purpose of analysis, we consider four initial plasma-$\beta$ parameters. In this analysis, we also fix the maximal spin value $a_k = 0.99$ for AGN. To set up the initial torus, we supply Keplerian flow angular momentum $ \lambda_{\rm K} = 7.21$ (see equation \ref{kep_ang}). Here, we compare the effect of the strength of the magnetic field on the torus evolution keeping all other parameters fixed. Initially, we compare density distribution on $(r-z)$ plane at a different simulation time as depicted in Figure \ref{Figure_2}. The first, second, third, and fourth rows are for $\beta_0$ = 10, 50, 100, and 1000, respectively. Also, the different columns are for different times of torus evolution $t_1 = 500 t_g$, $t_2 = 5000 t_g$, $t_3 = 10500 t_g$ and $t_4 = 17500 t_g$, respectively. In the first column, we observe that the initial torus is formed around AGN for all the magnetic field configurations. With the decrease of plasma-$\beta$ parameters. i.e., with the increase of magnetic field in the disc, the MRI enhances Maxwell stress more rapidly. As a result, the transport of angular momentum amplifies with the increase of MRI and initial disc spreads throughout the computational domain. It is found that matter from the torus expands more easily and rapidly for a higher magnetic field compared to the lower magnetic field, depicted in Figure \ref{Figure_2}. It implies that the magnetic field plays an essential role in torus evolution around AGN. It is also observed that mass is ejected from the disc as outflow. Due to the mass escape from the disc, the torus gradually destroys and forms a uniformly dense disc with a minimum density of torus at the end time of the simulation (see Figure \ref{Figure_2}, first row, last column). It is to be mentioned that we check the quality factors for resolving MRI in our simulation model \citep{Hawley-etal11, Hawley-etal13}. We find that both the quality factors in radial and vertical directions are $Q_r, Q_z \gtrsim 15$, see Appendix \ref{MRI_resolve}. Therefore, our simulation model is able to resolve MRI very efficiently. 

In Figure \ref{Figure_3}, we represent the distribution of temperature ($T$), plasma-$\beta$ $(\beta)$, azimuthal magnetic field $(B_{\phi})$ and magnetization parameter $(\sigma_{\rm M})$ for different initial magnetic field strength ($\beta_0$) at time $t_3 = 10500 t_g$. Column first, second, third and four are for different initial magnetic field $\beta_0 = 10, 50, 100$ and 1000, respectively. The disc is heated up ($T \gtrsim 10^9$K) by releasing the gravitational energy and expands in the vertical direction more efficiently for a higher magnetic field compared to a low magnetic field. The temperature is distributed throughout the disc for a higher magnetic field compared to the lower one. We show the plasma-$\beta$ distribution in the second row of Figure \ref{Figure_3}. It is observed that gas pressure is dominated in the disc region. However, the plasma-$\beta$ distribution becomes low ($\beta < 1$) in the disc corona where the magnetic field buoyantly escapes from the disc to the disc corona for a high magnetic field compared to a low one. In fact, in a low magnetic field case, it is impossible to escape the magnetic field from the torus, and it remains confined within the torus. Further, we present the distribution of the azimuthal magnetic field ($B_{\phi}$) in the third row of Figure \ref{Figure_3}. We also observe that the azimuthal magnetic fields buoyantly escape from the disc to the disc corona. In general, the azimuthal magnetic fields are antisymmetric with respect to the equatorial plane, and they facilitate magnetic reconnection. This magnetic reconnection may trigger heating in the disc (see Figure \ref{Figure_3}, first row). Moreover, the azimuthal magnetic field occasionally changes its polarity and escapes from the disc. Finally, we present the distribution of the magnetization parameter ($\sigma_{\rm M}$) in the fourth row of Figure \ref{Figure_3}. We observe that the magnetization parameter ($\sigma_{\rm M}$) becomes low in the equatorial disc region and significantly increases ($\sigma_{\rm M} \sim 1$) away from the equatorial plane for all the magnetized case, i.e., $\beta_0 = 10, 50$ and 100. Moreover, we find that the matters escaping from the disc are magnetically driven as depicted by the distribution of plasma-$\beta$ and magnetization parameter in Figure \ref{Figure_3}. However, we do not observe very high $\sigma_{\rm M}$ values for our simulation, even in the highly magnetized case ($\beta_0 = 10$). High values of magnetization parameters usually refer to highly relativistic jets with high Lorentz factor, i.e., BZ jet \citep{Dihingia-etal21, Narayan-etal-22, Dihingia-etal22, Hong-Xuan-etal23}. In this work, we only find magnetized mass outflow with $\sigma_{\rm M} \sim 1$ but not relativistic jets. 


\subsubsection{Magnetic state of the accretion flow}

Here, we investigate the characteristics of the magnetic state of the accretion flow in our model. It is usually investigated two quantities, mass accretion rate $(\dot{M}_{\rm acc})$ and the normalized
magnetic flux threading the BH horizon ($\dot{\phi}_{\rm acc}$) \citep{Tchekhovskoy-etal11, Narayan-etal12}. In this work, the mass accretion rate is defined as the mass flux entering through the inner boundary ($r_{\rm in}$) towards the black hole and is defined as
\begin{align}\label{mass_acc_eqn}
\dot{M}_{\rm acc} = - 2\pi \int{\rho (r,z) r v_r dz}. 
\end{align}
Here, the `-ve' sign indicates the inward direction of mass flux. Additionally, we define dimensionless normalized magnetic flux threading to the black hole horizon $(\dot{\phi}_{\rm acc})$. The normalized magnetic flux usually refers to the ``MAD-ness" parameter. Therefore, the normalized magnetic flux is given by \citep{Tchekhovskoy-etal11, Narayan-etal12, Dihingia-etal21, Dhang-etal23} 
\begin{align}\label{mag_flux_acc_eqn}
\dot{\phi}_{\rm acc} = \frac{\sqrt{4 \pi}}{2}  \frac{\int{|B_r|_{r = r_{\rm in}} dz}}{\sqrt{\dot{M}_{\rm acc}}},
\end{align}
where $r_{\rm in}$ represents the inner boundary. Here, $B_r$ is the radial component of the magnetic field. We represent the temporal variation of mass accretion rate and the normalized magnetic flux in Figure \ref{Figure_4}a and Figure \ref{Figure_4}b, respectively. Here, we fix spin $a_k = 0.99$ and angular momentum $\lambda = \lambda_{\rm K}$ and vary initial magnetic field ($\beta_0$). We observe that the mass accretion rate lies in the sub-Eddington limit ($\dot{M}_{\rm acc}< \dot{M}_{\rm Edd}$) throughout the simulation run for all the cases, shown in Figure \ref{Figure_4}a. It is generally observed that the value of saturated magnetic flux is a good indicator of the characteristic of the magnetic state of the accretion flow. We find that magnetic flux continues to grow and becomes saturated after $t\sim 2000, 2500, 3000 t_g$ for $\beta_0 = 10, 50$ and 100, respectively. In the previous studies in GRMHD simulation, it is observed that the criterion to enter the MAD state is when the normalized magnetic flux reaches the critical value. Generally the threshold value of magnetic flux is $\dot{\phi}_{\rm acc}$ $\sim$ 15 for MAD state \citep{Tchekhovskoy-etal11, Narayan-etal12, Dihingia-etal21, Hong-Xuan-etal23}. In this work, we find that the saturated magnetic flux crosses the threshold value for MAD state for the high magnetic case model ($\beta_0 = 10$), shown in Figure \ref{Figure_4}b. We argue that the magnetic model $\beta_0 = 10$ is similar to the MAD state as generally found in GRMHD simulation. However, the saturated magnetic flux remains more or less below the threshold value of MAD for the case of $\beta_0 = 50$ and 100. These two models are high-magnetized SANE states. It is to be mentioned that the inner boundary is higher compared to the event horizon for $a_k =0.99$,i.e., $r_{\rm in}> r_{\rm H}$ in our model, where $r_{\rm H} = 1+\sqrt{1-a_k^2}$. We also observe that for low-magnetic flow ($\beta_0 = 1000$), the mass accretion rate attains saturation value after a long time ($t\geq 12500 t_g$), and the mass accretion rate is very low compared to the high magnetized flow. Moreover, the saturation magnetic flux value is much lower compared to the higher magnetic case and remains in the low magnetic SANE state. 


\subsubsection{Radial dependence of the flow variables}

Now, we investigate the comparison of the overall radial variation of various flow variables. We represent density $\rho$ in g cm$^{-3}$, temperature $T$ in K and radial velocity $v_r$ in units of speed of light in Figure \ref{Figure_5}a, \ref{Figure_5}b and \ref{Figure_5}c, respectively. Here all the variables are vertically space-averaged between $-2 r_g \leq z \leq 2 r_g$ and time-averaged over $10500 t_g \leq t \leq 12500 t_g$. The red, blue, magenta and green curves are for $\beta_0$ = 10, 50, 100, and 1000, respectively. The radial distribution of density follows $\rho \sim r^{1/2}$ except low magnetic flow $\beta_0$ = 1000 in the inner region of the disc ($r < 20 r_g)$. It shows steeper dependence on the radial coordinate. Similarly, temperature profile follows $T\sim r^{-1}$ and radial velocity also shows $v_r \sim r^{-3/2}$ dependence (see the appendix C). This is almost similar to the non-radiative convection dominated accretion flows (CDAF) \citep{Narayan-etal00, Machida-etal01, Igarashi-etal20}. The dashed black lines represent CDAF model dependence. The density declines inwards towards the horizon as the radial velocity increases inwards, depicted in Figure \ref{Figure_5}a and \ref{Figure_5}c. Consequently, temperature increases towards the horizon and forms a low-density, hot, RIAF-like flow, shown in Figure \ref{Figure_5}b. In Figure \ref{Figure_5}d and \ref{Figure_5}e, we represent the space averaged plasma-$\beta$ $(\beta)$ and magnetization parameter $(\sigma_{\rm M})$, respectively. The plasma-$\beta$ parameter decreases at the inner part of the disc near the black hole horizon ($\beta \sim 1$). It implies that the magnetic pressure increases near to the horizon for all the cases. On the hand, we observe that the magnetization parameter increases significantly near to the horizon. However, we find that the magnetization parameter is lower than unity ($\sigma_{\rm M} <1$) in the equatorial plane for all the cases. In MHD flows, the angular momentum transports outward region, and gas accretes inwards toward the horizon via MRI. Initially, we set all the torus with constant angular momentum ($\lambda$). For the representation, we supply constant Keplerian angular momentum $(\lambda_{\rm K} = 7.21)$ at the pressure maximum surface. We observe that the specific angular momentum becomes nearly Keplerian distribution for all the cases, depicted in Figure \ref{Figure_5}f. The black dotted curve represents the theoretical Keplerian angular momentum distribution for $a_k =0.99$ (see equation \ref{kep_ang}). Figure \ref{Figure_5}g  represents mass accretion rates in Eddington units ($\dot{M}_{\rm Edd}$). The mass accretion rates are calculated using equation (\ref{mass_acc_eqn}) where the integration is carried out over the vertical direction $-2r_g$ to $2r_g$. We observe that the mass accretion rate is always in the sub-Eddington range ($\dot{M} < \dot{M}_{\rm Edd}$) throughout the radial distance. Also, the mass accretion rate is more or less constant very near to the black hole ($r < 10 r_g$). Now we calculate normalized Reynolds stress $\alpha_{\rm gas}$ and normalized Maxwell stress $(\alpha_{\rm mag})$. The Reynolds stress is calculated as $\alpha_{\rm gas} = \frac{<\rho v_r \delta v_{\phi}>}{P_{\rm gas}}$ and the Maxwell stress is estimated as $\alpha_{\rm mag} = -\frac{<2B_r B_{\phi}>}{B^2}$ \citep{Hawley-00, Stone-Pringle01, Proga-Begelman-03}. We observed that Maxwell stress lies $\alpha_{\rm mag} < 0.65$ and Reynolds stress is $\alpha_{\rm gas} < 0.3$. It also confirms that Maxwell's stress is much stronger than Reynolds's stress. It implies that the outward transport of angular momentum is predominantly driven by Maxwell stress in MHD flows.  


\subsubsection{Mass outflows and luminosity}

Further, it is observed that the gaseous matter in the disc expands vertically above and below the torus due to MRI turbulence, and it carries a significant amount of magnetic field with it (see Figure \ref{Figure_2}, \ref{Figure_3}). This indicates the possibility of a magnetized mass outflow from the disc. In this work, one of the main purposes is to estimate mass outflow from the disc. One of the major criteria for mass outflow is that when the Bernoulli parameter is positive $Be \geq 0$,i.e., unbound flow \citep{Narayan-Yi94, Narayan-etal12, Penna-etal13}. In Newtonian hydrodynamics, the Bernoulli parameter $(Be)$ is the sum of the kinetic energy, potential energy, and enthalpy as shown in equation (\ref{Bernoulli_eqn}). In general, at a large distance from the black hole, the potential energy diminishes. Here, we subtract unity from total energy to eliminate the rest of mass-energy from the gas (see equation (\ref{Bernoulli_eqn})). In Figure \ref{Figure_6}a, we show the variation of the Bernoulli parameter ($Be$) with simulation time for $\beta_0$ = 10 (red), 50 (blue), 100 (magenta) and 1000 (green). We observe that initially $Be< 0$, i.e., bound or no outflows for $\beta_0 = 10$, 50, 100. With the growing MRI and increasing magnetic pressure (i.e., magnetic energy) in the disc, the Bernoulli parameter becomes positive $Be>0$, depicted in Figure \ref{Figure_6}a. However, for low magnetic case $(\beta_0 = 1000)$ $Be$ remains negative, shown in the inset figure of Figure \ref{Figure_6}a. It implies that mass outflow is not possible for low magnetic flow. We also observe a significant positive correlation between initial plasma-$\beta$ and Bernoulli parameter.

In this paper, we calculate the mass outflow rates by calculating the rate of outflowing matter through the outer $z$-boundary $(z = \pm z_{\rm out})$ as
\begin{align} \label{mass_outflow_eqn}
\dot{M}_{\rm out} = 2 \pi \int^{r_{\rm out}}_{r_{\rm in}} \left[\rho(r, z_{\rm out}) v_z (r, z_{\rm out}) - \rho(r, -z_{\rm out}) v_z (r, -z_{\rm out})\right] r dr,
\end{align}
where, $v_z$ is the vertical velocity as a function of $(r, z)$. Here, we ignore the mass loss from the $r_{\rm out}$ boundary. In Figure \ref{Figure_6}b, we compare the mass outflow rates $\dot{M}_{\rm out}$ in units of g s$^{-1}$ for different magnetic field strengths. Interestingly, we observe that mass outflow rates significantly increase with the increase of magnetic field. For example, the mass outflow rates roughly increase almost ten times if we decrease initial plasma-$\beta$ from 50 to 10. It implies that the mass outflow rates are purely magnetically driven in our model. Moreover, we find the quasi-periodic nature of mass outflow rates in the magnetized flow \citep{Okuda-etal19}. 

Now, we attempt to estimate the luminosity emanating from our model. If we consider here only ion-electron bremsstrahlung emission in a single
temperature approximation, the optical thickness  $\Delta \tau = \kappa \rho \Delta r$ across the mesh size $\Delta r$ for the present model is 
\begin{align}
\Delta \tau = 6 \times 10^{-24} \left(\frac{\rho}{10^{-12}}\right)^2 \left(\frac{T}{10^{13}}\right)^{-3.5} \left(\frac{\Delta r}{0.02 \times 1.5 \times 10^{13}}\right) << 1 ,
\end{align}
where the opacity $\kappa$ is given by the Kramers approximation corresponding
to the bremsstrahlung emission \citep{Kley-89}. Accordingly, the gas is fully optically thin
to the bremsstrahlung, and then the total luminosity due to the bremsstrahlung emission is given  as 
\begin{align}
L = \int e_{\rm ff} dV,
\end{align}
where, $e_{\rm ff}$ is the bremsstrahlung emission rate per unit volume. The integration is carried out all over the computational domain. However, in a realistic accretion flow, we cannot ignore synchrotron emission in a magnetized flow \citep{Dihingia-etal22, Okuda-etal23, Curd-Narayan23}. We find that the flow in our model is optically thick to the monochromatic synchrotron emission in some radial zones for some frequency ranges in the radio to IR band. Then, the synchrotron luminosity is only estimated correctly by solving the radiative transfer \citep{Okuda-etal23}. Therefore, in the present formalism, we may regard bremsstrahlung luminosity as the measure of luminosity for the correlation. In Figure \ref{Figure_6}c, we show the variation of luminosity with time for all the magnetic flow cases. We find that there is a completely steady nature of luminosity for low magnetic flow $\beta_0$ = 1000. It can also be visualized from the density and temperature distribution in Figure \ref{Figure_2} and Figure \ref{Figure_3}. The magnetic field in the high $\beta_0$ case is not significant enough to trigger the turbulent flow in the torus. The fluctuation is minimal in this case. However, with the increase of the magnetic field, turbulence is enhanced in the torus via MRI, and consequently, the luminosity increases with time. Also, we observe quasi-periodic variation in the luminosity for higher magnetic field cases due to the turbulent nature of the accretion flow. This quasi-periodic nature of luminosity variations in a magnetized flow can explain flaring events, and QPOs originated from AGNs \citep{Okuda-etal19, Okuda-etal22}. Moreover, we observe that total luminosity significantly increases nearly 100 times with the increase of magnetic field from $\beta_0$ = 50 to 10, similar to mass outflow rates. The simulation run ends when there is no structure of torus at all, i.e., minimum density ($\rho_{\rm min}$) covers the whole disc (see Figure \ref{Figure_2}, first row, last column). At the end time of the simulation, most of the gas escaped from the disc to create a sudden increase of mass outflow rates and luminosity, shown in Figure \ref{Figure_6}b,c for $\beta_0$ = 10. Moreover, it is to be noted that MRI is well-sustained for longer periods of evolution if we consider large initial torus size \citep{Hong-Xuan-etal23}.

\subsection{Effect of black hole spin}
\label{effect_spin}

In this section, we investigate the effect of black hole spin on the evolution of torus. Here, we use the same initial torus configuration for $\beta_0$ = 100 as mentioned in sub-section \ref{effect_mag} but only vary black hole spin as $a_k = 0.99, 0.80, 0.50$, and 0.0. i.e.,  maximally spinning to a non-spinning black hole. It is observed that the initial torus size is slightly decreased with the lowering of the black hole spin \citep{Utsumi-etal22}. Because with the decrease of black hole spin, the total angular momentum of the system decreases. In Figure \ref{Figure_7}, we show density $\rho$, temperature $T$, plasma-$\beta$ and magnetization parameter $(\sigma_{\rm M})$ distribution at time $t_3 = 10500 t_g$. The first, second, third, and fourth columns are for spin $a_k = 0.99$, 0.80, 0.50, and non-spinning black hole $a_k=0.0$, respectively. We observe more or less similar trends for density, temperature, plasma-$\beta$, and magnetization parameter distribution for spinning or non-spinning black holes. We find that magnetized mass outflow is possible for rotating as well as non-rotating black holes, as depicted in Figure \ref{Figure_7}. Further, we observe no significant effect of black hole spin on the accretion process of the torus as such as pointed out by \citet{Hong-Xuan-etal23}. 

To investigate the effect of spin on the torus evolution rigorously, we compare the Bernoulli parameter, mass outflow rates and luminosity variation by varying spin, depicted in Figure \ref{Figure_8}a, \ref{Figure_8}b and \ref{Figure_8}c, respectively. The red, blue, magenta and green curves are for spin $a_k = 0.99, 0.80, 0.50$, and 0.0, respectively. We do not find any significant correlation of spin on the Bernoulli parameter, shown in Figure \ref{Figure_8}a. Moreover, we observe absolutely no correlation between spin ($a_k$) and mass outflow ($\dot{M}_{\rm out}$) as well as between spin and luminosity $(L)$. In this regards, a very weak correlation between mass outflow rates and spin has been observed based on analytical investigation \citep{Aktar-etal15}. This is because the mass outflow happens throughout the disc, and the effect of black hole spin diminishes with the increase of the distances from the event horizon.

\subsection{Effect of flow angular momentum}
\label{effect_ang}

In the same spirit, we investigate the effect of flow angular momentum on the torus evolution. The angular momentum plays an essential role in forming the initial torus around the black hole. In this regard, we consider the same initial configuration of torus set up as sub-section \ref{effect_mag} and \ref{effect_spin}. But, we only vary the specific angular momentum in the sub-Keplerian range ($\lambda < \lambda_{\rm K}$). Here, we fix the spin of the black hole as $a_k =0.99$. Now, we vary specific angular momentum as $\lambda = \lambda_{\rm K}, 7.00, 6.80$ and 6.60, where $\lambda_{\rm K} = 7.21$. The initial torus size decreases significantly with the decrease of angular momentum, and beyond a critical $\lambda$, there is no possibility of torus formation. It is obvious that angular momentum provides a repulsive centrifugal force against attractive gravity force to form a torus. It is found that there it is not possible to form torus beyond $\lambda < 6.60$ for this initial and boundary conditions. In Figure \ref{Figure_9}, we represent density $(\rho)$, temperature $(T)$, plasma-$\beta$ ($\beta$) and magnetization parameter ($\sigma_{\rm M}$) distribution similar as Figure \ref{Figure_7}. Here, the first, second, third, and fourth columns are for angular momentum $\lambda = \lambda_{\rm K}$, 7.00, 6.80 and 6.60, respectively. It is observed that the initial torus size is much smaller in $\lambda = 6.60$ compared to $\lambda = \lambda_{\rm K}$. Further, we represent the variation of angular momentum at different times of evolution in Figure \ref{Figure_10}. The red, blue, magenta and green curves are for angular momentum $\lambda = \lambda_{\rm K}, 7.00, 6.80$, and 6.60, respectively. The dotted black curves are theoretically calculated specific Keplerian angular momentum (see equation \ref{kep_ang}). We observe that with time evolution, the angular momentum distribution becomes near Keplerian distribution for all the cases, even if for the sub-Keplerian flow. Moreover, we investigate the effect of angular momentum on the Bernoulli parameter, mass outflow rates and luminosity, depicted in Figure \ref{Figure_11}a, \ref{Figure_11}b and Figure \ref{Figure_11}c, respectively. We also observe no correlation between angular momentum and mass outflow rates and luminosity similar to the black hole spin in our model.

\section{Discussions and Conclusions}
\label{conclusion}

In this paper, we present the simulation results in two-dimensional MHD accretion flows around spinning AGN. We use PLUTO code to simulate MHD flows \citep{Migone-etal07}. To mimic the general relativistic effects of spinning black holes, we adopt effective Kerr potential introduced by \citet{Dihingia-etal18b}. The advantage of this Kerr potential is that we investigate accretion flows in multi-dimension and with higher spatial resolution for highly spinning black holes without doing expensive and complex general-relativistic simulations. In this work, we adopt axisymmetric, two-dimensional torus evolution around spinning AGN, considering non-resistive and non-radiative (RIAF) approximation. The initial magnetic field in the torus is configured following \citet{Hawley-etal02} (see sub-section \ref{Mag_config}). The toroidal magnetic field is developed in the disc due to shear, and the poloidal magnetic field is amplified via MRI \citep{Hawley-00, Hawley-etal02} (see Figure \ref{Figure_2}). MRI ignites the non-linear turbulence in the accretion flow, and the flow becomes in an MHD turbulence state. As a result, Maxwell's stress transport angular momentum outwards, and mass accretion happens. In this work, we examine the effect of the magnetic field, the spin of the black, and the flow angular momentum on the torus evolution. 

In Figure \ref{Figure_2} and \ref{Figure_3}, we represent the distribution of density, temperature, plasma-$\beta$, azimuthal magnetic field, and magnetization parameter by varying initial magnetic field strengths. We find that the magnetic field plays a pivotal role in the torus evolution. We observe that gas and magnetic field can easily escape from the disc for highly magnetized disc compared to low magnetized flow, as depicted in Figure \ref{Figure_3} of $\beta$ and $\sigma_{\rm M}$ distribution. We also investigate the magnetic state of the accretion flow for our model. For the purpose of analysis, we calculate normalized magnetic flux $(\dot{\phi}_{\rm acc})$ with time \citep{Tchekhovskoy-etal11, Narayan-etal12, Dihingia-etal21}. We observe that the model $\beta_0 = 10$ is similar to the MAD, shown in Figure \ref{Figure_4}b. On the other hand, model $\beta_0 = 50$ and $\beta_0 = 100$ are highly magnetized SANE and $\beta_0 = 1000$ is in the low magnetized SANE state. The radial distribution of density, temperature, and radial velocity follow non-radiative CDAF distribution in magnetized flow, depicted in Figure \ref{Figure_5}a, b, c. We observe that the magnetic pressure and magnetic energy increases with the radial distance towards the horizon as shown in the radial variation of $\beta$ and $\sigma_{\rm M}$ in Figure \ref{Figure_5}d and Figure \ref{Figure_5}e, respectively. Also, we find that the angular momentum distribution is nearly Keplerian for all the cases; see Figure \ref{Figure_5}f. Figure \ref{Figure_5}g indicates that the accretion flow remains sub-Eddington limit throughout the disc. Also, the variation of Reynolds's stress and Maxwell's stress confirms that Maxwell's stress is more prominent compared to Reynolds's stress in magnetized flow, depicted in Figure \ref{Figure_5}h and \ref{Figure_5}i. Further, we examine the effect of the magnetic field on the mass outflow rates and luminosity emanating from the disc. To check the unbound matter as mass outflow as far as energy is concerned, we also calculate the Bernoulli parameter with time as depicted in Figure \ref{Figure_6}a \citep{Narayan-Yi94, Narayan-etal12, Penna-etal13}. We find a positive correlation between initial magnetic field and Bernoulli parameter as shown in Figure \ref{Figure_6}a. We also observe a significant positive correlation between magnetic field strengths and mass outflow and as well as luminosity, shown in Figure \ref{Figure_6}a,b,c. Interestingly, we find an almost steady nature of luminosity variation for less magnetized flow. On the other hand, the quasi-periodic nature of luminosity variation is observed in magnetized flow. Therefore, the magnetized flow may be very useful to explain the QPOs observed for various black hole sources \citep{Okuda-etal19, Okuda-etal22}.  

Further, we examine the effect of black hole spin on the torus evolution in magnetized flow. To investigate that, we plot the variation of Bernoulli parameter, mass outflow rates and luminosity by varying black hole spin, depicted in Figure \ref{Figure_8}. Interestingly, we have not found any correlation between black hole spin and mass outflow rates for magnetized flow \citep{Aktar-etal15}. Further, we observe that the black hole spin has no prominent role in the accretion dynamics \citep{Hong-Xuan-etal23}. Similarly, we investigate the effect of flow angular momentum on the magnetized flow. It is observed that initial torus formation is strongly dependent on angular momentum, shown in Figure \ref{Figure_9}. Beyond a critical angular momentum, torus formation is impossible. We also find no correlation between angular momentum and mass outflow rates from our MHD simulation model, depicted in Figure \ref{Figure_11}.  

It is to be emphasized that our simulation model belongs to the semi-relativistic regime by adopting effective Kerr potential to mimic the general relativistic effects around spinning black holes. One of the limitations of the semi-relativistic model is that it is unable to explain highly relativistic jets, which are commonly observed around AGNs or XRBs. The general consensus is that highly relativistic jets are originated via BZ process \citep{Blandford-Znajek77} around spinning black holes. In this regard, \citet{McKinney-Gammie-04}, for the first time, explored relativistic jets generated via BZ mechanism based on GRMHD simulation. Later, a growing number of independent simulation studies confirms that BZ jets are generic phenomena in GRMHD simulations around highly spinning black holes to explain the relativistic jets \citep{Tchekhovskoy-etal11, Narayan-etal12, Dihingia-etal21}, and reference therein. Therefore, one needs to model accretion flows considering GR effects around spinning black holes to address relativistic jets. We hope to investigate the accretion-jets mechanism by incorporating the GRMHD simulation scheme around spinning black holes in the future.   

In this work, we consider two-dimensional MHD flows in an axisymmetric assumption. To get the complete picture, we need to investigate global three-dimensional simulation studies. Moreover, the radiation transport mechanism always plays a crucial role in explaining state transition for black holes. Also, radiation flux can drive mass outflow to the relativistic limit. i.e., jets formation. Further, the radiation mechanism in the MHD flows may explain the CLAGNs \citep{Igarashi-etal20}. We plan to investigate global three-dimensional radiation-dominated MHD flows around spinning black holes in the future. 

\section*{Acknowledgments}

We sincerely thank the anonymous referee for the valuable suggestions and comments that helped us to improve the manuscript. This work is supported by the National Science and Technology Council of Taiwan through grant NSTC 111-2811-M-007-033, 111-2112-M-007-037, and by the Center for Informatics and Computation in Astronomy (CICA) at National Tsing Hua University through a grant from the Ministry of Education of Taiwan. The simulations and data analysis have been carried out on the CICA Cluster at National Tsing Hua University. We want to thank Indu Kalpa Dihingia for valuable comments and discussions during the preparation of the manuscript.

\section*{Data Availability}

The data and code underlying this article will be shared on reasonable request to the corresponding author.

\bibliographystyle{mnras}
\bibliography{refs}

\appendix

\section{Numerical Convergence Test}
\label{convergence_test}

\begin{figure}
	\begin{center}
		\includegraphics[width=0.5\textwidth]{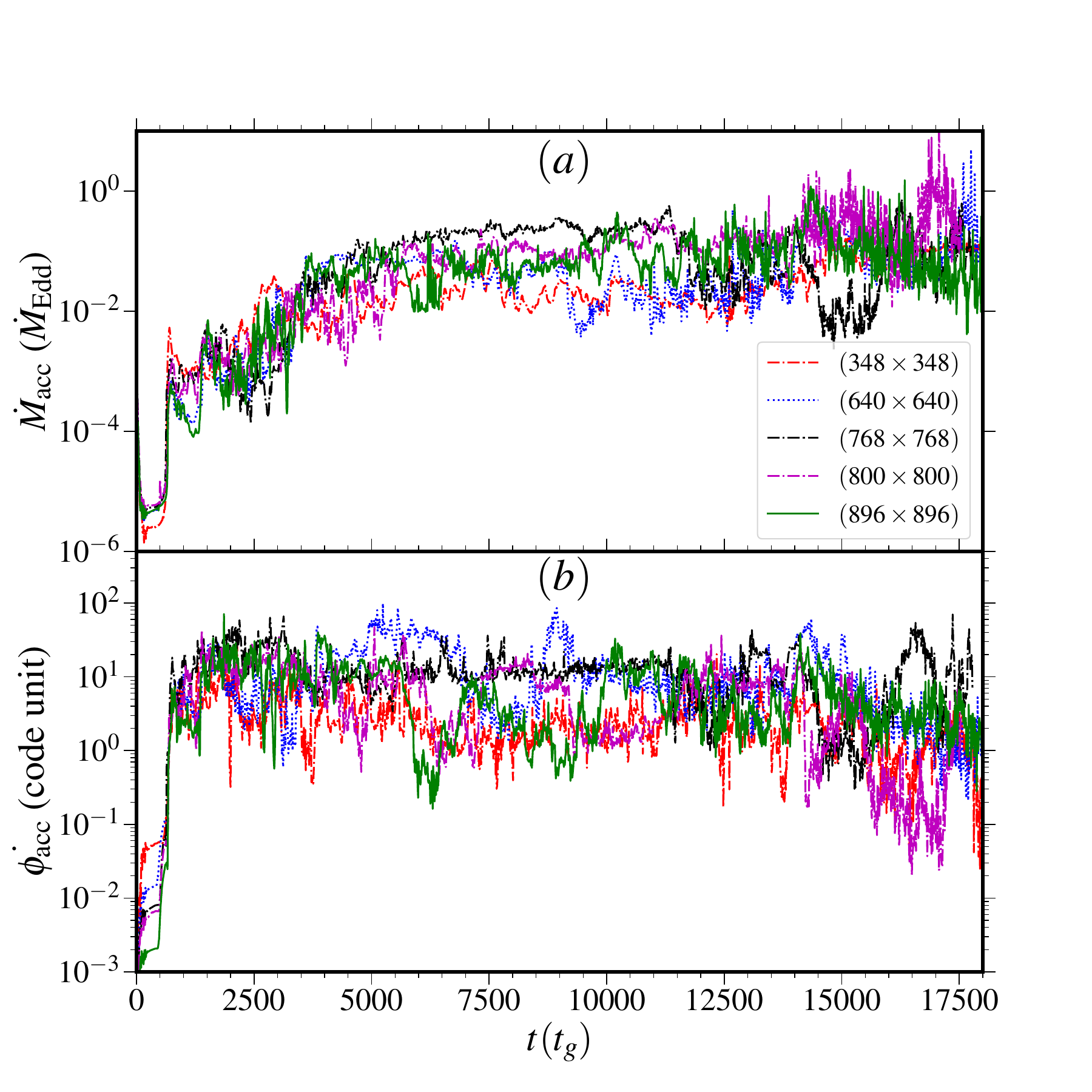} 
	\end{center}
	\caption{Time evolution of mass accretion rate $(\dot{M}_{\rm acc})$ and normalized magnetic flux $(\dot{\phi}_{\rm acc})$ for different simulation resolution. Red (dashed-dotted), blue (dotted), black (dashed-dotted), magenta (dashed-dotted) and green (solid) are for resolution ($348\times348$), ($640\times640$), ($768\times768$), ($800\times800$) and ($896\times896$), respectively.}
	\label{Figure_12}
\end{figure}

\begin{figure}
	\begin{center}
		\includegraphics[width=0.5\textwidth]{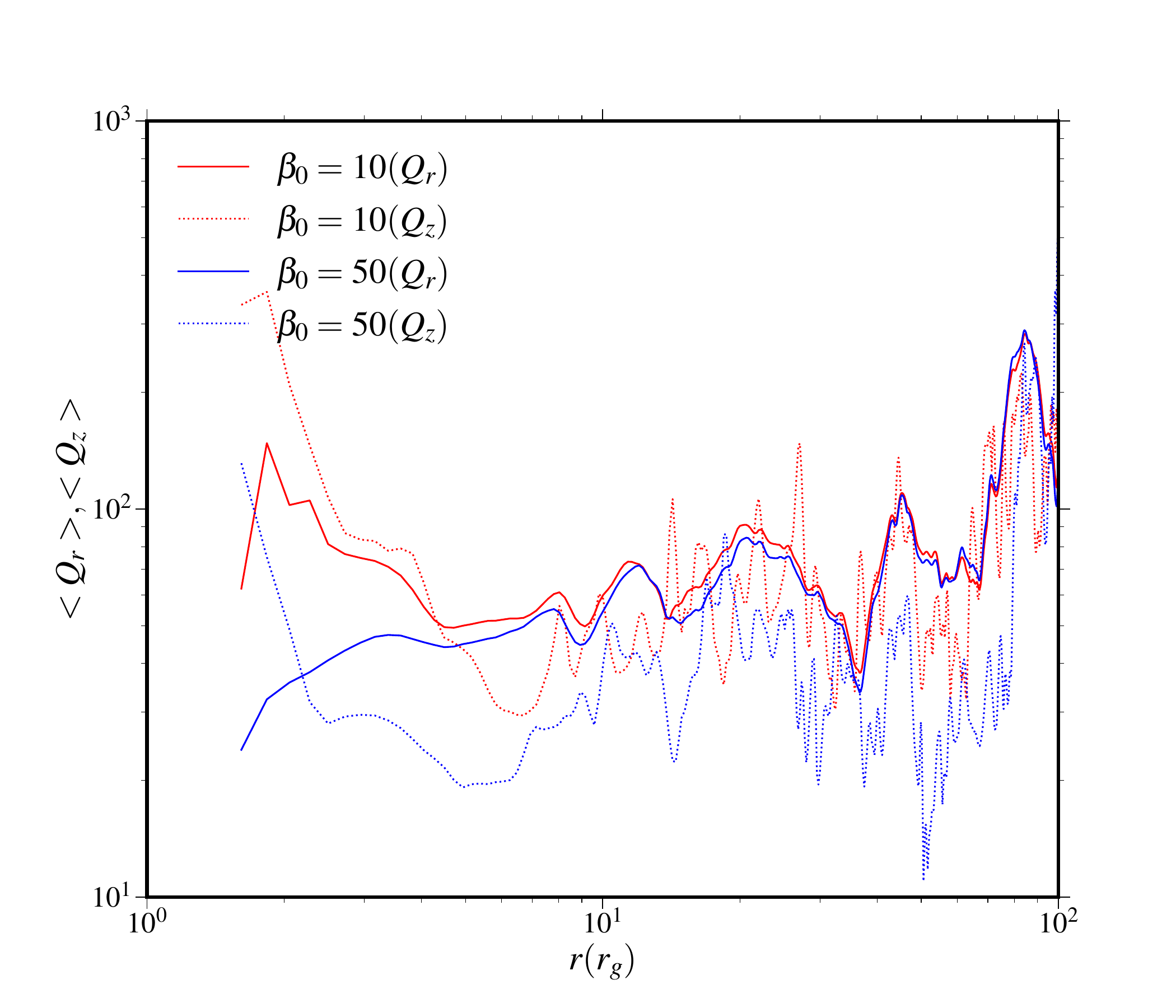} 
	\end{center}
	\caption{Time-averaged radial variation of MRI quality factor $Q_r$ (solid) and $Q_z$ (dotted). The quality factor are space-averaged vertically over $-5r_g<z<5r_g$. The red and blue curves are for $\beta_0 = 10$ and 50, respectively.}  
	\label{Figure_13}
\end{figure}

Ideally, the convergence test of the numerical model is compared with the exact analytical solution. However, there is a very limited exact solution available for fluid equations. Also, the simulation method is based on the discretization of the computational domain as well as various approximation methods. Here, we perform a convergence test for our simulation model by calculating two important parameters such as mass accretion rate $(\dot{M}_{\rm acc})$ and normalized magnetic flux $(\dot{\phi}_{\rm acc})$ in time using equation (\ref{mass_acc_eqn}) and (\ref{mag_flux_acc_eqn}) by varying different simulation resolution, shown in Figure \ref{Figure_12}. We vary the computational resolution as $348\times348$ (Red dashed-dotted), $640\times640$ (blue dotted), $768\times768$ (black dashed-dotted), $800\times800$ (magenta dashed-dotted) and $896\times896$ (green solid), respectively. We observe that with the increase of resolution, the mass accretion rate and normalized magnetic flux are settled down in mean values,i.e., convergence values for ($896\times896$). Therefore, we adopt the high resolution ($896\times896$) for all the simulation runs in this paper, keeping in mind the numerical accuracy and long-term simulation run.

\section{Resolution of the MRI }
\label{MRI_resolve}

In magnetized accretion flow, MRI is one of the essential triggering mechanisms to transport angular momentum. Here, we check whether the flow is subjected to MRI in our simulation model. In order to have MRI-driven accretion, it is necessary to resolve the fastest-growing MRI mode. To quantify the criteria to resolve the MRI, we define the characteristic wavelength of the fastest growing mode of MRI as $\lambda_{\rm MRI} = 2 \pi V_{\rm A}/\Omega$. Here, $V_{\rm A}$ is the Alfv\'en speed, and $\Omega$ is the angular velocity. The quality factor for resolving MRI is defined as \citep{Hawley-etal11, Hawley-etal13}
\begin{align}
Q_r = \frac{2 \pi V_{{\rm A}r}}{\Omega \Delta r},
\end{align}
\begin{align}
Q_z = \frac{2 \pi V_{{\rm A}z}}{\Omega \Delta z},
\end{align}
where, $V_{{\rm A}r}$ and $V_{{\rm A}z}$ are the radial and vertical component of Alfv\'en speed, respectively. $\Delta r$ and $\Delta z$ are the grid sizes in the radial and vertical directions, respectively. In general, it is shown that the quantitative values are $Q_r \gtrsim 10$ and $Q_z \gtrsim 10$ to sufficiently well-resolved MRI \citep{Hawley-etal11, Hawley-etal13}. In Figure \ref{Figure_13}, we represent the time-averaged radial variation of quality factors $(Q_r, Q_z)$ of MRI. Here, we consider the time average over $t=(7000-10000)t_g$ and vertical average over $-5r_g<z<5 r_g$. We show the variation of the quality factor for magnetized flow $\beta_0 = 10$ (red) and 50 (blue), depicted in Figure \ref{Figure_13}. Solid and dashed curves are for $Q_r$ and $Q_z$, respectively. We observe $Q_r$ $\gtrsim 15$ and $Q_z$ $\gtrsim 15$ for our simulation result. It clearly indicates that our simulation model is able to resolve MRI in good agreement.

\section{Radial dependence of flow variables in CDAF Model}
\label{CDAF_eqn}

The self-similar solution has been introduced to investigate accretion flow around black holes in the literature \citep{Spruit-etal87, Narayan-Yi94, Narayan-etal00}. Self-similarity aims to obtain a dimensionless system of equations whose primary essence is to scale physical quantities with local flow variables. In this regard, the Keplerian angular velocity and linear velocity can be written as $\Omega_{\rm K} = \left(\frac{GM_{\rm BH}}{r^3}\right)^{1/2}$ and $v_{\rm K} = r\Omega_{\rm K} = \left(\frac{GM_{\rm BH}}{r}\right)^{1/2}$, respectively. Here, $G$, $M_{\rm BH}$, and $r$ are the gravitational constant, the mass of the black hole, and cylindrical radius, respectively. Therefore, the self-similar scaling of angular velocity, sound speed, and scale height can be written as 
\begin{align}
\Omega = \Omega_0 \Omega_{\rm K} \propto r^{-3/2} 
\end{align}
\begin{align}
c_s^2 = c_0^2 v_{\rm K}^2 \propto r^{-1},
\end{align}
\begin{align}
H = c_s/\Omega_{\rm K} = c_0 r,
\end{align}
where $\Omega_0$ and $c_0$ are the dimensionless constant and can be determined. Now, we assume the density variation as 
\begin{align}
\rho = \rho_0 r^{-\alpha}.
\end{align}
and the pressure is scaled as 
\begin{align}
P = \rho c_s^2 \propto r^{-1-\alpha}.
\end{align}
Therefore, the radial dependence of temperature can be written
\begin{align}
T \propto r^{-1}.
\end{align}
Now, the mass accretion rate is obtained from the mass conservation equation as
\begin{align}
\dot{M} = 2 \pi \rho v r H.
\end{align}
To make the mass conservation dimensionless, i.e., independent of radial distance, then the velocity scale is 
\begin{align}
v  \propto r^{\alpha-2}.
\end{align}
\citet{Narayan-Yi94} indicates that there are two kinds of solutions possible: (i) advection-dominated accretion flows (ADAF) when $\alpha = 3/2$ and (ii) convection-dominated accretion flows (CDAF) when $\alpha = 1/2$. In the CDAF model, the angular momentum transfers inward via convection with very small viscosity \citep{Narayan-Yi94, Narayan-etal00}.


\label{lastpage}
\end{document}